\documentclass[nofootinbib,preprintnumbers]{revtex4}

\usepackage{graphicx}
\usepackage{dcolumn}
\usepackage{amsmath,amssymb,epsfig}
\usepackage{paralist}
\usepackage{comment}
\usepackage{graphicx}
\usepackage{multirow}
\usepackage{wrapfig}

\usepackage{wasysym}
\newcommand{\mathleftmoon}[0]{{\text{\normalfont\leftmoon}}}

\usepackage{hyperref}

\renewcommand{\vec}[1]{\boldsymbol{\mathrm{#1}}}

\newcommand{\editnote}[2]{}

\allowdisplaybreaks

\begin{document}

\title{Spherical harmonics representation of the gravitational phase shift}

\author{Slava G. Turyshev$^1$, Viktor T. Toth$^2$}

\affiliation{\vskip 3pt$^1$Jet Propulsion Laboratory, California Institute of Technology,\\
4800 Oak Grove Drive, Pasadena, CA 91109-0899, USA}

\affiliation{\vskip 3pt$^2$Ottawa, Ontario K1N 9H5, Canada}

\date{\today}

\begin{abstract}

We investigate the general relativistic phase of an electromagnetic wave as it propagates in the gravitational field of the Earth, which is modeled as an isolated, weakly aspherical gravitating body. We introduce coordinate systems to describe light propagation in the Earth's vicinity along with the relevant coordinate transformations, and discuss the transformations between proper and coordinate times. We represent the Earth's gravitational field using Cartesian symmetric trace-free (STF) mass multipole moments. The light propagation equation is solvable along the trajectory of a light ray to all STF orders $\ell$. Although we focus primarily on the quadrupole ($\ell=2$), octupole ($\ell=3$), and hexadecapole ($\ell=4$) cases, our approach is valid to all orders. We express the STF moments via spherical harmonic coefficients of various degree and order, $C_{\ell k}, S_{\ell k}$. The result is the gravitational phase shift expressed in terms of the spherical harmonics. These results are new. We also consider contributions due to tides and the Earth's rotation. We estimate the characteristic magnitudes of each term of the resulting overall gravitational phase shift. The terms assessed are either large enough to impact current-generation clocks or will become significant as future-generation clocks offer greater sensitivity. Our formulation is useful for many practical and scientific applications, including space-based time and frequency transfers, relativistic geodesy and navigation, as well as quantum communication links and space-based tests of fundamental physics.

\end{abstract}


\maketitle

\section{Introduction}
\label{sec:intro}

The accuracy of time and frequency measurements improved significantly in the recent past. Presently, atomic frequency standards play a vital part in a growing number of technological and scientific endeavors. Atomic clocks based on microwave transitions currently define the Syst\`eme Internationale (the International System of Units, or SI) second and are used extensively for network synchronization and satellite-based navigation systems, such as the Global Positioning System (GPS). Performance of the onboard clocks on global navigational satellite system (GNSS) satellites is improving.  The frequency stability of the current generation of GPS clocks is now better than $10^{-14}$ \cite{Ai-etal:2021,Chen-etal:2021}.

Optical frequency references based on Doppler-free spectroscopy of molecular iodine are seen as a promising candidate for a future GNSS optical clock \cite{Schuldt-etal:2021}. Compact and ruggedized setups have been developed, showing frequency instabilities at the $10^{-15}$ level for averaging times between 1~s and $10^4$~s. Optical atomic clocks have demonstrated stability of $4.8\times 10^{-17}/\sqrt{\tau}$, for an averaging time $\tau$ (in seconds). The ability to resolve frequency shifts below $10^{-18}$ over such short timescales will affect a wide range of applications for clocks in quantum sensing and fundamental physics \cite{Oelker-etal:2019,Schkolnik-etal:2023}.
As a result, future generations of GNSS will benefit from optical technologies. Especially optical clocks could provide backup or completely replace the currently used microwave clocks, potentially improving GNSS position determination thanks to their greater frequency stability. Furthermore, optical clock technologies, in combination with
optical intersatellite links, enable new GNSS architectures, e.g., by the synchronization of distant optical frequency references within the constellation using time and frequency transfer.

As a result, future time and frequency transfer links based on the new generations of clocks will require a general relativistic description of the light propagation in the gravitational field of the extended Earth.  For practical reasons,  the description of the Earth external gravity field is given in terms of spherical harmonics. Until recently, it was sufficient to include the contribution of the gravitational mass monopole represented by the well-known Shapiro time delay. However, with the increased accuracy of modern measurements, accounting for the Earth's quadrupole moment,  $J_2$,  became important \cite{Turyshev:2014dea}. This was accomplished by treating the Earth as a rotating axisymmetric body characterized only by zonal harmonics of even order, yielding the quadrupole phase shift \cite{Klioner:1991SvA}.

If greater accuracy is required, accounting for the quadrupole contribution is not sufficient \cite{Turyshev:2012nw}. It is necessary to account for additional spherical harmonics. For some applications of precision clocks and optical communication links, in addition to $J_2$, the tesseral harmonics of order $\ell=2$ must also be considered, along with additional low-order zonal harmonics of the Earth's gravitational field. Accurate estimates of these quantities are available to high degree and order in the form of modern models of the Earth's gravitational field. For instance, the EGM2008 Earth gravitational model is complete to degree and order 2159 \cite{Pavlis-etal:2012}. However, due to technical reasons, a description of gravitational phase shift in terms of spherical harmonics is not available. This represents an interesting challenge that must be addressed.

Motivated by this challenge, in the present paper we develop a model for proper time to coordinate time transformations and the gravitational phase shift in terms of spherical harmonics that offer the advanced representation of the Earth gravitational field.  We focus on the formulation of a relativistic model for the gravitational phase shift, accurate to $\delta f/f=1\times 10^{-16}$ in frequency stability and $1\times 10^{-12}$ s (i.e., 1~picosecond (ps)) in time resolution. We rely on a previously developed theory of relativistic proper reference frames for a system of extended gravitating bodies and the motion of light and test particles in the vicinity of an extended body \cite{Turyshev-Toth:2013}. We use methods and tools developed for the GRAIL, GRACE-FO and ACES missions \cite{Turyshev:2012nw,Turyshev:2014dea,Turyshev-Toth:2016} as well as the techniques that we recently developed for treating gravitational lensing phenomena by arbitrarily shaped weakly aspherical gravitating bodes \cite{Turyshev-Toth:2021-STF-moments,Turyshev-Toth:2022-platonic}.

This paper is organized as follows: In Sec.~\ref{sec:coord-transform} we introduce coordinate systems, coordinate transformations between various coordinates used for GPS and discuss the transformations between proper time and coordinate time. In Sec.~\ref{sec:light-prop} we discuss light propagation in the vicinity of the extended Earth and derive a general relativistic solution for the phase of an EM wave. For that we represent the Earth's external gravitational field using as set of Cartesian symmetric trace-free (STF) mass multipole moments ${\cal T}^{a_1...a_\ell}$ of various orders $\ell$. In Sec.~\ref{sec:phase-moments}, we study the phase shift introduced by the lowest-order multipole moments, including quadrupole ($\ell=2$), octupole ($\ell=3$), and hexadecapole ($\ell=4$).
We express our results in terms of spherical harmonics of the appropriate degree and order, $C_{\ell k}, S_{\ell k}$.  We also consider contributions due to lunar and solar tides and that of the Earth's rotation.  Using these new results, we evaluate the magnitudes of the relevant terms for various terrestrial GPS applications and consider an update to the light time equation.  We conclude with a set of recommendations and an outlook in Sec.~\ref{sec:sonc}.

For convenience, we put some additional details in appendices: In Appendix~\ref{sec:stf-sph-harm} we discuss the correspondence between the STF mass moments, ${\cal T}^{<ab>}$,  and spherical harmonics, $\big\{C_{2k},S_{2k}\big\}$, and present a practical way to establish the correspondence between them. In Appendix~\ref{sec:cases}, we compute useful relations for deriving contributions from low STF mass moments, including quadrupole, octupole, and hexadecapole. In Appendix~\ref{app:axisymm}, we derive gravitational phase shift for the same orders, but in the specific case of an axisymmetric gravitational field. Finally, in Appendix~\ref{app:analest}, we present analytical estimates for various terms in the gravitational phase shift relevant for terrestrial GPS applications.

\section{Coordinate systems and Proper-to-coordinate time transformations}
\label{sec:coord-transform}

Precision tracking of an Earth orbiting spacecraft relies on three standard coordinate systems: the Geocentric Coordinate Reference System (GCRS), which is centered at the Earth's center of mass and is used to track orbits in the vicinity of the Earth; the Topocentric Coordinate Reference System (TCRS), which is used to provide the positions of objects on the surface of the Earth, such as DSN ground stations; and the Satellite Coordinate Reference System (SCRS), which is needed for proper-to-coordinate time transformations. Definition and properties of TCRS together with useful details on relativistic time-keeping in the solar system are given in \cite{Turyshev-etal:2012}. The SCRS was discussed in \cite{Turyshev-etal:2012} and in \cite{Turyshev:2014dea} in the context of the GRAIL  and  GRACE-Follow-On missions, correspondingly.

Here we investigate a need for an updated formulation for the standard general relativistic models for spacetime coordinates and spacecraft equations of motion \cite{Petit-Luzum:2010} in connection to various terrestrial applications of the GNSS.
We begin our discussion by describing the GCRS, the reference system in which the gravitational field of the Earth is presented using measured spherical harmonic coefficients.

\subsection{Geocentric Coordinate Reference System}
\label{sec:GCRS}

In the vicinity of the Earth, we utilize the standard non-rotating coordinate system known as the Geocentric Coordinate Reference System (GCRS). Centered at the Earth's center of mass, the GCRS is used to track orbits in the vicinity of the Earth \cite{Turyshev:2012nw,Turyshev-Toth:2013}. It is also called the Earth-Centered Earth-Fixed (ECEF) coordinate system \cite{Ashby-2003}. We denote the coordinates in this reference frame as $\{x^m_{\rm E}\}\equiv(x^0=ct, {\vec x})$ and present the metric tensor $g^{\rm E}_{mn}$ in the following form\footnote{The notational conventions employed in this paper are those used in \cite{Landau-Lifshitz:1988}. Letters from the second half of the Latin alphabet, $m, n,...=0...3$ denote spacetime indices. Greek letters $\alpha, \beta,...=1...3$ denote spatial indices. The metric $\gamma_{mn}$ is that of Minkowski spacetime with $\gamma_{mn}={\rm diag}(+1,-1,-1,-1)$ in the Cartesian representation. We employ the Einstein summation convention with indices being lowered or raised using $\gamma_{mn}$. We use powers of $G$ and negative powers of $c$ as bookkeeping devices for order terms. Other notations are explained as they occur.}:
{}
\begin{eqnarray}
g^{\rm E}_{00}&=& 1-\frac{2}{c^2}w_{\rm E}+\frac{2}{c^4}w^2_{\rm E}+O(c^{-6}), \quad
g^{\rm E}_{0\alpha}= -\gamma_{\alpha\lambda}\frac{4}{c^3}w^\lambda_{\rm E}+O(c^{-5}),\quad
g^{\rm E}_{\alpha\beta}= \gamma_{\alpha\beta}+\gamma_{\alpha\beta}\frac{2}{c^2} w_{\rm E}+O(c^{-4}),~~~
\label{eq:gab-E}
\end{eqnarray}
where the scalar gravitational potential $w_{\rm E}$ is formed as a linear superposition of the gravitational potential $U_{\rm E}$ of the isolated Earth and the tidal potential $u^{\tt tidal}_{\rm E}$ produced by all the solar system bodies (excluding the Earth itself), evaluated at the origin of the GCRS:
{}
\begin{eqnarray}
w_{\rm E}(\vec x)&=&U_{\rm E}(\vec x)+ u^{\tt tidal}_{\rm E}(\vec x)+{\cal O}(c^{-3}).~~~
\label{eq:pot_loc-w_0}
\end{eqnarray}

The Earth's gravitational potential is well known and determined by the relativistic mass density inside the Earth, $\sigma(t,{\vec x'})$ (see discussion in \cite{Turyshev:2014dea}). With  $G$ being the universal gravitational constant this potential has the form
{}
\begin{eqnarray}
U_{\rm E}(\vec x)&=&G\int \frac{\sigma(t,{\vec x'})d^3x'}{|{\vec x}-{\vec x'}|}+{\cal O}(c^{-4}).
\label{eq:pot_w_0}
\end{eqnarray}

Outside the Earth ($r>R_\oplus$), the planet's gravitational potential $U_{\rm E}$ may be expanded in terms of spherical harmonics. In this case, at a particular location with spherical coordinates $(r\equiv|{\vec x}|,\psi,\theta)$ (where $\psi$ is the longitude and $\theta$ is the colatitude, which is $0$ at the pole and ${\textstyle\frac{\pi}{2}}$ at the equator) the Earth's potential $U_{\rm E}$ in (\ref{eq:pot_w_0}) is given as
{}
\begin{eqnarray}
U_{\rm E}(\vec x)&=&
\frac{GM_\oplus}{r}\Big\{1+\sum_{\ell=2}^\infty\sum_{k=0}^{+\ell}\Big(\frac{R_\oplus}{r}\Big)^\ell P_{\ell k}(\cos\theta)\big(C_{\ell k}\cos k\psi+S_{\ell k}\sin k\psi\big)\Big\}=\nonumber\\
&=&\frac{GM_\oplus}{r_{\rm }}\Big\{1-\sum_{\ell=2}^\infty\Big(\frac{R_\oplus}{r_{\rm }}\Big)^\ell J_\ell P_{\ell0}(\cos\theta)+\sum_{\ell=2}^\infty\sum_{k=1}^{+\ell}\Big(\frac{R_\oplus}{r}\Big)^\ell P_{\ell k}(\cos\theta)\big(C_{\ell k}\cos k\psi+S_{\ell k}\sin k\psi\big)\Big\},~~~
\label{eq:pot_w_0sh}
\end{eqnarray}
where $M_\oplus$ and $R_\oplus$ are the Earth's mass and equatorial radius, respectively, while $P_{\ell k}$ are the associated Legendre-polynomials \cite{Abramovitz-Stegun:1965}. The values $C_{\ell k}$ and $S_{\ell k}$ are the spherical harmonic coefficients that characterize contributions of the gravitational field of the Earth beyond the monopole potential. Of these, $J_\ell=-C_{\ell 0}$ are the zonal harmonic coefficients. Largest among these is $J_2=1.082635854\times 10^{-3}$, with all other spherical harmonic coefficients at least a factor of $\sim10^3$ times smaller \cite{Kozai:1966,Gasposchkin:1977,Lambeck:1988,JPL-EarthGrav:2021} (see Table~\ref{tab:sp-harmonics} for details).

\begin{table*}[t!]
\vskip-15pt
\caption{Some of the Earth's spherical gravitational coefficients up to degree and order $\ell=4$, with $GM_\oplus=398\,600.4415~{\rm km}^3{\rm s}^{-2}, R_\oplus=6\,378.13630~{\rm km}$ \cite{Tapley-1996,Montenbruck-Gill:2012}. Also, values of some additional lower order zonal harmonics are given as
$C_{50}= 2.28\times 10^{-7}$,
$C_{60}=-5.39\times 10^{-7}$,
$C_{70}= 3.51\times 10^{-7}$,
$C_{80}= 2.03\times 10^{-7}$,
$C_{90}= 1.19\times 10^{-7}$,
$C_{10\,0}=2.48\times 10^{-7}$.
\label{tab:sp-harmonics}}
\begin{tabular}{|c| c c c c c |}\hline
$C_{\ell k}$  & $k=0$  &  1 & 2  & 3& 4 \\\hline
$\ell=0$ & +1& &  & &\\
1 & 0.00 & 0.00&  & & \\
2 & $-1.0826359\times 10^{-3}$ & 0.00& ${+1.5745}\times 10^{-6}$ &   & \\
3 & $\phantom{000}{+2.5324}\times 10^{-6}$ & $+2.1928\times 10^{-6}$ &
$\phantom{0}{+3.090}\times 10^{-7}$ & ${+1.006}\times 10^{-7}$&
 \\
4 & $\phantom{000}{+1.6193}\times 10^{-6}$ & $\phantom{0}{-5.087}\times 10^{-7}$ &
$\phantom{00}{+7.84}\times 10^{-8}$ & $\phantom{0}{+5.92}\times 10^{-8}$ &
$-3.98\times 10^{-9}$
 \\
\hline
\hline
$S_{\ell k}$  & $k=0$  &  1 & 2  & 3& 4 \\\hline
$\ell=0$ & 0.00& &  & &\\
1 & 0.00 & 0.00&  & & \\
2 & 0.00 & $\phantom{0}{+1.54}\times 10^{-9}$ & $-9.039\times 10^{-7}$ &   & \\
3 & 0.00 & $+2.680\times 10^{-7}$ &
$-2.114\times 10^{-7}$ & $+1.972\times 10^{-7}$&
 \\
4 & 0.00 & $-4.494\times 10^{-7}$ &
$+1.482\times 10^{-7}$ & $\phantom{0}{+1.20}\times 10^{-8}$ &
$+6.53\times 10^{-9}$
 \\
\hline
\end{tabular}
\end{table*}

Insofar as the tidal potential $u^{\tt tidal}_{\rm E}$ is concerned, for GPS observables it is sufficient to keep only its Newtonian contribution (primarily due to the Sun and the Moon) which can be given as usual:
{}
\begin{eqnarray}
u^{\tt tidal}_{\rm E}(\vec x)&=&\sum_{b\not={\rm E}}\Big(U_b({\vec r}^{}_{b{\rm E}}+\vec{x})-U_b({\vec r}^{}_{b{\rm E}}) - \vec{x}\cdot {\vec \nabla} U_b ({\vec r}^{}_{b{\rm E}})\Big)\simeq\sum_{b\not={\rm E}}\frac{GM_b}{2r^3_{b{\rm E}}}\Big(3(\vec{n}^{}_{b{\rm E}}\cdot\vec{x})^2-\vec{x}^2\Big)+
{\cal O}\Big(\frac{x^3}{r^4_{b{\rm E}}},c^{-2}\Big),
\label{eq:u-tidal-E}
\end{eqnarray}
where $U_b$ is the Newtonian gravitational potential of body $b$, $\vec{r}_{b{\rm E}}$ is the vector connecting the center of mass of body $b$ with that of the Earth, and $\vec \nabla U_b$ denotes the gradient of the potential. We present only the largest term in the tidal potential, which is of ${\cal O}(r^{-3}_{b{\rm E}})$; however, using the explicit form of this potential on the left side of  Eq.~(\ref{eq:u-tidal-E}), one can easily evaluate this expression to any order needed to solve a particular problem.

Finally, the contribution of the Earth's rotation is captured by the vector harmonic  potential, $w^\alpha_{\rm E}$, defined as:
{}
\begin{eqnarray}
w^\alpha_{\rm E}(\vec x)
&=&G\int \frac{\sigma^\alpha(t,\vec{x}')d^3x'}{|\vec{x}-\vec{x}'|}+{\cal O}(c^{-2})=-\frac{GM_{\rm E}}{2r^3}[{\vec x}\times{\vec S}_\oplus]^\alpha+{\cal O}(r^{-3}, c^{-2}),
\label{eq:pot_loc-w_a+}
\end{eqnarray}
where $\sigma^\alpha(t,\vec{x}')$ is the relativistic current density of the matter distribution inside the rotating Earth. Also, in (\ref{eq:pot_loc-w_a+}) we explicitly account only for the largest rotational moment,  ${\vec S}_\oplus\simeq 9.8\times 10^8~{\rm m}^2$/s, which is the Earth's spin moment (angular momentum per unit of mass). Contributions of other vector harmonics due to the rotating Earth are negligible (however, if needed, they may be easily incorporated using the approach developed in this paper).

The metric tensor (\ref{eq:gab-E}) with the gravitational potentials (\ref{eq:pot_loc-w_0})--(\ref{eq:pot_loc-w_a+}) represents spacetime in the GCRS, which we choose to formulate the relativistic model for timing and frequency observables. Further technical details on the  formulation of the GCRS are given in \cite{Soffel:2003cr,Turyshev:2012nw,Turyshev-Toth:2013,Turyshev:2014dea}.

Refs.~\cite{Turyshev-Toth:2013,Turyshev-etal:2012}, show that the proper time, $\tau$, read by a clock in harmonic coordinates of the SCRS denoted here  by $\{y^m\}=\{cy^0, {\vec y}\}$, and coordinates of GCRS $\{x^m\}=\{ct, {\vec x}\}$, to sufficient accuracy, are  given by
{}
\begin{eqnarray}
\frac{d\tau}{dt}&=& 1-c^{-2}\Big[{\textstyle\frac{1}{2}}{\vec v}^2+
U_{\rm E}+u^{\tt tidal}_{\rm E}
\Big]+{\cal O}({c^{-4}}),
\label{eq:trans++y0}
\end{eqnarray}
where ${\vec v}$ is the velocity of the transmitter (or receiver) in GCRS and $U_{\rm E}$ is the Newtonian gravitational potential of the Earth (\ref{eq:pot_w_0}), evaluated at that location. Contribution of the tidal potential, $u^{\tt tidal}_{\rm E}$, varies depending on the distance from the Earth; it reaches the value above $\sim 1.71\times 10^{-15}$ for geostationary orbits.
Note that this expression does not contain the term due to  the Earth's rotation, $w^\alpha_{\rm E}$, from (\ref{eq:pot_loc-w_a+}), as its contribution is much below 1 ps (as estimated by (\ref{eq:phiS-ph-est})).
The $c^{-4}$ terms in Eq.~(\ref{eq:trans++y0}) are of ${\cal O}(v^4/c^4)\simeq 10^{-19}$ and are negligible for our purposes. For a complete post-Newtonian form of these transformations, including $c^{-4}$ terms, and their explicit derivation, consult Ref.~\cite{Turyshev-Toth:2013}.

The numerical applications made in here concern time and frequency transfer from between a GPS  and LEO spacecraft orbiting the Earth at the altitudes $h_{\tt GPS} = 20,200$~km and $h_{\tt LEO} = 200$~km to a ground station located at `C'.  As mentioned earlier, we consider  experimental uncertainties $1 \times 10^{-16}$ for frequency transfer and  at the level of 1~ps for time transfer.

\subsection{Topocentric Coordinate Reference System (TCRS): proper and coordinate times}
\label{sec:TCRS}

First, we consider a ground-based receiver located at GCRS coordinates
$\{x^m_C\}\equiv(cx^0_{\rm C},{\vec x}_{\rm C})$.
The proper time $\tau_{\rm C}$, kept by a clock located at the GCRS coordinate position ${\vec R}_{\rm C}(t)$, and moving with the coordinate velocity ${\vec v}_{\rm C} = d{\vec x}_{\rm C}/dt=[{\boldsymbol\omega}^{}_{\rm E}\times{\vec R}_{\rm C}]$, where ${\boldsymbol\omega}_{\rm E}$ is the angular rotational velocity of the Earth at $C$, is determined by
{}
\begin{equation}
\frac{d\tau_{\rm C}}{dt}= 1-\frac{1}{c^2}\Big[{\textstyle\frac{1}{2}}\vec{v}_{\rm C}^2+U_{\rm E}({\vec x}_{\rm C})+ \sum_{b\not={\rm E}}\frac{GM_b}{2r^3_{b{\rm E}}}\Big(3(\vec{n}_{b{\rm E}}\cdot\vec{x}_{\rm C})^2-\vec{x}_{\rm C}^2\Big)\Big]+{\cal O}({x^3_{\rm C},c^{-4}}),
\label{eq:proper-t-C-comp}
\end{equation}
where $\vec{n}_{b{\rm E}}$ is a unit spatial vector in the body-Earth direction, i.e., $\vec{n}_{b{\rm E}}=\vec{r}_{b{\rm E}}/|\vec{r}_{b{\rm E}}|$, where $\vec{r}_{b{\rm E}}$ is the vector connecting body $b$ with the Earth.

The first two terms in (\ref{eq:proper-t-C-comp}) are due to the  geocentric velocity of a ground station and the Newtonian potential at its location. Assuming a uniform diurnal rotation of the Earth, so that ${\textstyle\frac{1}{2}}\vec{v}_{\rm C}^2={\textstyle\frac{1}{2}}\omega_\oplus^2 R^2_{\rm C}(\theta)\sin^2\theta$, we evaluate the magnitudes of the largest contributions produced by these terms, evaluated at the Earth's equator $R_{\rm C}(\frac{\pi}{2})=R_\oplus$:
{}
\begin{eqnarray}
c^{-2}{\textstyle\frac{1}{2}}\vec{v}_{\rm C}^2&=& \frac{1}{2c^2} \omega_\oplus^2 R^2_\oplus
\lesssim
1.20 \times 10^{-12},
\label{eq:tau-vel}
\qquad
c^{-2}U_{\rm E} = \frac{1}{c^2}\frac{GM_\oplus}{R_\oplus} \lesssim
6.95 \times 10^{-10}.
\label{eq:tau-U}
\end{eqnarray}
Thus, both of these terms are very large and must be kept in the model. In addition, as we will see below, one would have to account for several terms in the spherical harmonics expansion of the Earth gravity potential.

The last term within the square brackets in Eq.~(\ref{eq:proper-t-C-comp}) is the sum of the Newtonian tides due to other bodies (mainly the Sun and the Moon) at the clock location ${\vec x}_{\rm C}$. These terms are small for the ground station, being of order
{}
\begin{eqnarray}
c^{-2}u^{\tt tidal}_{\rm E\odot}
&\simeq& \frac{GM_\odot R_\oplus^2}{2 {\rm AU}^3c^2}\Big(3(\vec{n}_{\odot{\rm E}}\cdot\vec{n}_{\rm C})^2-1\Big)\lesssim
1.79 \times 10^{-17},
\label{eq:tidalTS}\\
c^{-2}u^{\tt tidal}_{\rm E{\mathleftmoon}}
&\simeq& \frac{Gm_\mathleftmoon R_\oplus^2}{2 r^3_{\mathleftmoon\hskip -1pt{\rm  E}}c^2}\Big(3(\vec{n}_{{\mathleftmoon\hskip -1pt{\rm  E}}}\cdot\vec{n}_{\rm C})^2-1\Big)\lesssim
4.90 \times 10^{-17},
\label{eq:tidalTm1}
\end{eqnarray}
where for the Moon we used its distance from the Earth at the perigee\footnote{\url{https://en.wikipedia.org/wiki/Lunar_distance_(astronomy)}} of $r_{\mathleftmoon\hskip -1pt{\rm  E}}=356,500$~km.
Thus, both largest tidal contributions are currently negligible as far as the  definition of the TCRS for GPS is concerned.

Therefore, at the accuracy required for GPS, it is sufficient to keep only the first two terms in Eq.~(\ref{eq:proper-t-C-comp}) when defining the relationship between the proper time $\tau_{\rm C}$ and the coordinate time ${\rm TCG}=t$ (temps-coordonn\'ee g\'eocentrique). In other words, Eq.~(\ref{eq:proper-t-C-comp}) yields the differential equation that relates the rate of the proper $\tau_{\rm C}$ time, as measured by an receiver's clock on the surface of the Earth, so that ${\textstyle\frac{1}{2}}\vec{v}_{\rm C}^2={\textstyle\frac{1}{2}}\omega_\oplus^2 R^2_{\rm C}(\theta)\sin^2\theta$, to the time in GCRS, $t$,
{}
\begin{equation}
\frac{d\tau_{\rm C}}{dt} = 1 -\frac{1}{c^2}\Big[
{\textstyle\frac{1}{2}}\vec{v}_{\rm C}^2
+ U_{\rm E}({\vec x}_{\rm C})\Big]+{\cal O}(c^{-4}).
\label{eq:proper-coord-t+}
\end{equation}
At the level of accuracy required for GPS, it is important to account in Eq.~(\ref{eq:proper-coord-t+}) for the oblateness (nonsphericity) of the Earth's Newtonian potential, which is given in the form of Eq.~(\ref{eq:pot_w_0}). In fact, when we model the Earth's gravity potential, we need to take into account quadrupole and higher moments, time-dependent terms due to tides, as well as the tidal displacement of the ground-based receiver. Substituting in Eq.~(\ref{eq:proper-coord-t+}) potential $U_{\rm E}$ from (\ref{eq:pot_w_0sh}), evaluated at the Earth's equator $r=R_{\rm C}(\frac{\pi}{2})=R_\oplus$, we have:
{}
\begin{eqnarray}
\frac{d\tau_{\rm C}}{dt}&=&
1-\frac{1}{c^2}\Big[{\textstyle\frac{1}{2}}\omega_\oplus^2 R^2_\oplus+
\frac{GM_\oplus}{R_\oplus}\Big(1-\sum_{\ell=2}^\infty J_\ell P_{\ell0}(0)+\sum_{\ell=2}^\infty\sum_{k=1}^{+\ell}P_{\ell k}(0)(C_{\ell k}\cos k\phi+S_{\ell k}\sin k\phi)\Big)
\Big].~~~~~
\label{eq:prop-coord-time1}
\end{eqnarray}

The largest contribution to $d\tau_{\rm C}/dt$, of course, comes from the velocity and mass monopole terms, which are estimated to produce an effect of the order of $c^{-2}({\textstyle\frac{1}{2}}\omega_\oplus^2 R^2_\oplus+G M_\oplus/R_\oplus)\sim 6.97\times 10^{-10}$. The quadrupole term produces a contribution of the order of $c^{-2}G M_\oplus J_2/(2R_\oplus)\sim 3.76\times 10^{-13}$.
Contributions of other zonal harmonics ranging from $-c^{-2}3G M_\oplus J_4/(8R_\oplus)\sim 4.22\times 10^{-16}$ (from $J_4$) to $c^{-2}5G M_\oplus J_6/(16R_\oplus)\sim 1.04\times 10^{-16}$ (from $J_6$) and to $c^{-2}35G M_\oplus J_8/(128R_\oplus)\sim 1.01\times 10^{-16}$ (from $J_8$).

We also need to include contributions from the leading $\ell=2$ coefficients $C_{22}$ and $S_{22}$, which are of order $c^{-2}(G M_\oplus /R_\oplus)P_{22}(0)C_{22}\cos 2\phi \sim 3.28\times 10^{-15}\cos 2\phi $ and $c^{-2}(G M_\oplus /R_\oplus)P_{22}(0)S_{22}\sin 2\phi \sim 1.89\times 10^{-15}\cos 2\phi $. Some of the low-order tesseral harmonics, $C_{\ell k}$ and $S_{\ell k}$, are also responsible for the terms of the order of $\sim10^{-16}$, and thus, they should also  be included, up to at least $\ell=8$. Although individual contributions of these and other terms are quite small, their cumulative effect may be noticeable even at the level of up to $1\times 10^{-15}$. (This is especially important for the ACES mission on the ISS, which will operate clocks accurate to $1\times 10^{-16}$ in low Earth's orbit with altitude of $\sim$400 km.) However, the constant rate is typically absorbed for each clock during its  synchronization with the network of clocks, leaving only periodic terms, which are uncertain at the level of $\sim10^{-17}$. Therefore, keeping only the leading terms, Eq.~(\ref{eq:prop-coord-time1}) takes the form:
{}
\begin{eqnarray}
\frac{d\tau_{\rm C}}{dt}=
1-\frac{1}{c^2}\Big[{\textstyle\frac{1}{2}}\omega_\oplus^2 R^2_\oplus &+&
\frac{GM_\oplus}{R_\oplus}\Big(1+{\textstyle\frac{1}{2}}J_2-{\textstyle\frac{3}{8}}J_4+{\textstyle\frac{5}{16}}J_6-{\textstyle\frac{35}{128}}J_8+P_{22}(0)\big(C_{22}\cos 2\phi+S_{22}\sin 2\phi\big)+\nonumber\\
&+&
\sum_{\ell=3}^{8}\sum_{k=1}^{+\ell}P_{\ell k}(0)(C_{\ell k}\cos k\phi+S_{\ell k}\sin k\phi)\Big)
\Big]+{\cal O}(5.83\times 10^{-17}),
\label{eq:prop-coord-time-J2J8}
\end{eqnarray}
where the error bound is set by the contribution from $J_{10}$ and some of the low-order tesseral harmonics. Keeping only the $\ell=2$ terms, this expression can be truncated to
{}
\begin{eqnarray}
\frac{d\tau_{\rm C}}{dt}&=&
1-\frac{1}{c^2}\Big[{\textstyle\frac{1}{2}}\omega_\oplus^2 R^2_\oplus+
\frac{GM_\oplus}{R_\oplus}\Big(1+{\textstyle\frac{1}{2}}J_2+3(C_{22}\cos 2\phi+S_{22}\sin 2\phi)\Big)
\Big]+{\cal O}(2.28\times 10^{-15}\cos\phi),
\label{eq:prop-coord-time-J2}
\end{eqnarray}
where the error bound is set by the contribution from $C_{31}$ coefficient (see Table~\ref{tab:sp-harmonics}).

In the past, clock synchronization relied on a definition of Earth's geoid. In this case, for a clock situated on the surface of the Earth, the relativistic correction term appearing in Eq.~(\ref{eq:proper-coord-t+}) is given at the needed precision by
{}
\begin{equation}
\frac{{\vec v}^2_{\rm C}}{2} + U_{\rm E}({\vec x}_{\rm C})= W_0-\int_0^{h_{\rm C}}\hskip -7pt g\,dh,
\label{eq:proper-coord-t++}
\end{equation}
where $W_0 = 6.2636856\times 10^7$ m$^2$/s$^2$ is the Earth's potential at the reference geoid while $g$ denotes the Earth's acceleration (gravitational plus centrifugal), and where $h_{\rm C}$ is the clock's altitude above the reference geoid. However, as we mention above, this definition of terrestrial time is problematic when considering accuracies below $10^{-17}$ because of the uncertainties in the realization of the geoid at this level.

In practice, time measurements are based on averages of clock and frequency measurements on the Earth surface \cite{Moyer:2003}. Therefore it was decided to dissociate the definition of TT from the geoid while maintaining continuity with the previous definition. For this purpose, the time coordinate called Terrestrial Time (TT) is defined. TT is related to TCG$=t$ linearly by definition:
{}
\begin{equation}
\frac{dt_{\rm TT}}{dt}=1-L_{\rm G}.
\end{equation}
IAU Resolution B1.9 (2000) turned $L_G$ into a defining constant with its value fixed to $L_{\rm G}=6.969290134\times 10^{-10}$.

This definition accounts for the secular term due to the Earth's potential when converting between TCG and the time measured by an idealized clock on the Earth geoid \cite{Moyer:2003,Klioner:2008,Klioner-etal:2010,Kopeikin-book-2011}. Using Eq.~(\ref{eq:proper-coord-t+}), we also have
{}
\begin{equation}
\frac{d\tau_{\rm C}}{dt_{\rm TT}}=\frac{d\tau_{\rm C}}{dt}\frac{dt}{dt_{\rm TT}}=1+L_{\rm G}-\frac{1}{c^2}\Big[{\textstyle\frac{1}{2}}{\vec v}^2_{\rm C}
+ U_{\rm E}({\vec x}_{\rm C})\Big]+{\cal O}(c^{-4}).
\label{eq:synch}
\end{equation}
Clearly, if the target clock synchronization is of the order of  $\sim 10^{-15}$, the definition (\ref{eq:synch}) is rather clean with just a few terms given by (\ref{eq:prop-coord-time-J2}). This expression quickly becomes rather messy if a more precise synchronization is desired.

\subsection{Satellite Coordinate Reference System (SCRS)}
\label{sec:SCRS}

We can also determine the differential equation that relates the rate of the spacecraft proper $\tau_{\rm A}$ time, as measured by an on-board clock in Earth orbit with GCRS coordinates $\{y^m_{\rm A}\}=\{cy^0_{\rm A}, {\vec y}_{\rm A}\}$, to the time in GCRS, ${\rm TCG}=t$. Substituting in (\ref{eq:trans++y0}) potential $U_{\rm E}$ from (\ref{eq:pot_w_0sh}) and largest term for the tidael potential $u^{\tt tidal}_{\rm E}$ from (\ref{eq:u-tidal-E}), we have:
{}
\begin{eqnarray}
\frac{d\tau_{\rm A}}{dt}=
1-\frac{1}{c^2}\Big[{\textstyle\frac{1}{2}}{\vec v}^2_{\rm A} &+&
\frac{GM_{\rm E}}{r_{\rm A}}\Big(1-\sum_{\ell=2}^\infty\Big(\frac{R_{\rm E}}{r_{\rm A}}\Big)^\ell J_\ell P_{\ell0}(\cos\theta) + \sum_{\ell=2}^\infty\sum_{k=1}^{+\ell}\Big(\frac{R_{\rm E}}{r_{\rm A}}\Big)^\ell P_{\ell k}(\cos\theta)(C_{\ell k}\cos k\phi+S_{\ell k}\sin k\phi)\Big)+\nonumber\\
&+&
\sum_{b\not={\rm E}}\frac{GM_b}{2r^3_{b{\rm E}}}\Big(3(\vec{n}_{b{\rm E}}\cdot\vec{y}_{\rm A})^2-\vec{y}_{\rm A}^2\Big)\Big]+{\cal O}({y^3_{\rm A}/r^4_{b{\rm E}},c^{-4}}).
\label{eq:prop-coord-time}
\end{eqnarray}
We will evaluate the magnitude of the terms in this equations for two different orbits: LEO with altitude 200 km and GPS with altitude 20,200 km. We will keep in mind the anticipated frequency stability of $\Delta f/f= 1\times 10^{-16}$. We will use these numbers to evaluate the terms in (\ref{eq:prop-coord-time}).

\subsubsection{LEO clock: proper-to-coordinate time transformation}

In the case of the LEO orbit with altitude of $h_{\tt LEO}=200$~km, the largest contribution to $(d\tau_{\rm A}/dt)_{\tt LEO}$, of course, comes from the velocity and mass monopole terms, which are estimated to produce an effect of the order of $c^{-2}({\textstyle\frac{1}{2}}{\vec v}^2_{\rm A}+G M_{\rm E}/(R_\oplus +h_{\tt LEO})=c^{-2}3G M_{\rm E}/2(R_\oplus +h_{\tt LEO})\sim 1.01\times 10^{-9}$. Because of the larger contribution, this term is 1.45 times larger than for a receiver on the ground (see estimate presented above, just after Eq.~(\ref{eq:prop-coord-time1}).)

For the chosen LEO orbit, the quadrupole term produces contribution of the order of
\begin{eqnarray}
\frac{G M_\oplus R_\oplus^2}{c^2 (R_\oplus+h_{\tt LEO})^3}J_2P_{20}(\cos\theta)\lesssim 6.86\times 10^{-13},
\end{eqnarray}
which is large enough to be included in the model.
Contributions of other zonal harmonics are estimated as
{}
\begin{eqnarray}
\frac{G M_\oplus R_\oplus^3}{c^2 (R_\oplus+h_{\tt LEO})^4}J_3 P_{30}(\cos\theta)\lesssim 1.56\times 10^{-15}, ~~~~~~~~
\frac{G M_\oplus R_\oplus^4}{c^2 (R_\oplus+h_{\tt LEO})^5}J_4P_{40}(\cos\theta)\lesssim 9.65\times 10^{-16}, \\
\frac{G M_\oplus R_\oplus^5}{c^2 (R_\oplus+h_{\tt LEO})^6}J_5 P_{50}(\cos\theta)\lesssim 1.21\times 10^{-16}, ~~~~~~~~
\frac{G M_\oplus R_\oplus^6}{c^2 (R_\oplus+h_{\tt LEO})^7}J_6P_{40}(\cos\theta)\lesssim 3.03\times 10^{-16},\\
\frac{G M_\oplus R_\oplus^7}{c^2 (R_\oplus+h_{\tt LEO})^8}J_7 P_{70}(\cos\theta)\lesssim 2.71\times 10^{-16}, ~~~~~~~~
\frac{G M_\oplus R_\oplus^8}{c^2 (R_\oplus+h_{\tt LEO})^9}J_8P_{80}(\cos\theta)\lesssim 1.51\times 10^{-16},\\
\frac{G M_\oplus R_\oplus^9}{c^2 (R_\oplus+h_{\tt LEO})^{10}}J_9 P_{90}(\cos\theta)\lesssim 8.04\times 10^{-17}, ~~~~
\frac{G M_\oplus R_\oplus^{10}}{c^2 (R_\oplus+h_{\tt LEO})^{11}}J_{10}P_{10\,0}(\cos\theta)\lesssim 2.28\times 10^{-16}.
\end{eqnarray}
Note that to develop the estimates above we used approximated $P_{\ell0}$ by their largest value at $\theta=0$. In reality, these polynomials rarely take that value and thus $P_{\ell0}$ are much less than 1. On the other hand, although the contributions from $J_7,J_8,J_9,J_{10}$ are all on the order of 1 to 2 parts in $10^{-16}$, their cumulative effect may exceed the threshold of $10^{-16}$. Nevertheless, we recommend keeping only the contributions from $J_2,J_3$ and $J_4$.

The situation with tesseral harmonics is a bit more complicated as many of them produce contributions on the order of a few parts in $10^{-16}$. The largest among these are those due to $C_{22}$ and $S_{22}$:
{}
\begin{eqnarray}
\frac{G M_\oplus R_\oplus^2}{c^2 (R_\oplus+h_{\tt LEO})^3}P_{22}(\cos\theta)\Big\{C_{22}\cos 2\phi,~S_{22}\sin 2\phi \Big\} &\lesssim& \Big\{9.98\times 10^{-16}\cos 2\phi,  ~5.73\times 10^{-16}\sin 2\phi\Big\}.
\end{eqnarray}

The contributions from the terms of $\ell=3$ are
{}
\begin{eqnarray}
\frac{G M_\oplus R_\oplus^3}{c^2 (R_\oplus+h_{\tt LEO})^4}P_{31}(\cos\theta)\Big\{C_{31}\cos \phi, ~
S_{31}\sin \phi  \Big\}&\lesssim& \Big\{1.35\times 10^{-15}\cos \phi, ~ 1.65\times 10^{-16}\sin \phi\Big\}, \\
\frac{G M_\oplus R_\oplus^3}{c^2 (R_\oplus+h_{\tt LEO})^4}P_{32}(\cos\theta)\Big\{C_{32}\cos 2\phi,~
S_{32}\sin 2\phi \Big\}&\lesssim& \Big\{1.90\times 10^{-16}\cos2 \phi, ~ 1.30\times 10^{-16}\sin 2\phi\Big\},
\label{eq:cont-32}\\
\frac{G M_\oplus R_\oplus^3}{c^2 (R_\oplus+h_{\tt LEO})^4}P_{33}(\cos\theta)\Big\{C_{33}\cos 3\phi,~S_{33}\sin 3\phi \Big\} &\lesssim& \Big\{6.10\times 10^{-17}\cos 3\phi, ~ 1.21\times 10^{-16}\sin 3\phi\Big\}.
\label{eq:cont-33}
\end{eqnarray}

The contributions from the terms of $\ell=4$ are
{}
\begin{eqnarray}
\frac{G M_\oplus R_\oplus^4}{c^2 (R_\oplus+h_{\tt LEO})^5}P_{41}(\cos\theta)\Big\{C_{41}\cos \phi,~S_{41}\sin \phi\Big\} &\lesssim& \Big\{3.03\times 10^{-16}\cos \phi, ~ 2.67\times 10^{-16}\sin \phi\Big\}, \\
\frac{G M_\oplus R_\oplus^4}{c^2 (R_\oplus+h_{\tt LEO})^5}P_{42}(\cos\theta)\Big\{C_{42}\cos 2\phi, ~S_{42}\sin 2\phi \Big\} &\lesssim& \Big\{4.66\times 10^{-17}\cos2 \phi, ~ 8.82\times 10^{-17}\sin 2\phi\Big\}, \\
\frac{G M_\oplus R_\oplus^4}{c^2 (R_\oplus+h_{\tt LEO})^5}P_{43}(\cos\theta)\Big\{C_{43}\cos 3\phi,~ S_{43}\sin 3\phi\Big\} &\lesssim& \Big\{3.53\times 10^{-17}\cos 3\phi, ~ 7.16\times 10^{-18}\sin 3\phi\Big\},
\label{eq:cont-43}\\
\frac{G M_\oplus R_\oplus^4}{c^2 (R_\oplus+h_{\tt LEO})^5}P_{44}(\cos\theta)\Big\{C_{44}\cos 4\phi,~S_{44}\sin 4\phi\Big\} &\lesssim& \Big\{2.37\times 10^{-18}\cos 4\phi, ~ 3.88\times 10^{-18}\sin 4\phi\Big\}.
\end{eqnarray}

Although their individual contributions are quite small, the cumulative effect of these terms may easily reach the level of $\epsilon_{{\tt LEO}0}\approx 1.33\times 10^{-15}$. The constant rate $\epsilon_{{\tt LEO}0}$ would likely be absorbed in other terms during clock synchronization. What is important is the variability in  the entire error term $\epsilon_{\tt LEO}(t)=\epsilon_{{\tt LEO}0}+\delta\epsilon_{\tt LEO}(t)$, where the amplitude of the variable term $\delta\epsilon_{\tt LEO}(t)$ is due to seasonal changes in the Earth hydrosphere, crust, etc. and is expected to be of the order of  $\delta\epsilon_{\tt LEO}(t)\sim 3\times 10^{-17}$, resulting in the ultimate uncertainty in ${d\tau_{\rm A}}/{dt}$ at that level.

The last term within the square brackets in Eq.~(\ref{eq:prop-coord-time}) is the sum of the Newtonian tides due to the Sun and the Moon at the location of a clock in an Earth orbit. For a chosen LEO orbit, these terms are small:
{}
\begin{eqnarray}
c^{-2}u^{\tt tidal}_{\rm E\odot}
&\simeq& \frac{GM_\odot (R_\oplus+h_{\tt LEO})^2}{2 {\rm AU}^3c^2}\Big(3(\vec{n}_{\odot{\rm E}}\cdot\vec{n}_{\rm C})^2-1\Big)\lesssim
1.91 \times 10^{-17}, \\
c^{-2}u^{\tt tidal}_{\rm E{\mathleftmoon}}
&\simeq& \frac{Gm_\mathleftmoon (R_\oplus+h_{\tt LEO})^2}{2 r^3_{\mathleftmoon\hskip -1pt{\rm  E}}c^2}\Big(3(\vec{n}_{{\mathleftmoon\hskip -1pt{\rm  E}}}\cdot\vec{n}_{\rm C})^2-1\Big)\, \lesssim
5.21 \times 10^{-17},
\label{eq:tidalT}
\end{eqnarray}
and, thus, may be neglected.

Therefore, for LEO in equation (\ref{eq:prop-coord-time}) we must keep the following terms:
{}
\begin{eqnarray}
\frac{d\tau_{\rm A}}{dt}&=&
1-\frac{1}{c^2}\Big[\frac{{\vec v}^2_{\rm A}}{2}+
\frac{GM_{\rm E}}{r_{\rm A}}\Big(1-\sum_{\ell=2}^4\Big(\frac{R_{\rm E}}{r_{\rm A}}\Big)^\ell J_\ell P_{\ell0}(\cos\theta)+
\Big(\frac{R_{\rm E}}{r_{\rm A}}\Big)^2 P_{22}(\cos\theta)(C_{22}\cos 2\phi+S_{22}\sin 2\phi)+\nonumber\\
&&\hskip 50pt +\,
\Big(\frac{R_{\rm E}}{r_{\rm A}}\Big)^3 P_{31}(\cos\theta)(C_{31}\cos\phi+S_{31}\sin \phi)
\Big)\Big]+
{\cal O}(3.03\times 10^{-16}),~~~~~
\label{eq:prop-coord-time-J2-LEO}
\end{eqnarray}
where the size of the error term is set by $J_6$ and cumulative contribution of higher $\ell \geq 3$ gravitational harmonics. We truncated the error term at that level knowing there are other significant terms may be present (as evidenced by (\ref{eq:cont-32})--(\ref{eq:cont-43})) that may lead to a complicated modeling of the gravitational background -- the challenge that can be avoided with a higher spacecraft altitude.

Clearly, such a modeling accuracy is at the limit anticipated for the Deep Space Atomic Clocks (DSAC) with frequency stability at the level of $1\times 10^{-16}$ at 1 day (if DSAC is placed on a LEO). However, it is already insufficient for ESA's ACES mission on the International Space Station and DSAC, where clock accuracy is expected to be at the level of $1\times 10^{-16}$ at 1 day.  Furthermore, there are plans to fly an optical interferometer and highly-stable optical link as a part of the STE-QUEST mission \cite{Ahlers-etal:2022} and also a Space Optical Clock (SOC) mission \cite{Origlia-etal:2016} with frequency stability below $1\times 10^{-17}$, for which a new and more detailed model may be required.

\subsubsection{GPS clock: proper-to-coordinate time transformation}

Now we consider GPS orbits with $h_{\tt GPS}=20,200$~km and estimate the sizes of all the terms entering (\ref{eq:prop-coord-time}). Clearly, the largest contribution to $(d\tau_{\rm A}/dt)_{\tt GPS}$, of course, comes from the velocity and mass monopole terms, which are estimated to produce an effect of the order of
{}
\begin{eqnarray}
\frac{1}{c^2}
\Big({\textstyle\frac{1}{2}}{\vec v}^2_{\tt GPS}+\frac{G M_\oplus }{R_\oplus+h_{\tt GPS}}\Big)=\frac{3G M_\oplus }{2c^2(R_\oplus+h_{\tt GPS})}\sim 2.50\times 10^{-10},
\end{eqnarray}
or almost 4 times smaller than for LEO, but still providing a rather large contribution.

The quadrupole term produces contribution of the order of
{}
\begin{eqnarray}
\frac{G M_\oplus R_\oplus^2}{c^2 (R_\oplus+h_{\tt GPS})^3}J_2 P_{20}(\cos\theta)&\lesssim& 1.04\times 10^{-14},
\end{eqnarray}
which is significant. Contributions of tesseral harmonics with $\ell=2$ are estimated as
{}
\begin{eqnarray}
\frac{G M_\oplus R_\oplus^2}{c^2 (R_\oplus+h_{\tt GPS})^3}P_{22}(\cos\theta)\Big\{C_{22}\cos 2\phi ,~S_{22}\sin 2\phi \Big\}&\lesssim& \Big\{1.51\times 10^{-17}\cos 2\phi, ~ 8.69\times 10^{-18}\cos 2\phi\Big\},
\end{eqnarray}
and are small to provide a measurable contribution.
Contributions of other zonal harmonics ranging from $\lesssim 5.85\times 10^{-18}$ (from $J_3$), to $\lesssim 9.08\times 10^{-19}$ (from $J_4$),
to $\lesssim 2.79\times 10^{-20}$ (from $J_5$),
 to $\sim 1.72\times 10^{-20}$ (from $J_6$), which are clearly negligible and may be omitted.

Evaluating the contribution due to Newtonian tides in Eq.~(\ref{eq:prop-coord-time}) for a GPS orbit, we find the following:
{}
\begin{eqnarray}
c^{-2}u^{\tt tidal}_{\rm E\odot}
&\simeq& \frac{GM_\odot (R_\oplus+h_{\tt GPS})^2}{2 {\rm AU}^3c^2}\Big(3(\vec{n}_{\odot{\rm E}}\cdot\vec{n}_{\rm C})^2-1\Big)\lesssim
3.11 \times 10^{-16},
\label{eq:tidalT-Sun}\\
c^{-2}u^{\tt tidal}_{\rm E \mathleftmoon}
&\simeq& \frac{Gm_\mathleftmoon (R_\oplus+h_{\tt GPS})^2}{2 r^{*3}_{\mathleftmoon\hskip -1pt{\rm  E}}c^2}\Big(3(\vec{n}_{{\mathleftmoon\hskip -1pt{\rm  E}}}\cdot\vec{n}_{\rm C})^2-1\Big)\lesssim
8.50 \times 10^{-16},
\label{eq:tidalT-moon}
\end{eqnarray}
where $r^*_{\mathleftmoon\hskip -1pt{\rm  E}}$ is the Earth-moon distance at the shortest perigee of the lunar orbit.  Consequently, both of these terms must be kept in any model that aims to offer a frequency stability of ${\cal O}(10^{-16})$.

Therefore, for the case of atomic clocks on GPS spacecraft in the (\ref{eq:prop-coord-time}) we must keep only the quadrupole term as well as the two tidal terms:
{}
\begin{eqnarray}
\frac{d\tau_{\rm A}}{dt}&=&
1-\frac{1}{c^2}\Big[\frac{{\vec v}^2_{\rm A}}{2}+
\frac{GM_\oplus}{r_{\rm A}}\Big(1-J_2\Big(\frac{R_{\rm E}}{r_{\rm A}}\Big)^2
P_{20}(\cos\theta)\Big)+
\sum^{\rm S, m}_{b\not={\rm E}}\frac{GM_b}{2r^3_{b{\rm E}}}\Big(3(\vec{n}_{b{\rm E}}\cdot\vec{y}_{\rm A})^2-\vec{y}_{\rm A}^2\Big)\Big]+{\cal O}(6.34\times 10^{-17}),~~~~~
\label{eq:prop-coord-time-J2_GPS}
\end{eqnarray}
where the size of the error term is set by the magnitude of the next term in lunar tide, which is of the order of  $r_{\rm A}/r_{\mathleftmoon\hskip -1pt{\rm  E}}=(R_\oplus+h_{\tt GPS})/r_{\mathleftmoon\hskip -1pt{\rm  E}}\sim 0.075$ times smaller compared to the first term of lunar tide \cite{Kozai:1965}.  Also, we used a convenient notation for Legendre polynomial $P_{20}(\cos\theta)=(3z^2_{\rm A}-r^2_{\rm A})/2r^2_{\rm A}$.

As a result, the expression for proper to coordinate time transformation for LEO (\ref{eq:prop-coord-time-J2}) and GPS (\ref{eq:prop-coord-time-J2_GPS}) satellites explicitly includes the effects of the Earth's oblateness, $J_2$. The presence of this term in the equations of motion will lead to a perturbation of a Keplerian orbit. The effect of the quadrupole is large. Computing the perturbations to Keplerian orbital elements, we see that for the semi-major axis, $r_{\rm A}=a$, if the eccentricity is very small, the dominant contribution has a period twice the orbital period and has amplitude $3J_2 R^2_{\rm E}\sin^2 i/(2a)\simeq 1665$~m, assuming an orbital inclination of $i=55^\circ$. The effect of $J_2$ is significant and should be modeled for satellite clocks in low Earth orbit.

\subsubsection{Accounting for orbital perturbations due to Earth's oblateness $J_2$}

We may now account for the  presence of the $J_2$ term in (\ref{eq:prop-coord-time-J2}) and (\ref{eq:prop-coord-time-J2_GPS}). Perturbations of GPS orbits due to the Earth's quadrupole mass distribution contribute a significant fraction to the change in semi-major axis associated with the corresponding orbital change. We need to estimate the effect of Earth's quadrupole moment on the orbital elements of a Keplerian orbit and, as a result, on the changes in the frequency induced by such an orbital change.

Accounting for the perturbation in Keplerian orbital elements including the semi-major axis, $a$, eccentric anomaly, ${\cal E}={\cal M}+e\sin{\cal E}$ (with ${\cal M}$ being the mean anomaly), eccentricity, $e$, orbital radius, $r=a(1-e\cos{\cal E})$, we can compute perturbations to each of the terms $v_{\rm A}^2$,  in ${GM_{\rm E}}/{r_{\rm A}}$ and the quadrupole term in (\ref{eq:prop-coord-time-J2}) and (\ref{eq:prop-coord-time-J2_GPS}). The calculations involved are lengthy, but straightforward and are rather well-known \cite{Kouba:2004,Will-2014,Formichella:2021}. Here we present only the relevant result:
{}
\begin{eqnarray}
-\frac{{\vec v}^2_{\rm A}}{2c^2} &-&
\frac{GM_\oplus}{c^2r_{\rm A}}\Big(1-J_2\Big[\frac{R_\oplus}{r_{\rm A}}\Big]^2\frac{3z^2_{\rm A}-r^2_{\rm A}}{2r^2_{\rm A}}\Big)=\nonumber\\
&=&
\frac{3GM_\oplus}{2c^2a_0}-\frac{2GM_\oplus}{c^2a_0}e_0\cos{\cal E}_0-
\frac{7GM_\oplus R^2_\oplus J_2}{2c^2a^3_0}\big(1-{\textstyle\frac{3}{2}}\sin^2 i_0\big)-
\frac{GM_\oplus R^2_\oplus J_2\sin^2 i_0}{c^2a^3_0}\cos2(\omega_0+u),
\label{eq:perturb}
\end{eqnarray}
where $a_0=R_\oplus+h_{\rm GPS}$ is the semi-major axis, $i_0$ is the inclination, $\omega$ is the altitude of perigee and $u$ is the true anomaly. The subscript $0$ refers to an unperturbed quantity.

The first term, when combined with the reference potential at Earth's geoid, gives rise to the ``factory frequency offset'' and is estimated to be ${3GM_\oplus}/{2c^2a_0}\sim 2.50\times 10^{-10}$. The second term gives rise to the eccentricity effect with the magnitude that was evaluated to provide a contribution of $(2GM_\oplus/c^2a_0)e_0\sim 1.67\times 10^{-12}$, where eccentricity was taken to be $e_0=0.005$. The third term is a periodic contribution with estimated amplitude
{}
\begin{eqnarray}
\frac{7G M_\oplus R_\oplus^2 J_2}{2c^2 (R_\oplus+h_{\tt GPS})^3}\Big(1-{\textstyle\frac{3}{2}}\sin^2 i_0\Big)&\sim& 2.37\times 10^{-16},
\end{eqnarray}
which is at the limit of the anticipated accuracy.
This is for the nominal inclination of GPS orbits of $i_0=55^\circ$ such that the factor $(1 - {\textstyle\frac{3}{2}} \sin^2 i_0)=-6.52\times 10^{-3}$. The near-vanishing of this factor is pure coincidence for GPS.

The last term in Eq.~(\ref{eq:perturb}) has the amplitude
{}
\begin{eqnarray}
\frac{GM_\oplus R^2_\oplus J_2\sin^2 i_0}{c^2(R_\oplus+h_{\tt GPS})^3}=6.98\times10^{-15},
\label{eq:perturb2}
\end{eqnarray}
which is large enough to be considered when calculating frequency shifts due to orbit changes. The form of this term is similar to that which gives rise to the eccentricity correction, which is applied by GPS receivers. Considering only this periodic term, the additional time elapsed on the orbiting clock as it moves along its orbital path is given by
{}
\begin{eqnarray}
\Delta \tau_{J_2}&=&-\int_{\rm path}dt\Big[
\frac{GM_\oplus R^2_\oplus J_2\sin^2 i_0}{c^2a^3_0}\cos2(\omega_0+n_{\tt GPS}t)\Big]\simeq \nonumber\\
&\simeq&
-\sqrt{\frac{GM_\oplus}{a^3_0}}
\frac{R^2_\oplus J_2\sin^2 i_0}{c^2}\cos[2\omega_0+n_{\tt GPS}(t+t_0)]\sin[n_{\tt GPS}(t-t_0)],
\label{eq:perturbJ2a}
\label{eq:perturbJ2}
\end{eqnarray}
where, to a sufficient approximation, we have replaced the quantity $u$ in the integrand by $n_{\tt GPS}=\sqrt{GM_\oplus/a^3_0}$, which is the approximate mean motion of GPS satellites. This describes the periodic relativistic effect on the elapsed time of a clock on a GPS spacecraft due to Earth's quadrupole moment.

The correction that should be applied by the receiver is given by  (\ref{eq:perturbJ2a})  but with the opposite sign. The phase of this effect is zero when the satellite passes through Earth's equatorial plane going northwards. If not accounted for, this effect on a LEO clock time has the magnitude of $\Delta \tau_{J_2\tt LEO} = 5.79\times 10^{-10}\sin^2 i_0$~s, giving rise to a peak-to-peak periodic navigational error in position of approximately $2c\times \Delta \tau_{J_2\tt LEO} = 34.74\sin^2 i_0$~cm. The same effect on a GPS clock has the magnitude $\Delta \tau_{J_2\tt GPS} = 4.78\times 10^{-11}$~s and is responsible for a peak-to-peak periodic navigational error in position of about $2c\times \Delta \tau_{J_2\tt GPS} = 2.87$~cm. Therefore, these effects must be included in the model for high-accuracy orbit determination, especially for supporting for DSAC, ACES, STE-QUEST, and SOC missions in the near future.

\subsubsection{Accounting for tidal perturbations due to Sun and the Moon}

As we saw before, the tidal perturbations of the GPS orbit give rise of the sizable effects given by (\ref{eq:tidalT-Sun})--(\ref{eq:tidalT-moon}). We observe that $(\vec{n}_{\odot{\rm E}}\cdot\vec{n}_{\rm C})=\cos\big(\phi_{\odot 0}+(n_\oplus +n_{\tt GPS})t\big)$, where $n_\oplus $ and $n_{\tt GPS}$ are the orbital frequency of the Earth's sidereal motion and GPS spacecraft in orbit about Earth, correspondingly. Similarly, $(\vec{n}_{{\mathleftmoon\hskip -1pt{\rm  E}}}\cdot\vec{n}_{\rm C})=\cos\big(\phi_{\mathleftmoon\hskip -1pt 0}+(n_{\mathleftmoon} +n_{\tt GPS})t\big)$, where $n_{\mathleftmoon} $ is the orbital frequency of the Moon around the Earth.

With these definitions, we evaluate the contribution from the solar tide (\ref{eq:tidalT-Sun}) as
{}
\begin{eqnarray}
c^{-2}u^{\tt tidal}_{\rm E\odot}
&\simeq& \frac{GM_\odot (R_\oplus+h_{\tt GPS})^2}{4 c^2{\rm AU}^3}\Big(3\cos\Big[2\phi_{\odot 0}+2(n_\oplus +n_{\tt GPS})t\Big]+1\Big).
\label{eq:tidalT-S}
\end{eqnarray}
Assuming nearly circular orbits for the GPS satellites (i.e., $e=0.05$), we substitute this result into (\ref{eq:prop-coord-time-J2_GPS}) and evaluate the additional elapsed time measured by the orbiting clock due to $u^{\tt tidal}_{\rm E\odot}$ as the clock moves along it's orbital path:
{}
\begin{eqnarray}
\Delta \tau^{\rm tidal}_\odot&=&-\int_{t_0}^tdt\, c^{-2}u^{\tt tidal}_{\rm E\odot}\simeq
-\frac{GM_\odot (R_\oplus+h_{\tt GPS})^2}{4 c^2{\rm AU}^3}\Big(\Delta t+\frac{3}{n_{\tt GPS}}\cos\big[2\phi_{\odot 0}+n_{\tt GPS}(t+t_0)\big]\sin\big[n_{\tt GPS}\Delta t\big]\Big),
\label{eq:perturbJ2-tida}
\end{eqnarray}
where $\Delta t=t-t_0$ and we accounted for the fact that $n_\oplus \ll n_{\tt GPS}$. The term with the linear temporal drift here has the magnitude
{}
\begin{eqnarray}
 \frac{GM_\odot (R_\oplus+h_{\tt GPS})^2}{4 c^2{\rm AU}^3}\Delta t\simeq
7.79 \times 10^{-17}\Delta t,
\label{eq:tidalT3}
\end{eqnarray}
and, if not accounted for, in  6 hours (a half of the GPS orbital period), it could lead to clock desynchoniztion at the level of $1.68\times 10^{-12}$ s, which is significant for our purposes and should be included in the model. Next,  the periodic term in the expression (\ref{eq:perturbJ2-tida}) would lead to the  additional time elapsed on the orbiting clock that may be given by
{}
\begin{eqnarray}
\Delta \tau^{\rm tidal}_\odot&\simeq&
-\frac{3GM_\odot (R_\oplus+h_{\tt GPS})^2}{4 c^2{\rm AU}^3 n_{\tt GPS}}\cos\big[2\phi_{\odot 0}+n_{\tt GPS}(t+t_0)\big]\sin\big[n_{\tt GPS}\Delta t\big].
\label{eq:perturbJ2-tid}
\end{eqnarray}
If not accounted for, this effect on a GPS clock time has the magnitude of $\Delta \tau^{\rm tidal}_\odot = 1.67\times 10^{-12}$~s, giving rise to a peak-to-peak periodic navigational error in position of approximately $2c\times \Delta \tau^{\rm tidal}_\odot = 1.00$~mm, which is just above our threshold and needs to be accounted for. Such a correction (with an opposite sign) should be applied to the receiver.

Similarly, using (\ref{eq:tidalT-moon})  we evaluate the tidal contribution from the Moon:
{}
\begin{eqnarray}
\Delta \tau^{\rm tidal}_{\rm E \mathleftmoon}&=&-\int_{t_0}^tdt\,c^{-2}u^{\tt tidal}_{\rm E \mathleftmoon}\simeq\nonumber\\
&\simeq&
-\frac{Gm_\mathleftmoon (R_\oplus+h_{\tt GPS})^2}{4 c^2r^{*3}_{\mathleftmoon\hskip -1pt{\rm  E}}}\Big(\Delta t+\frac{3}{n_{\tt GPS}}\cos\big[2\phi_{\mathleftmoon\hskip -1pt 0}+n_{\tt GPS}(t+t_0)\big]\sin\big[n_{\tt GPS}\Delta t\big]\Big),
\label{eq:tidalTm}
\end{eqnarray}
where $r^*_{\mathleftmoon\hskip -1pt{\rm  E}}$ is the Earth-moon distance at the shortest value of the perigee of the lunar orbit and we again accounted for the fact that $n_\mathleftmoon \ll n_{\tt GPS}$. The  term with the linear temporal drift in this expression has the magnitude
{}
\begin{eqnarray}
\frac{Gm_\mathleftmoon (R_\oplus+h_{\tt GPS})^2}{4c^2r^{*3}_{\mathleftmoon\hskip -1pt{\rm  E}}}\Delta t\simeq
2.13 \times 10^{-16}\Delta t,
\label{eq:tidalT3m}
\end{eqnarray}
and is at the anticipated frequency stability limit. Furthermore, if not accounted for, in just 2 hours it could lead to clock desynchriiztion at the level of $ 1.53\times 10^{-12}$ s, which is important to consider in the clock model.

Considering the periodic term in  expression (\ref{eq:tidalTm}), the additional time elapsed on the orbiting clock is given by
{}
\begin{eqnarray}
\Delta \tau^{\rm tidal}_\mathleftmoon&=&
-\frac{3Gm_\mathleftmoon (R_\oplus+h_{\tt GPS})^2}{4 c^2r^{*3}_{\mathleftmoon\hskip -1pt{\rm  E}}n_{\tt GPS}}\cos\big[2\phi_{\mathleftmoon\hskip -1pt 0}+n_{\tt GPS}(t+t_0)\big]\sin\big[n_{\tt GPS}\Delta t\big].
\label{eq:perturbJ2mm}
\end{eqnarray}
Similarly to (\ref{eq:perturbJ2a}) and (\ref{eq:perturbJ2-tid}), the correction that should be applied by the receiver is the negative of this expression. If not accounted for, this effect on a GPS clock time has the magnitude of $\Delta \tau^{\rm tidal}_\mathleftmoon = 4.37\times 10^{-12}$~s for lunar perigee, giving rise to a peak-to-peak periodic navigational error in position of approximately $2c\times \Delta \tau^{\rm tidal}_\mathleftmoon= 2.62$~mm, which is significant at the anticipated level of accuracy.

\subsubsection{GPS clock: proper-to-coordinate time transformation: recommended formulation}

As a result of the analysis presented above, the recommended expression for the differential equation that relates the rate of the spacecraft proper $\tau_{\rm A}$ time, as measured by an on-board clock on a  GPS  spacecraft  to the time in GCRS, ${\rm TCG}=t$ that was derived as  (\ref{eq:prop-coord-time-J2_GPS}), should have the quadrupole term as well as the lunar and solar tidal terms:
{}
\begin{eqnarray}
\frac{d\tau_{\rm A}}{dt}&=&
1-\frac{1}{c^2}\Big[\frac{{\vec v}^2_{\rm A}}{2}+
\frac{GM_\oplus}{r_{\rm A}}\Big(1-J_2\Big(\frac{R_
\oplus}{r_{\rm A}}\Big)^2
P_{20}(\cos\theta)\Big)+\nonumber\\
&& \hskip 51pt +\,
\frac{Gm_\mathleftmoon}{r^3_{\mathleftmoon\hskip -1pt{\rm  E}}}r_{\rm A}^2 P_{20}\big((\vec{n}_{\mathleftmoon \hskip -1pt{\rm E}}\cdot\vec{n}_{\rm A})\big)+
\frac{GM_\odot}{r^3_{\odot{\rm  E}}}r_{\rm A}^2 P_{20}\big((\vec{n}_{\odot{\rm E}}\cdot\vec{n}_{\rm A})\big)
\Big]+{\cal O}(6.34\times 10^{-17}),~~~~~
\label{eq:prop-coord-time-J2_GPS*}
\end{eqnarray}
where $\vec{r}_{\rm A}=r_A\vec{n}_{\rm A},
\vec{r}_{\mathleftmoon \hskip -1pt{\rm E}}={r}_{\mathleftmoon \hskip -1pt{\rm E}}\vec{n}_{\mathleftmoon \hskip -1pt{\rm E}}$, and $\vec{r}_{\odot{\rm E}}={r}_{\odot{\rm E}}\vec{n}_{\odot{\rm E}}$ are the geocentric position vectors of a GPS satellite, the moon, and the Sun, correspondingly;  $\theta$ is the colatitude as defined by (\ref{eq:pot_w_0sh}). The size of the error term is set by the magnitude of the next term (i.e.,  of the order of ${\cal O}(x^3/r^4_{\mathleftmoon \hskip -1pt{\rm E}})$) in the lunar tidal contribution as given by (\ref{eq:u-tidal-E}) and (\ref{eq:prop-coord-time-J2_GPS}).  This error term contributes periodic signal with the magnitude of $4.35\times 10^{-13}$ s, which is too small to be part of the model.

Near the surface of the Earth, the expression beyond the monopole term is dominated by the quadrupole moment characterized by $J_2$. At higher orbits, however, as the effect of $J_2$ diminishes, contributions due to lunar and solar tides become more significant. For spacecraft going beyond geostationary orbits, e.g., spacecraft in lunar transfer orbits carrying accurate clocks, eventually it becomes necessary to switch to a more accurate representation of the lunar and solar gravitational fields, such as the representation in \cite{Turyshev:2012nw}.

\section{Light propagation in the vicinity of the Earth }
\label{sec:light-prop}

Now we need to consider the relativistic corrections to the light propagation in the presence of the extended Earth.  These effects are important as they contribute sizable time delays for the signal propagating in the vicinity of a massive body. Currently, only one of these effects is present in the relevant models of light propagation: The  Shapiro time delay \cite{Kopeikin-book-2011,Poisson-Will:2014} that is due to the gravitational monopole of a massive body. This effect is important for astronomy and spacecraft navigation in the solar system \cite{Will,Moyer:2003,MG2005}.

Our question here is whether or not the next order terms in the Earth's gravitational potential (i.e., as given by  (\ref{eq:pot_w_0sh}) and shown in Table~\ref{tab:sp-harmonics}) would play a significant role in models describing light propagation in the solar system. Specifically, are the low-order terms $\ell=2,3,4$ in (\ref{eq:pot_w_0sh}) important? Currently available models treat the Earth as an axisymmetric rotating body \cite{Klioner:1991SvA,Klioner-Kopeikin:1992,Zschocke-Klioner:2010} and they take into account only the Earth's zonal harmonics given  by, e.g.  the oblateness term $J_2$, but not the $\ell=2$ tesseral  spherical harmonics or higher order terms present in (\ref{eq:pot_w_0sh}).

In this section, we assess the contributions of the higher-order terms in the gravitational potential on the phase of an EM wave that propagates in the vicinity of an extended body.

\subsection{Gravitational phase shift of an EM wave}
\label{sec:geom-gr-phase}

The phase of an EM wave is a scalar function that is invariant under a set of general coordinate transformations. In the geometric optics approximation, the phase $\varphi$ is found as a solution to the eikonal equation \cite{Fock-book:1959,Landau-Lifshitz:1988,Sazhin:1998,Kopeikin:2009,Kopeikin-book-2011},
\begin{equation}
g^{mn}\partial_m\varphi\partial_n\varphi=0,
\label{eq:eq_eik}
\end{equation}
which is a direct consequence of Maxwell's equations. Its solution describes the wavefront of an EM wave propagating in curved spacetime. The solution's geometric properties are defined by the metric tensor $g_{mn}$, which is derived from Einstein's field equations. In the vicinity of the Earth, this tensor is given by Eqs.~(\ref{eq:gab-E})--(\ref{eq:u-tidal-E}).

To solve Eq.~(\ref{eq:eq_eik}), we introduce a covector of the electromagnetic wavefront in curved spacetime, $K_m = \partial_m\varphi$. We use $\lambda$ to denote an affine parameter along the trajectory of a light ray, which is orthogonal to the wavefront $\varphi$. The vector $K^m = dx^m/d\lambda = g^{mn}\partial_n\varphi$ is tangent to the light ray. Equation~(\ref{eq:eq_eik}) states that $K^m$ is null: $g_{mn}K^mK^n = 0$.

To find a solution of Eq.~(\ref{eq:eq_eik}), we expand the eikonal $\varphi$ with respect to the gravitational constant $G$, assuming that the unperturbed solution is a plane wave. The expansion may be given as
{}
\begin{equation}
\varphi(t,{\vec x}) = \varphi_0+\int k_m dx^m+\varphi_G (t,{\vec x})+{\cal O}(G^2),
\label{eq:eq_eik-phi}
\end{equation}
where $\varphi_0$ is an integration constant and $k_m = k(1, {\vec k})$ is a constant (with respect to the Minkowski metric) null vector (i.e., $\gamma_{mn}k^mk^n=0$) along the direction of propagation of the unperturbed EM plane wave. The wave direction is given by the vector ${\vec k}\equiv k^\epsilon$, which is the unit vector along the ray's path, $|{\vec k}|=1$. Furthermore, $k=\omega/c$, where $\omega$ is the constant angular frequency of the unperturbed wave, and $\varphi_G$ is the perturbation of the eikonal of first order in $G$, which is yet to be determined. Also, as a consequence of Eq.~(\ref{eq:eq_eik-phi}), the wave vector of an electromagnetic wave in curved spacetime, $K^m(t,{\vec x})$, admits a series expansion with respect to $G$ in the form
{}
\begin{equation}
K^m(t,{\vec x})=\frac{dx^m}{d\lambda}= g^{mn}\partial_n\varphi=k^m+k_G^m(t,{\vec x})+{\cal O}(G^2),
\label{eq:K}
\end{equation}
where $k^m_G(t,{\vec x})=\gamma^{mn}\partial_n\varphi_G(t,{\vec x})$ is the first order perturbation of the wave vector with respect to $G$.

To solve Eqs.~(\ref{eq:eq_eik}) and (\ref{eq:eq_eik-phi}) for $\varphi_G$ in the GCRS, we first substitute (\ref{eq:eq_eik-phi}) into (\ref{eq:eq_eik}).
Then, defining $h^{mn}=g^{mn}-\gamma^{mn}$ and keeping only first order terms in $G$, we obtain an ordinary differential equation to determine $\varphi_G$:
{}
\begin{equation}
\frac{d\varphi_G}{d\lambda}= -\frac{1}{2}h^{mn}k_mk_n +{\cal O}(G^2),
\label{eq:eq_eik-phi-lamb}
\end{equation}
where ${d\varphi_G}/{d\lambda}= k_m\partial^m\varphi_G+{\cal O}(G^2)$.  (Note, that Eq.~(\ref{eq:eq_eik-phi-lamb}) alternatively can also be obtained by integrating the null geodesic equation \cite{Kopeikin-book-2011}).
With $g_{mn}$ given by (\ref{eq:gab-E}), the equation to determine the phase of the EM wave as it propagates in the vicinity of the Earth takes the following form:
{}
\begin{equation}
\frac{d\varphi_G}{d\lambda}=
-k^2\Big\{\frac{2}{c^2}w_{\rm E}+\frac{2}{c^2}w^{\tt tidal}_{\rm E}+ \frac{4}{c^3}(k_\epsilon w^\epsilon_{\rm E})+{\cal O}(G^2)\Big\}.
\label{eq:eq_eik-phi-lamb*}
\end{equation}
This equation describes the gravitational phase shift introduced by various contributions to the effective gravity field of the GCRS (\ref{eq:gab-E}) with the potentials (\ref{eq:pot_loc-w_0})--(\ref{eq:pot_loc-w_a+}).  Below, we will integrate it along the light ray's trajectory.

\subsection{Parameterizing the light ray's trajectory}
\label{sec:eik-wfr}

To solve (\ref{eq:eq_eik-phi-lamb*}), we need to present the geometry of the problem and introduce our basic notations. Following \cite{Turyshev-Toth:2017,Turyshev-Toth:2021-multipoles}, we represent the light ray's trajectory, correct to the Newtonian order, as
{}
\begin{eqnarray}
\{x^m\}\equiv\Big(x^0=ct, ~~{\vec x}(t)\equiv {\vec r}(t)={\vec r}_{\rm 0}+{\vec k} c(t-t_0)\Big)+{\cal O}(G),
\label{eq:x-Newt0}
\end{eqnarray}
where $\vec{k}$ is a unit vector in the incident direction of the light ray's propagation path and $\vec r_0$ represents the point of emission that may be expressed as $\vec{k}=({\vec{r}-\vec{r}_0})/{|\vec{r}-\vec{r}_0|}$. Next, we define the impact parameter of the unperturbed trajectory of the light ray parameter $\vec b$ as
{}
\begin{eqnarray}
{\vec b}=[[{\vec k}\times{\vec r}_0]\times{\vec k}].
\label{eq:impact-par}
\end{eqnarray}
Next, we introduce the parameter $\tau=\tau(t)$ along the path of the light ray (see details in Appendix~B in \cite{Turyshev-Toth:2017}):
{}
\begin{eqnarray}
\tau &=&({\vec k}\cdot {\vec r})=({\vec k}\cdot {\vec r}_{0})+c(t-t_0),
\label{eq:x-Newt*=0}
\end{eqnarray}
which may be positive or negative. The parameter $\tau$ allows us to rewrite (\ref{eq:x-Newt0}) as
{}
\begin{eqnarray}
{\vec r}(\tau)&=&{\vec b}+{\vec k} \tau+{\cal O}(G),
\qquad {\rm with} \qquad r(\tau) \equiv |{\vec x}(\tau)|=\sqrt{b^2+\tau^2}+{\cal O}(G).
\label{eq:b0}
\end{eqnarray}

Using the result (\ref{eq:b0}) we determine that the following relations are valid to ${\cal O} (r_g)$:
{}
\begin{eqnarray}
r&=& \sqrt{b^2+\tau^2}, \qquad ~~\, r+({\vec k}\cdot{\vec r}) =\sqrt{b^2+\tau^2} +\tau.
 \label{eq:notat}
\end{eqnarray}
Based on these results, we present a useful relationship:
{}
\begin{eqnarray}
b^2=\big(r+({\vec k}\cdot{\vec r})\big)\big(r-({\vec k}\cdot{\vec r})\big) +{\cal O}(G).
\label{eq:b*0}
\end{eqnarray}

This representation allows us to express the Newtonian part of the wave vector $K^m$ presented by Eq.~(\ref{eq:K}) as follows:
$k^m= {dx^m}/{d\lambda} =k\big(1, {\vec k}\big)+{\cal O}(G)$, where the wave number $k$ is immediately derived as $k={d\tau}/{d\lambda}+{\cal O}(G)$ and $|{\vec k}|=1$. Keeping in mind that $k^m$ is constant, we establish an important relationship:
\begin{equation}
d\lambda= \frac{d\tau}{k}+{\cal O}(G),
\label{eq:eq_eik-relat}
\end{equation}
which we use to integrate (\ref{eq:eq_eik-phi-lamb}). This expression allows including contributions from all multipoles of the Earth's mass distribution, as given in Eq.~(\ref{eq:pot_w_0}).

With these definitions, we may now present the  solution to (\ref{eq:eq_eik-phi-lamb*}) (see relevant discussion in \cite{Turyshev-Toth:2021-multipoles}). The gravitational phase shift, $\varphi_G$, that is acquired by the EM wave as it propagates along its geodesic path from the point of emission at $\tau_0$  to the point of reception at $\tau$ on the background of the gravitational field (\ref{eq:gab-E})--(\ref{eq:u-tidal-E}) has the form:
{}
\begin{eqnarray}
\varphi_G(\vec x) =
-k\int^{\tau}_{\tau_0} \Big\{\frac{2}{c^2}\Big(U_{\rm E}(\tau')+ u^{\tt tidal}_{\rm E}(\tau')\Big)+ \frac{4}{c^3}(k_\epsilon w^\epsilon_{\rm E}(\tau'))
+{\cal O}(G^2)
\Big\}d\tau'
\equiv
\varphi_G^{\tt E}(\vec x)+\varphi_G^{\tt tidal}(\vec x)+\varphi_G^{\tt S}(\vec x)+{\cal O}(G^2).~
\label{eq:Psi+}
\end{eqnarray}

We may now integrate (\ref{eq:Psi+}) for each of the relativistic terms with the terms describing contributions from the Earth's gravitational potential, tidal gravity (primarily due to the Sun and the Moon), and Earth's rotation, correspondingly.

\subsection{The STF representation of the  Earth's gravitational potential}
\label{sec:potU}

Although the form of the Earth gravitational potential (\ref{eq:pot_w_0sh}) is effective for many applications in geodesy, it is not technically convenient when light propagation in a gravitational field in concerned. No closed form expressions are known for the integral (\ref{eq:Psi+}) are known for a potential in the form (\ref{eq:pot_w_0sh}), expressed in terms of spherical harmonics; indeed, no useful semi-analytical approximations exist in the general case either. Thus, alternative representations of $U(\vec x)$ are needed. In \cite{Turyshev-Toth:2021-multipoles}, we considered the case of axisymmetric bodies. In \cite{Turyshev-Toth:2021-STF-moments}, we considered a generic potential, expanding $U(\vec x)$ in terms of STF mass-moment tensors. Below, we discuss the the STF mass moment representation, which allows us to fully benefit from the spherical harmonics representation in the most general case. (In Appendix~\ref{app:axisymm}, we discuss the less general axisymmetric case, using it as a limiting case to verify our results.)

Considering a generic case, it was discussed in \cite{Thorne:1980,Blanchet-Damour:1986,Blanchet-Damour:1989,Kopeikin:1997,Poisson-Will:2014,Soffel-Han:2019} that the scalar gravitational potential (\ref{eq:pot_w_0}) may equivalently be given in terms of Cartesian spatial trace-free (STF) tensor moments in the following form:
{}
\begin{eqnarray}
U&=&GM\Big\{\frac{1}{r}+\sum_{\ell=2}^\infty \frac{(-1)^\ell}{\ell!} {\cal T}^{<a_1...a_\ell>}\frac{\partial^\ell}{\partial x^{<a_1...}\partial x^{a_\ell>}}\Big(\frac{1}{r}\Big)\Big\}+
{\cal O}(c^{-4}),
\label{eq:pot_stf}
\end{eqnarray}
where $r=|{\vec x}|$, $M$ is the mass  and ${\cal T}^{<a_1...a_\ell>}$ are the STF mass multipole moments of the body, defined as
{}
\begin{eqnarray}
M&=&\int_{\tt V} d^3{\vec x}\, \rho({\vec x}),\qquad
{\cal T}^{<a_1...a_\ell>}=\frac{1}{M}\int_{\tt V} d^3{\vec x}\, \rho({\vec x})\, x^{<a_1...a_\ell>},
\label{eq:mom}
\end{eqnarray}
where $x^{<a_1...a_\ell>}=x^{<a_1}x^{a_2...}x^{a_\ell>}$, the angle  brackets $<...>$ denote the STF operator, and ${\tt V}$ means the total volume of the isolated gravitating body under consideration. The dipole moment ${\cal T}^a$ is absent from this expansion (\ref{eq:pot_stf}), by virtue of the fact that the origin of the coordinates is assumed to coincide with the body's barycenter.

Using the identity \cite{Soffel-Han:2019},
{}
\begin{eqnarray}
\frac{\partial^\ell}{\partial x^{<a_1...}\partial x^{a_\ell>}}\Big(\frac{1}{r}\Big)&=& (-1)^\ell(2\ell-1)!!\frac{\hat n_{<a_1...a_\ell>}}{r^{\ell+1}},
\label{eq:pot_stf-dir}
\end{eqnarray}
the potential (\ref{eq:pot_stf}) may be given in the following form:
{}
\begin{eqnarray}
U(\vec r)&=& GM\sum_{\ell\geq 0}\frac{(2\ell-1)!!}{\ell !}{\cal T}_L\frac{\hat n_L}{r^{\ell+1}}.
\label{eq:pot_w_0STF}
\end{eqnarray}
The first few terms of (\ref{eq:pot_w_0STF}) or, equivalently, (\ref{eq:pot_stf}), are given as
{}
\begin{eqnarray}
U(\vec r)&=&GM\Big\{\frac{1}{r}+ \frac{3{\cal T}^{<ab>}}{2r^5}x^ax^b +\frac{5{\cal T}^{<abc>}}{2r^7}x^ax^bx^c+\frac{35{\cal T}^{<abcd>}}{8r^9}x^ax^bx^cx^d+{\cal O}(r^{-6})\Big\}.
\label{eq:pot_w_0STF2}
\end{eqnarray}
This Cartesian multipole expansion of the Newtonian gravitational potential is equivalent to expansion in terms of spherical harmonics (\ref{eq:pot_w_0sh}) \cite{Thorne:1980,Blanchet-Damour:1986,Blanchet-Damour:1989,Kopeikin:1997,Mathis-LePoncinLafitte:2007}. In fact, this expression may be used to establish the correspondence between ${\cal T}^{<a_1...a_\ell>}$ and $C_{\ell k}$ and $S_{\ell k}$ from (\ref{eq:pot_w_0sh}) (see Appending~\ref{sec:stf-sph-harm} for details on how to establish this correspondence).

\subsection{Expressing the gravitational phase shift via STF mass moments}

Using the representations (\ref{eq:pot_w_0}), (\ref{eq:pot_stf}) or (\ref{eq:pot_st3}), it is convenient to present  the $U_{\rm E}$-dependent term that yields $\varphi_G^{\tt E}(\vec x)$ in the total gravitational phase shift in (\ref{eq:Psi+}) as
{}
\begin{eqnarray}
\frac{2U}{c^2}&=&
r_g\Big\{\frac{1}{r}+ \sum_{\ell=2}^\infty \frac{(-1)^\ell}{\ell!} {\cal T}^{<a_1...a_\ell>}\frac{\partial^\ell}{\partial x^{<a_1...}\partial x^{a_\ell>}}\Big(\frac{1}{r}\Big)\Big\},
\label{eq:U-c2]}
\end{eqnarray}
where $r_g=2GM/c^2$ is the Schwarzschild radius of the body (in our case this is the Earth, but our discussion is generic.) As such, this form is valid for any deviation from spherical symmetry in the gravitational field.

We may then generalize expression ${\vec \nabla}={\nabla}_b+{\vec k}\,d/d\tau +{\cal O}(r_g)$ and write
{}
\begin{eqnarray}
\frac{\partial^\ell}{\partial x^{<a_1...}\partial x^{a_\ell>}}&\equiv&
{\vec \nabla}^{<a_1....}{\vec \nabla}^{a_\ell>}=\sum_{p=0}^\ell \frac{\ell!}{p!(\ell-p)!}k_{<a_1}...k_{a_p} \partial_{a_{p+1}}... \partial_{a_\ell>} \frac{\partial^p}{\partial \tau^p}+{\cal O}(r_g),
\label{eq:derivA}
\end{eqnarray}
where a new shorthand notation $\partial_a\equiv \partial/\partial {\vec b}^a$ has been used and $\tau$ is defined by (\ref{eq:x-Newt*=0}).

 Using this representation (\ref{eq:derivA}), we can compute the relevant integral (with $r=\sqrt{b^2+\tau^2}$ and $r_0=\sqrt{b^2+\tau_0^2}$, as discussed in Sec.~\ref{sec:eik-wfr}):
{}
\begin{eqnarray}
 \int^{\tau}_{\tau_0} \frac{\partial^\ell}{\partial x^{<a_1...}\partial x^{a_\ell>}}\Big(\frac{1}{r}\Big) d\tau'&=&
  \sum_{p=0}^\ell \frac{\ell!}{p!(\ell-p)!}k_{<a_1}...k_{a_p} \partial_{a_{p+1}}... \partial_{a_\ell>} \Big\{\frac{\partial^{p}}{\partial \tau^{p}}\ln \Big(\frac{\sqrt{b^2+\tau^2}+\tau}{b}\Big)
\Big\}\Big|^\tau_{\tau_0} =\nonumber\\
 &&\hskip -140pt\,=
 \partial_{<a_1}... \partial_{a_\ell>} \ln  \frac{\sqrt{b^2+\tau^2}+\tau}{\sqrt{b^2+\tau^2}+\tau_0}+
  \sum_{p=1}^\ell \frac{\ell!}{p!(\ell-p)!}k_{<a_1}...k_{a_p} \partial_{a_{p+1}}... \partial_{a_\ell>}
  \Big\{\frac{\partial^{p-1}}{\partial \tau^{p-1}}  \frac{1}{\sqrt{b^2+\tau^2}}-\frac{\partial^{p-1}}{\partial \tau_0^{p-1}}  \frac{1}{\sqrt{b^2+\tau_0^2}}\Big\}.
\label{eq:int+1*}
\end{eqnarray}

As a result, the gravitational eikonal phase shift $\varphi^{\tt E}_G$ from (\ref{eq:Psi+}) takes the form\footnote{Result (\ref{eq:eik-phA2}) was independently derived in  \cite{Kopeikin:1997} where one can also find the phase contribution due to vector spherical harmonics. In the Earth's gravity field such harmonics are small, providing contributions below the expected level of the measurement accuracy (Sec.~\ref{sec:intro}). Thus, beyond the spin term with $\ell=1$ (\ref{eq:pot_loc-w_a+}), higher order contributions of the vector harmonics were not considered in this paper.}:
{}
\begin{eqnarray}
\varphi^{\tt E}_G(\vec r)
&=& -kr_g\Big\{\ln \frac{\sqrt{b^2+\tau^2}+\tau}{\sqrt{b^2+\tau_0^2}+\tau_0}+
\sum_{\ell=2}^\infty
\frac{(-1)^\ell}{\ell!} {\cal T}^{<a_1...a_\ell>} \Big\{
 \partial_{<a_1}... \partial_{a_\ell>}\ln \frac{\sqrt{b^2+\tau^2}+\tau}{\sqrt{b^2+\tau_0^2}+\tau_0}+\nonumber\\
 &&\hskip 20pt\,+
  \sum_{p=1}^\ell \frac{\ell!}{p!(\ell-p)!}k_{<a_1}...k_{a_p} \partial_{a_{p+1}}... \partial_{a_\ell>}
  \Big\{\frac{\partial^{p-1}}{\partial \tau^{p-1}}  \frac{1}{\sqrt{b^2+\tau^2}}-\frac{\partial^{p-1}}{\partial \tau_0^{p-1}}  \frac{1}{\sqrt{b^2+\tau_0^2}}\Big\}\Big\}\Big\},
\label{eq:eik-phA2}
\end{eqnarray}
or, equivalently, using (\ref{eq:notat}) we have
{}
\begin{eqnarray}
\varphi^{\tt E}_G(\vec r,\vec r_0)
&=&   -kr_g \Big(\ln \Big[\frac{r+(\vec k\cdot \vec r)}{r_0+(\vec k\cdot \vec r_0)}\Big]+\sum_{\ell=2}^\infty
\frac{(-1)^\ell}{\ell!} {\cal T}^{<a_1...a_\ell>} {\cal I}_{a_1... a_\ell}(\vec r,\vec r_0)\Big)+{\cal O}(r^2_g).
\label{eq:eik-phA+}
\end{eqnarray}

The first term in this expression is the well-known Shapiro phase shift. The next term is the contribution to the gravitational phase delay from the STF gravitational multipoles to any order  $\ell$. The quantity ${\cal I}_{a_1... a_\ell}(\vec r,\vec r_0)={\cal I}_{a_1... a_\ell}(\vec r)-{\cal I}_{a_1... a_\ell}(\vec r_0)$ is the projection operator of the $\ell$-th order along the light ray's trajectory:
{}
\begin{eqnarray}
{\cal I}_{a_1... a_\ell}(\vec r,\vec r_0)
&=&\Big\{
 \partial_{<a_1}... \partial_{a_\ell>}\ln k\Big({\sqrt{b^2+\tau^2}+\tau}\Big)+  \sum_{p=1}^\ell \frac{\ell!}{p!(\ell-p)!}k_{<a_1}...k_{a_p} \partial_{a_{p+1}}... \partial_{a_\ell>}
 \frac{\partial^{p-1}}{\partial \tau^{p-1}}  \frac{1}{\sqrt{b^2+\tau^2}}\Big\}\Big|^\tau_{\tau_0}.~~~~
\label{eq:eik-ph2*}
\end{eqnarray}

Expression (\ref{eq:eik-phA}) together with (\ref{eq:eik-ph2*})  is a key result. It demonstrates that with the help of the STF tensor formalism, it is possible to evaluate contributions to the gravitational phase shift, {\em to all orders} beyond the Shapiro phase shift, due to the mass multipole moment contributions of the gravitating body.

\subsection{Rotation to the STF moments to the light ray coordinate system}
\label{sec:compute-shift}

The main objective of this manuscript is to develop the functional form of the gravitational phase shift $\varphi^{\tt E}_G$ in terms of the spherical harmonics. As (\ref{eq:eik-phA+}) suggests, the total gravitational phase shift induced by all the multipoles $\ell \geq2$ is a sum of the individual shifts $\varphi_\ell$ inducted by multipoles at each particular order $\ell$, where $\varphi_\ell$ have the form
{}
\begin{eqnarray}
\varphi_\ell(\vec r, \vec r_0)
 &=&   -kr_g \frac{(-1)^\ell}{\ell!} {\cal T }^{<a_1...a_\ell>} {\cal I}_{a_1... a_\ell}(\vec r,\vec r_0).
\label{eq:eik-phA}
\end{eqnarray}
In what follows we will demonstrate how to compute individual terms in this sum.

First, we recognize that the GCRS is defined by the unit basis vectors $\vec{e}_{x}$ (the prime meridian) and $\vec e_{z}\equiv \vec{s}$ (the Earth's rotation axis), yielding the following set of base vectors:
{}
\begin{eqnarray}
\vec{e}_{x}, \qquad \vec{e}_{y}=[\vec e_z\times \vec e_x], \qquad \vec{e}_{z}\equiv \vec s.
\end{eqnarray}
Next, the propagation direction of the EM wave is defined by $\vec k$ and the relationship between the EM wave's trajectory and the Earth is given by the vector impact parameter $\vec{b}$ (introduced by (\ref{eq:note-b})):
{}
\begin{eqnarray}
\vec{k}=\frac{\vec{r}-\vec{r}_0}{|\vec{r}-\vec{r}_0|},\qquad
\vec{b}=[[{\vec k}\times{\vec r}_0]\times{\vec k}]
\equiv b\, \vec m \qquad\Rightarrow \qquad\vec m=\frac{[[{\vec k}\times{\vec r}_0]\times{\vec k}]}{|[[{\vec k}\times{\vec r}_0]\times{\vec k}]|}.
\label{eq:k+b}
\end{eqnarray}
Thus, once the GCRS positions of the emitter, $\vec{r}_0$, and receiver, $\vec{r}$,  are known,  everything else is computable.

\begin{figure}
\includegraphics[scale=1]{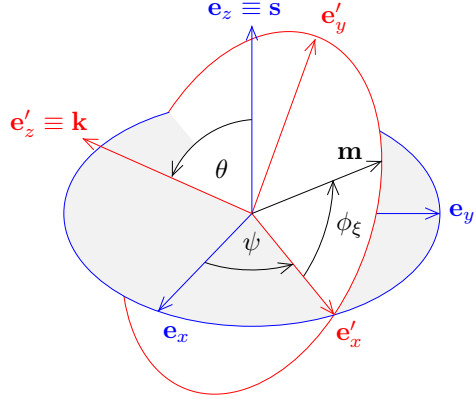}
\caption{\label{fig:rotate}Multipole moments are usually provided with respect to a body-centric coordinate reference system, depicted here using the Cartesian $\vec{e}_x$, $\vec{e}_y$ and $\vec{e}_z$ axes. For the Earth, $\vec{e}_z\equiv \vec{s}$ corresponds to the direction of the North pole, whereas $\vec{e}_x$ points in the direction of the prime meridian and $\vec{e}_y=[\vec e_z\times \vec e_x]$ spans the equatorial plane with $\vec{e}_x$. The signal propagates in the direction of the wavevector $\vec{k}$, represented in our Cartesian reference frame by the $\vec{e}'_z\equiv \vec k$ axis. The $\vec{e}'_x\vec{e}'_y$-plane contains the unit impact parameter vector $\vec{m}$ that is normal to $\vec{k}$.}
\end{figure}

To simplify the computations, we introduce a coordinate system associated with the propagating ray of light. Reading off figure~\ref{fig:rotate}, with knowledge of $\vec k$, we obtain the unit vectors defining a coordinate system associated with the direction of transmission:
\begin{equation}
\vec{e}'_x    = \frac{[\vec{e}_z\times\vec{e}'_z]}{|[\vec{e}_z\times\vec{e}'_z]|},
\qquad
\vec{e}'_y= [\vec{e}'_z\times {\vec{e}'_x}],
 \qquad
\vec{e}'_z \equiv \vec k.
\end{equation}

Together, the vectors $\vec{k}$ and $\vec{b}$ allow us to define a rotated coordinate system, where the $z$-axis is aligned with the direction of propagation of the EM wave given by vector $\vec{k}$, while the $xy$-plane is perpendicular to it (see Fig.~\ref{fig:rotate}). In this coordinate system, the vectors $\vec{k}'$, $\vec{b}'$, are given as below
{}
\begin{eqnarray}
\vec k'&=&\big(0,0,1\big),
\label{eq:note-k}
\qquad
{\vec b}'=b\big(\cos\phi_\xi,\sin \phi_\xi,0\big)=b\, \vec m',
\label{eq:note-b}
\end{eqnarray}
where $b=|\vec b|$ from (\ref{eq:k+b}) and
the orientation angle of the impact parameter, $\phi_\xi$, is given by
{}
\begin{eqnarray}
\cos \phi_\xi  = (\vec{m}\cdot\vec{e}'_x), \qquad
\sin \phi_\xi    = (\vec{m}\cdot\vec{e}'_y).
\end{eqnarray}

The two coordinate systems are related by a rotation where $R_a^b$ is the rotation matrix, given as
{}
\begin{eqnarray}
R(\theta,\psi)=R_3(\psi)R_1(\theta)=
\begin{pmatrix}
\cos\psi& \sin\psi& 0\\
-\sin\psi& \cos\psi& 0 \\
0 & 0& 1
\end{pmatrix}
\begin{pmatrix}
1& 0& 0\\
0& \cos\theta& \sin\theta \\
0 & -\sin\theta& \cos\theta
\end{pmatrix}
\equiv R_a^b,
  \label{eq:rot}
\end{eqnarray}
where $R_z(\psi)$ is a right-handed rotation by $\psi$ around the $z$-axis ($\vec{e}_z$-axis), $R_1(\theta)$  is a right-handed rotation by $\theta$ about the $x$-axis ($\vec{e}'_x$-axis) of a Cartesian coordinate frame; see Fig.~\ref{fig:rotate}. The angles $\theta$ and $\psi$ are determined from the following equations:
{}
\begin{eqnarray}
\cos \theta  = (\vec{e}'_z\cdot\vec{e}_z), \qquad
 \sin \theta   = (\vec{e}'_y\cdot\vec{e}_z), \qquad
\cos \psi  = (\vec{e}'_x\cdot\vec{e}_x),\qquad
 \sin \psi  = (\vec{e}'_x\cdot\vec{e}_y).
 \label{eq:thet-psi}
\end{eqnarray}

With the introduction of these rotation matrices (\ref{eq:rot}), the geometry of the problem is fully defined. Technically, it is easier to compute the components of the projection operator  in  (\ref{eq:eik-ph2*}) in the primed coordinate system that is aligned with  the light ray, yielding ${\cal I}'_{a_1... a_\ell}$. Two representations of this operator, in the GCRS, ${\cal I}_{a_1... a_\ell}$, and its counterpart in the light ray coordinates, ${\cal I}'_{a_1... a_\ell}$, are related by a simple rotation with the rotation matrix  $R_a^b$ from (\ref{eq:rot}):
{}
\begin{align}
{\cal I}{}_{<b_1...b_\ell>}={\cal I}'{}_{<a_1...a_\ell>}R^{a_1}_{b_1}... R^{a_\ell}_{b_\ell},
\label{eq:rot3Ia}
\end{align}
where ${\cal I}_{a_1... a_\ell}$ is expressed in terms of $\vec k$ and $\vec m$ defined in GCRS by (\ref{eq:k+b}) and ${\cal I}'_{a_1... a_\ell}$ depends on $\vec k'$ and $\vec m'$ defined in the light ray coordinate system that was introduced by (\ref{eq:note-b}).

Expression (\ref{eq:rot3Ia}) allows us to write the term with tensorial inner product  in (\ref{eq:eik-phA}) as
{}
\begin{align}
{\cal T }^{<a_1...a_\ell>} {\cal I}_{a_1... a_\ell}=
{\cal T}^{<a_1...a_\ell>} R^{b_1}_{a_1}... R^{b_\ell}_{a_\ell}\,{\cal I}'_{b_1... b_\ell} \equiv
{\cal T }'{}^{<a_1...a_\ell>} {\cal I}'_{a_1... a_\ell},
\label{eq:rot3a+}
\end{align}
where ${\cal T}'{}^{<a_l...a_\ell>}$ are the components of the STF mass moment tensor projected on the light ray trajectory
{}
\begin{align}
{\cal T}'{}^{<a_1...a_\ell>}={\cal T}^{<b_1...b_\ell>} R^{a_1}_{b_1}... R^{a_\ell}_{b_\ell}.
\label{eq:rot3a}
\end{align}

Typically, the STF tensor mass moments ${\cal T}^{<a_1...a_\ell>}$ are expressed in terms of the spherical harmonic coefficients that, in turn, are expressed using the Cartesian ECEF coordinates of the GCRS. Expression (\ref{eq:rot3a}) allows us to compute the components of ${\cal I}_{a_1... a_\ell}$ in the rotated coordinate system that is aligned with the direction of propagation.

Through this rotation procedure with matrix (\ref{eq:rot}) and angles (\ref{eq:thet-psi}), we are able to express the STF mass moments with respect to a cylindrical coordinate system in which the $z$-axis coincides with the direction of signal propagation. Moreover, as a non-degenerate linear relationship exists between STF tensor components and spherical harmonic coefficients (see Appendix~\ref{sec:stf-sph-harm}) for details), this procedure also allows us to directly compute spherical harmonic coefficients with respect to this new coordinate system in a three-step process:
\begin{inparaenum}[1)]
\item Convert spherical harmonics to the STF representation;
\item Rotate the STF tensor using (\ref{eq:rot3a}),
\item Invert the equation relating STF tensor components and spherical harmonics and solve for the latter.
\end{inparaenum}
This procedure is powerful and straightforward, also computationally inexpensive, allowing us to express the multipole mass moments of the gravitating body in a coordinate reference frame of arbitrary orientation, including the orientation defined by the direction of signal transmission.

\section{Phase shift introduced by the lowest order multipole moments}
\label{sec:phase-moments}

To demonstrate the practical utility of our results, we now compute several low-order terms in (\ref{eq:eik-ph2*}), for $\ell=2,3,4$. In Appendix~\ref{sec:cases}, we compute the corresponding derivatives with respect to the vector impact parameter, which are present in (\ref{eq:eik-ph2*}).  Below, we present the results for the eikonal phase shift for the quadrupole ($\ell=2$), octupole ($\ell=3$) and hexadecapole ($\ell=4$) STF multipole moments.

\subsection{Quadrupole moment}
\label{sec:quad-mom}

\subsubsection{The structure of the quadrupole phase shift}

In the $\ell=2$ quadrupole case, applying (\ref{eq:eik-phA}) leads to the following expression for the gravitational eikonal phase shift  $\varphi_2(\vec r,\vec r_0)$ induced by the quadrupole STF mass moment,  ${\cal T}^{<ab>}$:
{}
\begin{eqnarray}
\varphi_2(\vec r,\vec r_0)
&=&-{\textstyle\frac{1}{2}} kr_g
{\cal T}^{<ab>}
{\cal I}_{ab}(\vec r,\vec r_0),
\label{eq:delta-eik-gen*}
\end{eqnarray}
where ${\cal I}_{ab}(\vec r,\vec r_0)$ is the $\ell=2$ light ray trajectory projection operator given by (\ref{eq:eik-ph2*}), that is,
{}
\begin{eqnarray}
{\cal I}_{ab}(\vec r,\vec r_0)&=&-
\Big\{
\Big(2m_am_b+k_ak_b\Big)
\frac{1}{r\big(r+({\vec k}\cdot{\vec r})\big)}
+
\Big(k_ak_b-m_am_b\Big)
\frac{({\vec k}\cdot{\vec r})}{r^3}+
\Big(k_am_b+k_bm_a\Big)
\frac{b}{r^3}
\Big\}\Big|^r_{r_0},
\label{eq:delta-eik-proj*}
\end{eqnarray}
where we used the derivatives (\ref{eq:dab51}) and (\ref{eq:dab61}) and $\vec k$ and $\vec m$ are defined by (\ref{eq:k+b}). After some rearrangement,  expression (\ref{eq:delta-eik-proj*}) may be presented  is the following equivalent form:
{}
\begin{eqnarray}
{\cal I}_{ab}(\vec r,\vec r_0)&=&
\Big\{
\Big(2m_am_b+k_ak_b\Big)\Big(
\frac{1}{r\big(r+({\vec k}\cdot{\vec r})\big)}-\frac{({\vec k}\cdot{\vec r})}{2r^3}\Big)
+
\Big(k_am_b+k_bm_a\Big)
\frac{b}{r^3}+
{\textstyle\frac{3}{2}}k_ak_b
\frac{({\vec k}\cdot{\vec r})}{r^3}
\Big\}\Big|^r_{r_0}.
\label{eq:delta-eik-proj*+}
\end{eqnarray}
The convenience of the form (\ref{eq:delta-eik-proj*+}) is due to the fact that in the light ray's coordinate system that we use, $\vec{k}$ is a unit vector in the $e'_z$-axis direction, whereas $\vec{m}$ is a unit vector in the perpendicular $e'_xe'_y$-plane. This simplifies various inner products in (\ref{eq:delta-eik-gen*}) which, relying on (\ref{eq:rot3a+})--(\ref{eq:rot3a}) with parameterization (\ref{eq:note-b}) take the form:
{}
\begin{eqnarray}
{\cal T}^{<ab>}
\Big(2m_am_b+k_ak_b\Big)
={\cal T}'{}^{<ab>}
\Big(2m'_am'_b+k'_ak'_b\Big)
&=&
\Big({\cal T}'_{11}-{\cal T}'_{22}\Big)\cos2\phi_\xi+2{\cal T}'_{12}\sin2\phi_\xi,
\label{eq:delta-mm}
\end{eqnarray}
where $\vec k'$ and $\vec m'$ are from (\ref{eq:note-b}) and we specifically emphasized the use of (\ref{eq:rot3a+}), a reminder to the reader that this scalar-valued tensor product is not dependent on the choice of coordinate system in which it is calculated, allowing us to express the product in this simple form, using the values of the STF tensor and the projection operator in the rotated light-ray coordinate system. Similarly,
{}
\begin{eqnarray}
{\cal T}^{<ab>}
\Big(k_am_b+k_bm_a\Big)&=&2{\cal T}'_{13}\cos\phi_\xi+2{\cal T}'_{23}\sin\phi_\xi,\label{eq:delta-km}\\
{\cal T}^{<ab>}
k_ak_b&=&{\cal T}'_{33},
\label{eq:delta-kk}
\end{eqnarray}
 where we relied on the trace-free nature of ${\cal T}^{<ab>}$, valid in any representation, hence  ${\cal T}'_{11}+{\cal T}'_{22}+{\cal T}'_{33}=0$.

As a result, using these expressions (\ref{eq:delta-mm})--(\ref{eq:delta-kk}) and representing $\vec{m}$ using from (\ref{eq:note-b}), expression (\ref{eq:delta-eik-gen*})
takes the form
{}
\begin{eqnarray}
\varphi_2(\vec r,\vec r_0)
&=& {\textstyle\frac{1}{2}}kr_g
\bigg\{\Big\{\Big({\cal T}'_{11}-{\cal T}'_{22}\Big)\cos2\phi_\xi+
 2{\cal T}'_{12}\sin2\phi_\xi\Big\}\Big(
 \frac{1}{r\big(r+({\vec k}\cdot{\vec r})\big)}-\frac{({\vec k}\cdot{\vec r})}{2r^3}\Big)
+\nonumber\\
 &&\hskip 14pt+\,
2\Big\{{\cal T}'_{13} \,\cos\phi_\xi+
{\cal T}'_{23}\,\sin\phi_\xi\Big\}
\frac{b}{r^3}+3{\cal T}'_{33}
\frac{({\vec k}\cdot{\vec r})}{2r^3}\bigg\}\Big|^r_{r_0}.
\label{eq:phase-sh-quad-prime}
\end{eqnarray}
Thus, as the light travels from a transmitter to a receiver, it samples the gravitational filed along its path. That field is represented by the STF mass moments ${\cal T}'{}^{<ab>}$ that  are related to their GCRS values via (\ref{eq:rot3a}) .

\subsubsection{Rotating the quadrupole mass moment}

What is left is to express ${\cal T}'{}^{<ab>}$  in accordance with (\ref{eq:rot3a}) for $\ell=2$ that has the form
{}
\begin{eqnarray}
{\cal T}'{}^{<ab>}&=&{\cal T}^{<ij>}  R^a_i  R^b_j,
\label{eq:t2}
\end{eqnarray}
where the relationship between ${\cal T}^{<ab>}$ and spherical harmonic coefficients expressed in the same coordinate reference frame are given by comparing (\ref{eq:pot_stf}) against (\ref{eq:pot_w_0sh+}). We work this out explicitly for $\ell=2$ in Appendix~\ref{sec:stf-sph-harm}, yielding the matrix components of ${\cal T}^{<ab>}$ in the GCRS reference frame in the form (\ref{eq:sp-harm2*}):
{}
\begin{eqnarray}
{\cal T}_{11}&=&\Big(-{\textstyle\frac{1}{3}}C_{20}+2C_{22}\Big)R^2, \qquad  {\cal T}_{12}=2S_{22}R^2, \nonumber\\
{\cal T}_{22}&=&\Big(-{\textstyle\frac{1}{3}}C_{20}-2C_{22}\Big)R^2, \qquad {\cal T}_{13}=C_{21}R^2,\nonumber\\
{\cal T}_{33}&=&{\textstyle\frac{2}{3}} C_{20} R^2, \qquad\qquad \qquad\quad~~
{\cal T}_{23}=S_{21}R^2,
\label{eq:sp-harm2*=}
\end{eqnarray}
or, using the relations between ${\cal T}^{<ab>}$ and $C_{2k}, S_{2k}$ from (\ref{eq:sp-harm2*=}), we may express ${\cal T}'{}^{<ab>}$ as
{}
\begin{eqnarray}
R^{-2}{\cal T}'_{11}
&=& -\textstyle\frac{1}{3}\, C_{20}+2\Big(C_{22}\cos2\psi-S_{22}\sin2\psi\Big),\nonumber\\
R^{-2}{\cal T}'_{22}
&=& \textstyle\frac{1}{3}\Big(3\sin^2\theta-1\Big)\, C_{20}-\sin2\theta\Big(C_{21}\sin\psi+S_{21}\cos\psi\Big)+2\cos^2\theta\Big(S_{22}\sin2\psi-C_{22}\cos2\psi\Big),\nonumber\\
R^{-2}{\cal T}'_{33}
&=&
{\textstyle\frac{1}{3}}\big(3\cos^2\theta-1\big)C_{20}+\sin2\theta\Big(C_{21}\sin\psi+S_{21}\cos\psi\Big)+2\sin^2\theta\Big(S_{22}\sin2\psi-C_{22}\cos2\psi\Big),\nonumber\\
R^{-2}{\cal T}'_{12}
&=& -
\sin\theta\Big(C_{21}\cos\psi-S_{21}\sin\psi\Big)+2\cos\theta\Big(S_{22}\cos2\psi+C_{22}\sin2\psi\Big),\nonumber\\
R^{-2}{\cal T}'_{13}
&=&
\cos\theta\Big(C_{21}\cos\psi-S_{21}\sin\psi\Big)+2\sin\theta\Big(S_{22}\cos2\psi+C_{22}\sin2\psi\Big),\nonumber\\
R^{-2}{\cal T}'_{23}
&=& -
{\textstyle\frac{1}{2}}\sin2\theta\,
C_{20}+\cos2\theta\Big(C_{21}\sin\psi+S_{21}\cos\psi\Big)-\sin2\theta\Big(C_{22}\cos2\psi-S_{22}\sin2\psi\Big),
\label{eq:phase-sh-prime}
\end{eqnarray}
where angles $\theta$ and $\psi$ are fixed for each transmitter-receiver configuration and are given by (\ref{eq:thet-psi}).

We define the rotated spherical harmonic coefficients $\big\{C'_{2k},S'_{2k}\big\}$ by substituting primed in place of unprimed terms in (\ref{eq:sp-harm2*=}) and then solving the resulting system of equations. For $\ell=2$, this results in the following relations between $\big\{C'_{2k},S'_{2k}\big\}$ and ${\cal T}'{}^{<ab>}$:
{}
\begin{eqnarray}
C'_{20} =  {\textstyle\frac{3}{2}} R^{-2}{\cal T}'_{33},
\qquad C'_{21} &=& R^{-2}{\cal T}'_{13},
\qquad C'_{22} = {\textstyle\frac{1}{4}} R^{-2}\Big({\cal T}'_{11}-{\cal T}'_{22}\Big),
\qquad
S'_{21} = R^{-2}{\cal T}'_{23},
\qquad \,
S'_{22} = {\textstyle\frac{1}{2}} R^{-2}{\cal T}'_{12}.~~~
  \label{eq:spijT}
\end{eqnarray}

Due to the tensorial nature of (\ref{eq:delta-eik-gen*}), both expressions (\ref{eq:phase-sh-quad-prime}) and (\ref{eq:phase-sh-quad-prime-h+}) demonstrate the form invariance of the gravitational phase (\ref{eq:eik-ph2*})--(\ref{eq:eik-phA}). The structure of the expression for the gravitational phase is the same in any new rotated coordinates, thus for any direction of signal propagation. Furthermore, the relationship between the STF tensor mass moments and  spherical harmonics, ${\cal T}^{<ab>}\Leftrightarrow \big\{C_{2k},S_{2k}\big\}$, given by (\ref{eq:sp-harm2*=}), is also the same in any new coordinates, ${\cal T}'{}^{<ab>}\Leftrightarrow \big\{C'_{2k},S'_{2k}\big\}$. This form invariance of the phase and relevant relations between the moments and harmonics exist at any STF order $\ell$.  This property may be used to establish expressions for the spherical harmonics $C'_{\ell k},S'_{\ell k}$ at any order $\ell$ and will be demonstrated below for $\ell=2,3,4$.

\subsubsection{Quadrupole phase in terms of spherical harmonics}

Results (\ref{eq:phase-sh-prime}) and (\ref{eq:spijT}) allow us to express (\ref{eq:phase-sh-quad-prime})
 in terms of the rotated spherical harmonic coefficients $C'_{2k},S'_{2k}$. Using $\vec n=\vec r/r$, this expression has the form
{}
\begin{eqnarray}
\varphi_2(\vec r,\vec r_0)
&=& kr_g\Big(\frac{R_\oplus}{b}\Big)^2 \bigg\{2\Big\{C'_{22}\cos2\phi_\xi+ S'_{22}\sin2\phi_\xi\Big\}\Big\{
 1-({\vec k}\cdot{\vec n})\Big(1+{\textstyle\frac{1}{2}}\frac{b^2}{r^2}\Big)\Big\}
+\nonumber\\
 &&\hskip 51pt+\,
\Big\{C'_{21} \cos\phi_\xi+
S'_{21}\sin\phi_\xi\Big\}
\frac{b^3}{r^3}+
{\textstyle\frac{1}{2}} C'_{20}
\frac{b^2}{r^2}({\vec k}\cdot{\vec n})\bigg\}\Big|^r_{r_0},
\label{eq:phase-sh-quad-prime-h+}
\end{eqnarray}
 where the relationship between the spherical harmonic coefficients in the GCRS reference frame vs. their value at the OCS which is associated with the light ray, yield the following form for $C'_{2k}$ and $S'_{2k}$:
{}
\begin{eqnarray}
C'_{20} &=&  {\textstyle\frac{1}{4}} \Big(1 + 3 \cos 2 \theta\Big)C_{20}  +  {\textstyle\frac{3}{2}} \sin 2\theta \Big(C_{21} \sin\psi + S_{21} \cos\psi \Big) - 3 \sin^2\theta \Big(C_{22} \cos2 \psi - S_{22} \sin2 \psi\Big),\nonumber\\
C'_{21} &=&  \cos\theta\Big(C_{21} \cos\psi - S_{21} \sin\psi\Big) +  2 \sin\theta\Big(S_{22} \cos 2 \psi + C_{22} \sin2 \psi\Big),\nonumber\\
C'_{22} &=&  -{\textstyle\frac{1}{4}} \sin^2\theta\, C_{20}  + {\textstyle\frac{1}{4}} \sin2\theta\Big(C_{21} \sin\psi+S_{21} \cos\psi\Big) + {\textstyle\frac{1}{4}}\Big(3 + \cos2 \theta\Big) \Big(C_{22} \cos2 \psi - S_{22} \sin2 \psi\Big),
\nonumber\\
S'_{21} &=&  - {\textstyle\frac{1}{2}}  \sin2\theta \,C_{20}+
\cos2\theta\Big( C_{21} \sin\psi+S_{21} \cos\psi\Big) - \sin2\theta\Big(C_{22} \cos2 \psi-S_{22} \sin2 \psi\Big),\nonumber\\
S'_{22} &=& -{\textstyle\frac{1}{2}} \sin\theta\Big(C_{21} \cos\psi-S_{21} \sin\psi \Big) +  \cos\theta\Big(S_{22} \cos 2 \psi + C_{22} \sin2 \psi\Big).
  \label{eq:spij}
\end{eqnarray}

Equation (\ref{eq:phase-sh-quad-prime-h+}), together with (\ref{eq:spij}) is the most general form of eikonal phase shift induced by a mass quadrupole moment, $\ell=2$. Following the same procedure, we can obtain similar expressions for higher orders of spherical harmonics with $\ell \geq 3$. Some of the relevant expressions for the STF tensors and its relations to harmonic coefficients $C_{\ell k}, S_{\ell k}$ in (\ref{eq:pot_w_0}) for orders $\ell=3,4$ are given in \cite{Turyshev-Toth:2021-STF-moments}.  These results may be extended to arbitrary orders $\ell$.

Expressions (\ref{eq:spij}) relate the values of the spherical harmonics coefficients that are sampled by the propagating EM wave, $C'_{\ell k}, S'_{\ell k}$, to  those that are typically reported in in the standard GCRS coordinates, $C_{\ell k}, S_{\ell k}$. For that we performed the rotation of the standard harmonics onto the direction of the light propagation given by $\vec k$.

We note that the change of spherical harmonics under a linear transformation of coordinates has been studied in the past, with a rich history (see, e.g., \cite{Schmidt1899}). In our case, the transformation rules that correspond to rotating the north pole axis to coincide with $\vec{k}$ were accomplished with relative ease as an additional benefit of the use of the STF tensor formalism, which we invoked primarily in order to integrate the eikonal equation. The process, as we shall see below, is easily generalized to higher order harmonics.

\subsubsection{The case of an axisymmetric body}

We know that, in the case of an axisymmetric gravitating body, all of the spherical harmonic coefficients accept for $C_{20}$ vanish, namely $C_{21}=C_{22}=S_{21}=S_{22}=0$. In this case, Eq.~(\ref{eq:phase-sh-quad-prime-h+}), with the help of (\ref{eq:spij}), takes the familiar form:
{}
\begin{eqnarray}
\varphi_2(\vec r,\vec r_0)
&=& {\textstyle\frac{1}{2}}kr_gC_{20}R^2_\oplus
\Big\{-\sin^2\theta\cos2\phi_\xi
\frac{1}{r\big(r+({\vec k}\cdot{\vec r})\big)}+
\nonumber\\
 &&\hskip 55pt+\,
\Big(\cos^2\theta-
\sin^2\theta\sin^2\phi_\xi\Big)
\frac{({\vec k}\cdot{\vec r})}{r^3}-
\sin2\theta\sin\phi_\xi
\frac{b}{r^3}\Big\}\Big|^r_{r_0}.
\label{eq:phase-sh-axisym0}
\end{eqnarray}
Alternatively, using $J_2=-C_{20}$ and relying on the definitions of the unit vectors $\vec{b}$, $\vec{k}$ and $\vec{m}$, we present (\ref{eq:phase-sh-axisym0}) as
{}
\begin{equation}
\varphi_2(\vec r,\vec r_0)
= -{\textstyle\frac{1}{2}}kr_gJ_2R_\oplus^2 \Big\{\Big(2({\vec s}\cdot{\vec m})^2+({\vec s}\cdot{\vec k})^2-1\Big)
\frac{1}{r\big(r+({\vec k}\cdot{\vec r})\big)}+
\big(({\vec s}\cdot{\vec k})^2-({\vec s}\cdot{\vec m})^2\big)\frac{(\vec k\cdot\vec r)}{r^3}+
2({\vec s}\cdot{\vec m}) ({\vec s}\cdot{\vec k}) \frac{b}{r^3}\Big\}\Big|^r_{r_0},
\label{eq:phase-sh-axisym}
\end{equation}
which is exactly the $J_2$ part of (\ref{eq:phiE0}).

\subsection{Octupole moment}
\label{sec:oct-mom}

\subsubsection{The structure of the octupole phase shift}

Setting $\ell=3$ in (\ref{eq:eik-phA}) and (\ref{eq:eik-ph2*}), we use the result for the two types of derivatives (\ref{eq:dab53}) and (\ref{eq:dab63}). We derive the eikonal phase shift, $\varphi_3(\vec r,\vec r_0)$, introduced by the octupole STF mass moment, ${\cal T}^{<abc>}$, which may be given as
{}
\begin{eqnarray}
\varphi_3(\vec r,\vec r_0)
&=&{\textstyle\frac{1}{6}} kr_g
{\cal T}{}^{<abc>}
{\cal I}_{abc}(\vec r,\vec r_0),
\label{eq:delta-eik-octu*}
\end{eqnarray}
where ${\cal I}_{abc}(\vec r,\vec r_0)$ is  light ray trajectory projection operator of the order of $\ell=3$ that is given as
{}
\begin{eqnarray}
{\cal I}_{abc}(\vec r,\vec r_0)&=&
\Big\{\Big(4m_am_bm_c+3k_ak_bm_c\Big) \frac{1}{b}\Big(\frac{2}{r\big(r+(\vec k\cdot \vec r) \big)}-\frac{(\vec k\cdot\vec r)}{r^3}\Big)+\nonumber\\
&+&3\Big(3k_ak_bm_c-m_am_bm_c\Big)b\frac{(\vec k\cdot\vec r)}{r^5}+
3\Big(3k_am_bm_c-k_ak_bk_c\Big)\frac{b^2}{r^5}+
\frac{5k_ak_bk_c}{r^3}
\Big\}\Big|^r_{r_0}.
\label{eq:delta-eik-octu=*}
\end{eqnarray}
Similarly to (\ref{eq:delta-eik-proj*+}), we re-arrange  (\ref{eq:delta-eik-octu*}) to separate individual projection operators
{}
\begin{eqnarray}
{\cal I}_{abc}(\vec r,\vec r_0)&=& 2\Big\{\Big(4m_am_bm_c+3k_ak_bm_c\Big) \Big\{\frac{1}{b}\Big(\frac{1}{r\big(r+(\vec k\cdot \vec r) \big)}-\frac{(\vec k\cdot\vec r)}{2r^3}\Big)-
{\textstyle\frac{3}{8}}b\frac{(\vec k\cdot\vec r)}{r^5}\Big\}+\nonumber\\
&+&
{\textstyle\frac{9}{2}}\Big(k_am_bm_c+{\textstyle\frac{1}{2}}k_ak_bk_c\Big)
\frac{b^2}{r^5}+{\textstyle\frac{45}{8}}k_ak_bm_c\,b\frac{(\vec k\cdot\vec r)}{r^5}+
{\textstyle\frac{5}{2}}k_ak_bk_c\frac{1}{r^3}\Big(1-{\textstyle\frac{3}{2}}\frac{b^2}{r^2}\Big)
\Big\}\Big|^r_{r_0},
\label{eq:delta-eik-octu}
\end{eqnarray}
where, again, $\vec k, \vec m$ are given by (\ref{eq:note-b}). Note that there is no need to explicitly STF this quantity as it will be acting on the STF tensor ${\cal T}^{<abc>}$ in (\ref{eq:delta-eik-octu*}).

Again, the form (\ref{eq:delta-eik-octu}) is convenient because it simplifies various inner products in (\ref{eq:delta-eik-octu*}) when expressed in the light ray's coordinate system:
{}
\begin{eqnarray}
{\cal T}{}^{<abc>}
\Big(4m_am_bm_c+3k_ak_bm_c\Big)&=&
\Big({\cal T}'_{111}-3{\cal T}'_{122}\Big)\cos3\phi_\xi+\Big(3{\cal T}'_{112}-{\cal T}'_{222}\Big)\sin3\phi_\xi,
\label{eq:delta-mm3}\\
{\cal T}{}^{<abc>}\Big(k_am_bm_c+{\textstyle\frac{1}{2}}k_ak_bk_c\Big)&=&{\textstyle\frac{1}{2}}\Big({\cal T}'_{113}-{\cal T}'_{223}\Big)\cos2\phi_\xi+{\cal T}'_{123}\sin2\phi_\xi,
\label{eq:delta-kmm3}\\[5pt]
{\cal T}{}^{<abc>}k_ak_bm_c&=&{\cal T}'_{133} \cos\phi_\xi+{\cal T}'_{233}\sin\phi_\xi,
\label{eq:delta-km3}\\[5pt]
{\cal T}{}^{<abc>} k_ak_bk_c&=&{\cal T}'_{333}.
\label{eq:delta-kk3}
\end{eqnarray}

 As a result, the octupole gravitational phase shift corresponding to $\ell=3$ from (\ref{eq:delta-eik-octu*}) takes the following  form:
{}
\begin{eqnarray}
\varphi_3(\vec r,\vec r_0)
&=&{\textstyle\frac{1}{3}} kr_g
\bigg\{\Big\{
\Big({\cal T}'_{111}-3{\cal T}'_{122}\Big)\cos3\phi_\xi+\Big(3{\cal T}'_{112}-{\cal T}'_{222}\Big)\sin3\phi_\xi\Big\}
\Big\{\frac{1}{b}\Big(\frac{1}{r\big(r+(\vec k\cdot \vec r) \big)}-\frac{(\vec k\cdot\vec r)}{2r^3}\Big)-{\textstyle\frac{3}{8}} b\frac{(\vec k\cdot\vec r)}{r^5}\Big\}+\nonumber\\
&+&
{\textstyle\frac{9}{4}}
\Big\{\Big({\cal T}'_{113}-{\cal T}'_{223}\Big)\cos2\phi_\xi+2{\cal T}'_{123}\sin2\phi_\xi\Big\}\frac{b^2}{r^5}+
{\textstyle\frac{45}{8}}\Big\{{\cal T}'_{133} \cos\phi_\xi+{\cal T}'_{233}\sin\phi_\xi\Big\}b\frac{(\vec k\cdot\vec r)}{r^5}
+\nonumber\\
&+&
{\textstyle\frac{5}{2}}{\cal T}'_{333} \frac{1}{r^3}\Big(
1-{\textstyle\frac{3}{2}}\frac{b^2}{r^2}\Big)
\bigg\}\Big|^r_{r_0}.
\label{eq:delta-eik-octu*all-prime}
\end{eqnarray}

\subsubsection{Rotating the octupole mass moment}

The next step is to express ${\cal T}'{}^{<abc>}$ present in (\ref{eq:delta-eik-octu*all-prime}) in terms of the spherical harmonics coefficients. For that, we  implement the rotation of ${\cal T}^{<abc>}$ in accordance with (\ref{eq:rot3a}):
{}
\begin{eqnarray}
{\cal T}'{}^{<abc>}&=&{\cal T}^{<ijk>}  R^a_i  R^b_j  R^c_k,
  \label{eq:rot3+}
\end{eqnarray}
 where ${\cal T}^{<ijk>} $ are the components of the Cartesian representation of the octupole mass tensor in the GCRS.  These components are known and are related to the  spherical harmonics $C_{3k},S_{3k}$. In Appendix~\ref{sec:stf-sph-harm}, we established the relationship between STF moments and spherical harmonics for $\ell=2$. The same approach may be used for $\ell=3$ to establish the  the correspondence between ${\cal T}^{<abc>}$ from (\ref{eq:pot_w_0STF}) and the spherical harmonics from (\ref{eq:pot_w_0sh})  and (\ref{eq:pot_w_0STF2}). The result given by
{}
\begin{eqnarray}
{\cal T}_{111}&=&\Big({\textstyle\frac{3}{5}}C_{31}-6C_{33}\Big)R^3,\qquad\quad
{\cal T}_{112}=\Big({\textstyle\frac{1}{5}}S_{31}-6S_{33}\Big)R^3,
\qquad\quad
{\cal T}_{113}=\Big(-{\textstyle\frac{1}{5}}C_{30}+2C_{32}\Big)R^3,
\nonumber\\
{\cal T}_{122}&=&\Big({\textstyle\frac{1}{5}}C_{31}+6C_{33}\Big)R^3,\qquad\quad
{\cal T}_{222}=\Big({\textstyle\frac{3}{5}}S_{31}+6S_{33}\Big)R^3,
\qquad\quad
{\cal T}_{223}=\Big(-{\textstyle\frac{1}{5}}C_{30}-2C_{32}\Big)R^3,
\nonumber\\
{\cal T}_{123}&=&2S_{32}R^3, \qquad\quad
{\cal T}_{133}=-{\textstyle\frac{4}{5}}C_{31}R^3,  \qquad\quad
{\cal T}_{233}=-{\textstyle\frac{4}{5}}S_{31}R^3,  \qquad\quad
{\cal T}_{333}={\textstyle\frac{2}{5}}C_{30}R^3.
\label{eq:sp-harm3}
\end{eqnarray}
It is easy to check that the rank-3 STF tensor ${\cal T}^{<abc>}$ has seven independent components; the values of the remaining 20 components are determined by its symmetries and vanishing trace. The system of equations (\ref{eq:sp-harm3}) is redundant, with only seven independent equations relating the components of ${\cal T}^{<abc>}$ to the seven spherical harmonic coefficients in the case $\ell=3$.

We can now implement the rotation (\ref{eq:rot3+}) that in accordance with (\ref{eq:rot3a}) and obtain the following components of the STF mass moments  ${\cal T}'{}^{<abc>}$ expressed via spherical harmonics of GCRS system:
{}
\begin{eqnarray}
R^{-3} {\cal T}'_{111} &=& {\textstyle\frac{3}{5}} \Big(C_{31} \cos\psi - S_{31} \sin\psi\Big) +
  6 \Big(S_{33} \sin3 \psi - C_{33} \cos3 \psi\Big),\nonumber\\
R^{-3} {\cal T}'_{112} &=& {\textstyle\frac{1}{5}}\sin\theta C_{30} +
{\textstyle\frac{1}{5}} \cos\theta \Big(S_{31} \cos\psi + C_{31} \sin\psi\Big) +
  2 \sin\theta \Big(S_{32} \sin2 \psi - C_{32} \cos2 \psi\Big) -  \nonumber\\
&&-\,
  6 \cos\theta \Big(S_{33} \cos3 \psi + C_{33} \sin3 \psi\Big),   \nonumber\\
R^{-3} {\cal T}'_{113} &=& -{\textstyle\frac{1}{5}} \cos\theta C_{30} +
 {\textstyle\frac{1}{5}} \sin\theta \Big(S_{31} \cos\psi + C_{31} \sin\psi\Big) +
  2 \cos\theta \Big( C_{32} \cos2 \psi - S_{32} \sin2 \psi\Big) -  \nonumber\\
&&-\,
  6 \sin\theta \Big(S_{33} \cos3 \psi + C_{33} \sin3 \psi\Big),
 \nonumber\\
R^{-3} {\cal T}'_{122} &=&  {\textstyle\frac{4}{5}}  \sin^2\theta \Big(S_{31} \sin\psi - C_{31} \cos\psi\Big) - 2 \sin2 \theta \Big(S_{32} \cos2 \psi + C_{32} \sin2 \psi\Big) +
  \nonumber\\
&&+\,
  {\textstyle\frac{1}{5}}\cos^2\theta \Big(C_{31} \cos\psi - S_{31} \sin\psi\Big) +
  6 \cos^2\theta \Big(C_{33} \cos3 \psi - S_{33} \sin3 \psi\Big),  \nonumber\\
R^{-3} {\cal T}'_{222} &=&  {\textstyle\frac{1}{10}} \Big(1 + 5 \cos2 \theta\Big) \sin\theta C_{30} +  {\textstyle\frac{3}{10}} \Big(-3 + 5 \cos2 \theta\Big) \cos\theta \Big(S_{31} \cos\psi + C_{31} \sin\psi\Big) + \nonumber\\
 &&+\,
  3 \cos\theta \sin2 \theta \Big(C_{32} \cos2 \psi - S_{32} \sin2 \psi\Big) +
  6 \cos^3\theta \Big( S_{33} \cos3 \psi + C_{33} \sin3 \psi\Big), \nonumber\\
R^{-3} {\cal T}'_{223} &=& {\textstyle\frac{1}{10}}  C_{30} \cos\theta \Big(3 - 5 \cos2 \theta\Big) +
 {\textstyle\frac{1}{10}} \Big(7 + 15 \cos2 \theta\Big) \sin\theta (S_{31} \cos\psi + C_{31} \sin\psi\Big) -
\nonumber\\
&&-\,
 {\textstyle\frac{1}{2}} \Big (\cos\theta + 3 \cos3 \theta\Big) (C_{32} \cos2 \psi -
     S_{32} \sin2 \psi\Big) +
  6 \cos^2\theta \sin\theta \Big(S_{33} \cos3 \psi + C_{33} \sin3 \psi\Big), \nonumber\\
R^{-3} {\cal T}'_{123} &=& \sin\theta \cos\theta \Big(C_{31} \cos\psi -  S_{31} \sin\psi\Big) +
  2 \cos2 \theta \Big(C_{32}  \sin2 \psi + S_{32} \cos2 \psi\Big) +
  3 \sin2 \theta \Big(C_{33} \cos3 \psi - S_{33} \sin3 \psi\Big),
  \nonumber\\
R^{-3}  {\cal T}'_{133} &=& {\textstyle\frac{1}{10}} \Big(3 + 5 \cos2 \theta\Big) \Big(S_{31} \sin\psi - C_{31} \cos\psi\Big) + 2 \sin2 \theta \Big(S_{32} \cos2 \psi + C_{32} \sin2 \psi\Big) + \nonumber\\
&&+\,
  6 \sin^2\theta \Big(C_{33} \cos3 \psi - S_{33} \sin3 \psi\Big),  \nonumber\\
R^{-3} {\cal T}'_{233} &=& -{\textstyle\frac{1}{20}}  C_{30} \Big(\sin\theta + 5 \sin3 \theta\Big) +  {\textstyle\frac{1}{10}} \cos\theta \Big(7 - 15 \cos2 \theta\Big) \Big(S_{31} \cos\psi + C_{31} \sin\psi\Big) -  \nonumber\\
 &&-\,
\Big(1 + 3 \cos2 \theta\Big) \sin\theta \Big(C_{32} \cos2 \psi - S_{32} \sin2 \psi\Big) +
  6 \cos\theta \sin^2\theta \Big(S_{33} \cos3 \psi +  C_{33} \sin3 \psi\Big),
  \nonumber\\
R^{-3}{\cal T}'_{333} &=& {\textstyle\frac{1}{20}} C_{30}\Big (3 \cos\theta + 5 \cos 3 \theta\Big) -
 {\textstyle\frac{3}{10}}  \Big(3 + 5 \cos2 \theta\Big) \sin\theta \Big(S_{31} \cos\psi + C_{31} \sin\psi\Big) - \nonumber\\
&&-\,
6  \cos\theta \sin^2\theta \Big(C_{32} \cos2 \psi - S_{32} \sin2 \psi\Big) +
  6 \sin^3\theta \Big(S_{33} \cos3 \psi + C_{33} \sin3 \psi\Big).
\label{eq:phase-sh-octu}
\end{eqnarray}

The relations between $\big\{C'_{3k},S'_{3k}\big\}$ and ${\cal T}'{}^{<abc>}$ have the same structure as in (\ref{eq:sp-harm3}) and thus are given as
{}
\begin{eqnarray}
C'_{30} &=&  {\textstyle\frac{5}{2}} R^{-3}{\cal T}'_{333},
\qquad C'_{31} =  -{\textstyle\frac{5}{4}} R^{-3}{\cal T}'_{133},
\qquad C'_{32} = {\textstyle\frac{1}{4}} R^{-3}\Big({\cal T}'_{113}-{\cal T}'_{223}\Big),
\qquad C'_{33} = {\textstyle\frac{1}{24}} R^{-3}\Big(3{\cal T}'_{122}-{\cal T}'_{111}\Big),
\nonumber\\
S'_{31} &=&-{\textstyle\frac{5}{4}} R^{-3}{\cal T}'_{233},
\qquad \,
S'_{32} = {\textstyle\frac{1}{2}} R^{-3}{\cal T}'_{123},
\qquad~~~
S'_{33} =  {\textstyle\frac{1}{24}} R^{-3}\Big({\cal T}'_{222}-3{\cal T}'_{112}\Big).~~
  \label{eq:spijT3}
\end{eqnarray}

\subsubsection{Octupole phase in terms of spherical harmonics}

The expressions developed in the precessing section allow us to write (\ref{eq:delta-eik-octu*all-prime}) in terms of the spherical harmonic coefficients $C'_{3k},S'_{3k}$ as below
{}
\begin{eqnarray}
\varphi_3(\vec r,\vec r_0)
&=&
-kr_g \Big(\frac{R_\oplus}{b}\Big)^3 \bigg\{8\Big\{
C'_{33}\cos3\phi_\xi+S'_{33}\sin3\phi_\xi\Big\}\Big\{1-
(\vec k\cdot \vec n)\Big(1+
{\textstyle\frac{1}{2}}\frac{b^2}{r^2}+{\textstyle\frac{3}{8}} \frac{b^4}{r^4}\Big)\Big\}-\nonumber\\
&-&
3\Big\{C'_{32}\cos2\phi_\xi+S'_{32}\sin2\phi_\xi\Big\}\frac{b^5}{r^5}+
{\textstyle\frac{3}{2}} \Big\{C'_{31} \cos\phi_\xi+S'_{31}\sin\phi_\xi\Big\}\frac{b^4}{r^4}(\vec k\cdot\vec n)- {\textstyle\frac{1}{3}} C'_{30}\frac{b^3}{r^3}\Big(
1-{\textstyle\frac{3}{2}} \frac{b^2}{r^2}\Big)
\bigg\}\bigg|^r_{r_0},~~~
\label{eq:delta-eik-octu*all-prime-h=}
\end{eqnarray}
where, using (\ref{eq:spijT3}) and (\ref{eq:phase-sh-octu}), we find that $C'_{3 k}$ and $S'_{3 k}$ are related to their unrotated GCRS counterparts as
{}
\begin{eqnarray}
C'_{30} &=& {\textstyle\frac{1}{8}}  \Big(3 \cos\theta + 5 \cos3 \theta\Big) C_{30} - {\textstyle\frac{3}{4}}  \Big(3 + 5 \cos2 \theta\Big) \sin\theta \Big(S_{31} \cos\psi +
 C_{31} \sin\psi\Big) - \nonumber\\
 &&-\,
 15 \cos\theta \sin^2\theta\Big(C_{32} \cos2 \psi -
S_{32}\sin2 \psi\Big) +
15 \sin^3\theta \Big(S_{33} \cos3 \psi + C_{33} \sin3 \psi\Big), \nonumber\\
C'_{31} &=& {\textstyle\frac{1}{8}}  \Big(3 + 5 \cos2 \theta\Big) \Big(C_{31} \cos\psi - S_{31}  \sin\psi\Big) -
{\textstyle\frac{5}{2}} \sin2 \theta \Big(S_{32}\cos2 \psi + C_{32} \sin2 \psi\Big) - \nonumber\\
&&-\,
{\textstyle\frac{15}{2}}  \sin^2\theta\Big(C_{33} \cos3 \psi - S_{33} \sin3 \psi\Big), \nonumber\\
C'_{32} &=& -{\textstyle\frac{1}{4}}\cos\theta \sin^2\theta\, C_{30} -
  {\textstyle\frac{1}{8}}  \Big(1 + 3 \cos2 \theta\Big) \sin\theta \Big(S_{31} \cos\psi +
     C_{31} \sin\psi\Big) + \nonumber\\
     &&+\,
  {\textstyle\frac{1}{8}}  \Big(5 \cos\theta + 3 \cos3 \theta\Big) \Big(C_{32} \cos2 \psi -
     S_{32}\sin2 \psi\Big) -
 {\textstyle\frac{3}{8}} \Big (5 \sin\theta + \sin3 \theta\Big) \Big(C_{33} \sin3 \psi+S_{33} \cos3 \psi\Big), \nonumber\\
C'_{33} &=& -  {\textstyle\frac{1}{8}}   \sin^2\theta\Big(C_{31} \cos\psi - S_{31} \sin\psi\Big) - {\textstyle\frac{1}{4}}  \sin2 \theta \Big(C_{32} \sin2 \psi+S_{32} \cos2 \psi\Big) + \nonumber\\
 &&+\, {\textstyle\frac{1}{8}} \Big(5 + 3 \cos2 \theta\Big) \Big(C_{33} \cos3 \psi - S_{33} \sin3 \psi\Big), \nonumber\\
S'_{31} &=& {\textstyle\frac{1}{8}}  \sin\theta \Big(3 + 5\cos2 \theta\Big)C_{30} +
  {\textstyle\frac{1}{8}}  \cos\theta \Big(15 \cos2 \theta-7\Big) \Big(C_{31} \sin\psi+S_{31} \cos\psi\Big) - \nonumber\\
  &&-\,
  {\textstyle\frac{5}{8}}  \Big(\sin\theta - 3 \sin3 \theta\Big) \Big(C_{32} \cos2 \psi -
     S_{32}\sin2 \psi\Big) -
  {\textstyle\frac{15}{2}}  \cos\theta \sin^2\theta\Big(S_{33} \cos3 \psi +
     C_{33} \sin3 \psi\Big), \nonumber\\
S'_{32} &=& \cos2 \theta \Big(C_{32} \sin2 \psi+S_{32}\cos2 \psi \Big) +
  {\textstyle\frac{1}{4}}  \sin2 \theta \Big(C_{31} \cos\psi - S_{31} \sin\psi\Big) +
  {\textstyle\frac{3}{2}}  \sin2 \theta \Big(C_{33} \cos3 \psi - S_{33} \sin3 \psi\Big),
  \nonumber\\
S'_{33} &=& - {\textstyle\frac{1}{24}}   \sin^3\theta\, C_{30} -  {\textstyle\frac{1}{8}}  \cos\theta \sin^2\theta \Big(C_{31} \sin\psi+S_{31} \cos\psi\Big) +  {\textstyle\frac{1}{16}}  \Big(5 \sin\theta + \sin3 \theta\Big) \Big(C_{32} \cos2 \psi -S_{32} \sin2 \psi\Big) + \nonumber\\
&&+\,
 {\textstyle\frac{1}{16}}  \Big(15 \cos\theta + \cos3 \theta\Big) \Big(C_{33} \sin3 \psi+S_{33} \cos3 \psi\Big).
  \label{eq:spij3}
\end{eqnarray}

\subsubsection{The case of an axisymmetric body}

We know that, in the case of an axisymmetric gravitating body, all the spherical harmonic coefficients except $C_{30}$ vanish: $C_{31}=C_{32}=C_{33}=S_{21}=S_{22}=S_{33}=0$. In this case, expression (\ref{eq:delta-eik-octu*all-prime-h=}) with (\ref{eq:spij3}) takes the form
{}
\begin{eqnarray}
\varphi_3(\vec r,\vec r_0)
&=&{\textstyle\frac{1}{3}} kr_g C_{30}R^3_\oplus
\Big\{
-\sin\theta\sin\phi_\xi\Big(4\sin^2\theta\sin^2\phi_\xi -3\sin^2\theta\Big)\Big\{\frac{1}{b}\Big(\frac{1}{r\big(r+(\vec k\cdot \vec r) \big)}-\frac{(\vec k\cdot\vec r)}{2r^3}\Big)\Big\}+\nonumber\\
&&+\,
{\textstyle\frac{3}{2}}
\sin\theta\sin\phi_\xi\Big(\sin^2\theta\sin^2\phi_\xi-3\cos^2\theta\Big)b\frac{(\vec k\cdot\vec r)}{r^5}-\nonumber\\
&&-\,
{\textstyle\frac{3}{2}}\cos\theta\Big(\cos^2\theta-3\sin^2\theta \sin^2\phi_\xi\Big)\frac{b^2}{2r^5}-
{\textstyle\frac{1}{2}}\cos\theta\big(3-5\cos^2\theta\big)
\frac{1}{r^3}
\Big\}\Big|^r_{r_0}.
\label{eq:delta-eik-octu00*}
\end{eqnarray}
Again, relying on the definitions of the unit vectors $\vec m,\vec k,\vec s$  and using $J_3=-C_{30}$, we present (\ref{eq:phase-sh-axisym0}) as
{}
\begin{eqnarray}
\varphi_3(\vec r,\vec r_0)
&=& -{\textstyle\frac{1}{3}}kr_gJ_3R_\oplus^3\Big\{
(\vec s\cdot \vec m)\big(4({\vec s}\cdot{\vec m})^2+3({\vec s}\cdot{\vec k})^2-3\big)
\Big(\frac{1}{b}\Big(\frac{1}{r\big(r+({\vec k}\cdot{\vec r})\big)}-\frac{(\vec k\cdot \vec r)}{2r^3}\Big)-
\nonumber\\
&&\hskip -40pt -\,
{\textstyle\frac{3}{2}}(\vec s\cdot \vec m)
\big(({\vec s}\cdot{\vec m})^2-3({\vec s}\cdot{\vec k})^2\big)\frac{b(\vec k\cdot\vec r)}{r^5}-
{\textstyle\frac{3}{2}}(\vec s\cdot \vec k)\big(({\vec s}\cdot{\vec k})^2-3({\vec s}
\cdot{\vec m})^2\big)\frac{b^2}{r^5}-{\textstyle\frac{1}{2}}({\vec s}\cdot{\vec k})\big(3-5({\vec s}\cdot{\vec k})^2\big)\frac{1}{r^3}\Big\}\Big|^r_{r_0},~~~
\label{eq:phase-sh-axisymJ3}
\end{eqnarray}
which checks out nicely with the relevant $J_3$-part of (\ref{eq:phiE0}).

\subsection{Hexadecapole moment}
\label{sec:hexa-mom}

\subsubsection{The structure of the hexadecapole phase shift}

In the case when $\ell=4$, we use the derivatives (\ref{eq:dab54}), (\ref{eq:dab64})  and derive the eikonal phase shift, $\varphi_4(\vec r,\vec r_0)$ from (\ref{eq:eik-phA}), introduced by the hexadecapole STF moment, ${\cal T}^{<abcd>}$, in the form
{}
\begin{eqnarray}
\varphi_4(\vec r,\vec r_0)
&=&-{\textstyle\frac{1}{24}} kr_g
{\cal T}{}^{<ijkl>}
{\cal I}_{abcd}(\vec r,\vec r_0),
\label{eq:delta-eik-hexa*}
\end{eqnarray}
where ${\cal I}_{abcd}(\vec r,\vec r_0)$ is the $\ell=4$ ligh ray's trajectory projection operator that is given as
{}
\begin{eqnarray}
{\cal I}_{abcd}(\vec r,\vec r_0)&=&
-3\bigg\{\Big(8m_am_bm_cm_d+8k_ak_bm_cm_d+k_ak_bk_ck_d\Big)\frac{1}{b^2}\Big(\frac{2}{r\big(r+({\vec k}\cdot{\vec r})\big)}-\frac{(\vec k\cdot\vec r)}{r^3}\Big)+\nonumber\\
&&\hskip -60pt +\,
2\Big(4k_ak_bk_ck_d-3k_ak_bm_cm_d-3m_am_bm_cm_d\Big) \frac{(\vec k\cdot \vec r)}{r^5}+
5\Big(6k_ak_bm_cm_d-k_ak_bk_ck_d-m_am_bm_cm_d\Big) b^2\frac{(\vec k\cdot \vec r)}{r^7}+\nonumber\\
&+&
20\Big(k_am_bm_cm_d-k_ak_bk_cm_d\Big)\frac{b^3}{r^7} +28 k_ak_bk_cm_d \frac{b}{r^5}\Big)
\bigg\}\Big|^r_{r_0}.
\label{eq:delta-eik-hexa*=}
\end{eqnarray}
As we did for (\ref{eq:delta-eik-proj*+}) and (\ref{eq:delta-eik-octu}), we identically re-arrange the terms in (\ref{eq:delta-eik-hexa*=}) to make it more convenient for calculations
{}
\begin{eqnarray}
{\cal I}_{abcd}(\vec r,\vec r_0)&=&
-6\bigg\{\Big(8m_am_bm_cm_d+8k_ak_bm_cm_d+k_ak_bk_ck_d\Big)\times\nonumber\\
&&\hskip 60pt \times\,
\Big\{\frac{1}{b^2}\Big(\frac{1}{r\big(r+({\vec k}\cdot{\vec r})\big)}-\frac{(\vec k\cdot\vec r)}{2r^3}\Big)-{\textstyle\frac{3}{8}}\frac{(\vec k\cdot \vec r)}{r^5}-
{\textstyle\frac{5}{16}}b^2\frac{(\vec k\cdot \vec r)}{r^7}
\Big\}+\nonumber\\
&+&
10\Big(k_am_bm_cm_d+{\textstyle\frac{3}{4}} k_ak_bk_cm_d\Big)\frac{b^3}{r^7} +
{\textstyle\frac{35}{2}}\Big(k_ak_bm_cm_d +{\textstyle\frac{1}{2}} k_ak_bk_ck_d\Big)b^2\frac{(\vec k\cdot \vec r)}{r^7}+\nonumber\\
&+&
14 k_ak_bk_cm_d \frac{b}{r^5}\Big(1-{\textstyle\frac{5}{4}}\frac{b^2}{r^2} \Big)+
{\textstyle\frac{35}{8}} k_ak_bk_ck_d \frac{(\vec k\cdot \vec r)}{r^5}\Big(1-
{\textstyle\frac{5}{2}}  \frac{b^2}{r^2}\Big)
\bigg\}\Big|^r_{r_0}.
\label{eq:delta-eik-hexa}
\end{eqnarray}
Similarly to (\ref{eq:delta-eik-octu}),  there is no need to make ${\cal I}_{abcd}(\vec r,\vec r_0)$ to be an STF quantity as it will be acting on the STF tensor ${\cal T}^{<abcd>}$ in (\ref{eq:delta-eik-hexa*}), so that the presence of delta Kronecker symbols will not change the overall result.

The form (\ref{eq:delta-eik-hexa}) is convenient as it allows to express various inner products in (\ref{eq:delta-eik-hexa*}) in the light ray's system of coordinates and using (\ref{eq:rot3a}), yielding remarkably structured results:
{}
\begin{eqnarray}
{\cal T}{}^{<abcd>}
\Big(8m_am_bm_cm_d+8k_ak_bm_cm_d+k_ak_bk_ck_d\Big)&=&\nonumber\\
&&\hskip -80pt =\,
\Big({\cal T}'_{1111}+{\cal T}'_{2222}-6{\cal T}'_{1122}\Big)\cos4\phi_\xi+4\Big({\cal T}'_{1112}-{\cal T}'_{1222}\Big)\sin4\phi_\xi,
\label{eq:delta-mm4}\\
{\cal T}{}^{<abcd>}\Big(k_am_bm_cm_d+{\textstyle\frac{3}{4}} k_ak_bk_cm_d\Big)&=&
{\textstyle\frac{1}{4}}\Big({\cal T}'_{1113}-3{\cal T}'_{1223}\Big)\cos3\phi_\xi+
{\textstyle\frac{1}{4}}\Big(3{\cal T}'_{1123}-{\cal T}'_{2223}\Big)\sin3\phi_\xi,~~~~~
\label{eq:delta-kmm4}\\[5pt]
{\cal T}{}^{<abcd>}\Big(k_ak_bm_cm_d +{\textstyle\frac{1}{2}} k_ak_bk_ck_d\Big)&=&
{\textstyle\frac{1}{2}}\big({\cal T}'_{1133}-{\cal T}'_{2233}\Big)\cos2\phi_\xi+
{\cal T}'_{1223}\sin2\phi_\xi,
\label{eq:delta-km4}\\[5pt]
{\cal T}{}^{<abcd>} k_ak_bk_cm_d&=&{\cal T}'_{1333}\cos\phi_\xi+{\cal T}'_{2333}\sin\phi_\xi,
\label{eq:delta-kkm4}\\[5pt]
{\cal T}{}^{<abcd>} k_ak_bk_ck_d&=&{\cal T}'_{3333}.
\label{eq:delta-kk4}
\end{eqnarray}

Using the expressions (\ref{eq:delta-mm4})--(\ref{eq:delta-kk4}), as a result, we obtain the following compact form for the hexadecapole gravitational phase shift expressed via rotated $\ell=4$ STM mass moments:
{}
\begin{eqnarray}
\varphi_4(\vec r,\vec r_0)
&=&{\textstyle\frac{1}{4}} kr_g
\bigg\{\Big\{\Big({\cal T}'_{1111}+{\cal T}'_{2222}-6{\cal T}'_{1122}\Big)\cos4\phi_\xi+
4\Big({\cal T}'_{1112}-{\cal T}'_{1222}\Big)\sin4\phi_\xi\Big\}\times\nonumber\\
&&\hskip 70pt
\times\, \Big\{\frac{1}{b^2}\Big(\frac{1}{r\big(r+({\vec k}\cdot{\vec r})\big)}-\frac{(\vec k\cdot\vec r)}{2r^3}\Big)-{\textstyle\frac{3}{8}}
\frac{(\vec k\cdot \vec r)}{r^5}-{\textstyle\frac{5}{16}} b^2\frac{(\vec k\cdot \vec r)}{r^7}\Big\}+\nonumber\\
&&\hskip -60pt+\,
{\textstyle\frac{5}{2}} \Big\{\Big({\cal T}'_{1113}-3{\cal T}'_{1223}\Big)\cos3\phi_\xi+\Big(3{\cal T}'_{1123}-{\cal T}'_{2223}\Big)\sin3\phi_\xi\Big\}\frac{b^3}{r^7} -
{\textstyle\frac{35}{2}}\Big\{{\textstyle\frac{1}{2}}\Big({\cal T}'_{1133}-{\cal T}'_{2233}\Big)\cos2\phi_\xi+
{\cal T}'_{1223}\sin2\phi_\xi\Big\} b^2\frac{(\vec k\cdot \vec r)}{r^7}+\nonumber\\
&+&
14 \Big\{{\cal T}'_{1333}\cos\phi_\xi+{\cal T}'_{2333}\sin\phi_\xi\Big\} \frac{b}{r^5}\Big(1-
{\textstyle\frac{5}{4}} \frac{b^2}{r^2}\Big)+
{\textstyle\frac{35}{8}} {\cal T}'_{3333} \frac{(\vec k\cdot \vec r)}{r^5}\Big(1-{\textstyle\frac{5}{2}} \frac{b^2}{r^2}\Big)
\bigg\}\Big|^r_{r_0}.
\label{eq:delta-eik-hexa-pr}
\end{eqnarray}

\subsubsection{Rotating the hexadecapole mass moment}

We now need to establish the relationships between ${\cal T}^{<abcd>}$ and the spherical harmonics, $C_{4k}, S_{k}, k=0...4$, that are given with respect to the GCRS.
For that, as we did for the cases with $\ell=2,3$, we establish a correspondence between ${\cal T}^{<abcd>}$ from (\ref{eq:pot_w_0STF}) and the spherical harmonics coefficients  for $\ell=4$ that are present in (\ref{eq:pot_w_0sh}) (i.e., using the same procedure demonstrated in Appendix \ref{sec:stf-sph-harm}):
{}
\begin{eqnarray}
{\cal T}_{1111}&=&\Big({\textstyle\frac{3}{35}}C_{40}- {\textstyle\frac{12}{7}}C_{42}+24C_{44}\Big)R^4,\qquad\quad
{\cal T}_{2222}=\Big({\textstyle\frac{3}{35}}C_{40}+{\textstyle\frac{12}{7}}C_{42}+24C_{44}\Big)R^4,
\nonumber\\
{\cal T}_{1112}&=&\Big(-{\textstyle\frac{6}{7}}S_{42}+ 24S_{44}\Big)R^4,\qquad\quad~\,
{\cal T}_{1113}=\Big({\textstyle\frac{3}{7}}C_{41}-{\textstyle\frac{60}{7}}C_{43}\Big)R^4,
\qquad\quad
{\cal T}_{2223}=\Big({\textstyle\frac{3}{7}}S_{41}+{\textstyle\frac{60}{7}}S_{43}\Big)R^4,
\nonumber\\
{\cal T}_{1122}&=&\Big({\textstyle\frac{1}{35}}C_{40}- 24C_{44}\Big)R^4,
\qquad\quad~~~\,
{\cal T}_{1123}=\Big({\textstyle\frac{1}{7}}S_{41}-{\textstyle\frac{60}{7}}S_{43}\Big)R^4,
\qquad\quad\,\,
{\cal T}_{1133}=\Big(-{\textstyle\frac{4}{35}}C_{40}+{\textstyle\frac{12}{7}}C_{42}\Big)R^4,
\nonumber\\
{\cal T}_{2233}&=&\Big(-{\textstyle\frac{4}{35}}C_{40}-{\textstyle\frac{12}{7}}C_{42}\Big)R^4, \qquad\quad
{\cal T}_{1222}=\Big(-{\textstyle\frac{6}{7}}S_{42}- 24S_{44}\Big)R^4,
\qquad\,
{\cal T}_{1223}=\Big({\textstyle\frac{1}{7}}C_{41}+{\textstyle\frac{60}{7}}C_{43}\Big)R^4,
\nonumber\\
{\cal T}_{3333}&=&{\textstyle\frac{8}{35}}C_{40}R^4, \qquad\quad
{\cal T}_{1233}={\textstyle\frac{12}{7}}S_{42}R^4,  \qquad\quad
{\cal T}_{1333}=-{\textstyle\frac{4}{7}}C_{41}R^4,  \qquad\quad
{\cal T}_{2333}=-{\textstyle\frac{4}{7}}S_{41}R^4.
\label{eq:sp-harm4}
\end{eqnarray}
As these quantities are the non-vanishing components of the hexadecapole  STF mass moment tensor, out of the fifteen terms ${\cal T}^{<abcd>}$ in (\ref{eq:sp-harm4}), only nine are independent.

We can now rotate ${\cal T}^{<abcd>}$ and obtain ${\cal T}'{}^{<abcd>}$ in the light ray's system of coordinates, using in (\ref{eq:rot3a}):
{}
\begin{eqnarray}
{\cal T}'{}^{<abcd>}&=&{\cal T}^{<ijkl>}  R^a_i  R^b_j  R^c_k  R^d_l.
  \label{eq:rot4+}
\end{eqnarray}
This allows us to derive the following components of the STF mass moments in the rotated coordinate system:
{}
\begin{eqnarray}
R^{-4}{\cal T}'_{1111} &=&{\textstyle\frac{3}{35}}
 C_{40} + {\textstyle\frac{12}{7}} \Big(S_{42} \sin2 \psi - C_{42}\cos 2 \psi\Big) +
  24 \Big(C_{44} \cos4\psi - S_{44} \sin4 \psi\Big),\nonumber\\
R^{-4}{\cal T}'_{1112} &=&
{\textstyle\frac{3}{7}}\sin\theta\Big(S_{41} \sin\psi - C_{41} \cos\psi\Big) +
{\textstyle\frac{60}{7}}
 \sin\theta \Big(C_{43} \cos 3 \psi - S_{43} \sin3 \psi\Big) -
  {\textstyle\frac{6}{7}}\cos\theta \Big(C_{42} \sin2 \psi + S_{42}\cos 2 \psi\Big) + \nonumber\\
  &&+\,
  24 \cos\theta  \Big(S_{44} \cos4\psi + C_{44} \sin4 \psi\Big), \nonumber\\
R^{-4}{\cal T}'_{1113} &=&
{\textstyle\frac{6}{7}} \cos\theta \Big(C_{41} \cos\psi - S_{41} \sin\psi\Big) +
 {\textstyle\frac{60}{7}}  \cos\theta \Big(S_{43} \sin3 \psi - C_{43} \cos 3 \psi\Big) -
  {\textstyle\frac{6}{7}} \sin\theta\Big(C_{42} \sin2 \psi + S_{42}\cos 2 \psi\Big) + \nonumber\\
  &&+\,
  24 \sin\theta\Big(S_{44} \cos4\psi + C_{44} \sin4 \psi\Big),
 \nonumber\\
R^{-4}{\cal T}'_{1122} &=&
 {\textstyle\frac{1}{70}} C_{40} \Big(-3 + 5\cos 2 \theta\Big) -
  {\textstyle\frac{1}{7}} \sin2 \theta\Big(C_{41} \sin\psi + S_{41} \cos\psi\Big) +
  {\textstyle\frac{12}{7}}  \sin^2\theta\Big(C_{42}\cos 2 \psi - S_{42} \sin2 \psi\Big) + \nonumber\\
  &&+\,
 {\textstyle\frac{60}{7}}  \sin2 \theta\Big(C_{43} \sin3 \psi + S_{43} \cos 3 \psi\Big) -
  24 \cos^2\theta \Big(C_{44} \cos4\psi - S_{44} \sin4 \psi\Big),
\nonumber\\
R^{-4}{\cal T}'_{1123} &=&
 {\textstyle\frac{1}{7}} C_{40} \cos\theta \sin\theta+
  {\textstyle\frac{1}{7}}\cos 2 \theta\Big(S_{41} \cos\psi + C_{41} \sin\psi\Big) -
  {\textstyle\frac{6}{7}}\sin2 \theta\Big(C_{42}\cos 2 \psi - S_{42} \sin2 \psi\Big) - \nonumber\\
  &&-\,
 {\textstyle\frac{60}{7}}\cos 2 \theta\Big(S_{43} \cos 3 \psi + C_{43} \sin3 \psi\Big) +
  12 \sin2 \theta\Big(-C_{44} \cos4\psi + S_{44} \sin4 \psi\Big),
  \nonumber\\
R^{-4}{\cal T}'_{1133} &=& -{\textstyle\frac{1}{70}} C_{40} \Big(3 + 5\cos 2 \theta\Big) +
  {\textstyle\frac{1}{7}} \sin2 \theta\Big(S_{41} \cos\psi + C_{41} \sin\psi\Big) +
  {\textstyle\frac{12}{7}} \cos^2\theta \Big(C_{42}\cos 2 \psi - S_{42} \sin2 \psi\Big) - \nonumber\\
  &&-\,
 {\textstyle\frac{60}{7}} \sin2 \theta\Big(S_{43} \cos 3 \psi + C_{43} \sin3 \psi\Big) -
  24 \sin^2\theta \Big(C_{44} \cos4\psi - S_{44} \sin4 \psi\Big),     \nonumber\\
R^{-4}{\cal T}'_{2222} &=&
 {\textstyle\frac{1}{280}} C_{40} \Big(9 - 20\cos 2 \theta+ 35 \cos 4 \theta\Big) +
  {\textstyle\frac{1}{14}} \Big(2 \sin2 \theta- 7 \sin4 \theta\Big) \Big(S_{41} \cos\psi +
     C_{41} \sin\psi\Big) + \nonumber\\
 &&+\,
  {\textstyle\frac{6}{7}}\cos^2\theta \Big(-5 + 7\cos 2 \theta\Big) \Big(C_{42}\cos 2 \psi -
     S_{42} \sin2 \psi\Big) -
  {\textstyle\frac{240}{7}} \cos^3\theta \sin\theta\Big(S_{43} \cos 3 \psi +
     C_{43} \sin3 \psi\Big) + \nonumber\\
     &&+\,
  24 \cos^4\theta \Big(C_{44} \cos4\psi - S_{44} \sin4 \psi\Big),
 \nonumber\\
R^{-4}{\cal T}'_{2223} &=&
 {\textstyle\frac{1}{56}} C_{40} \Big(-2 \sin2 \theta+ 7 \sin4 \theta\Big) - {\textstyle\frac{1}{14}} \Big(\cos2 \theta- 7 \cos 4 \theta\Big) \Big(S_{41} \cos\psi + C_{41} \sin\psi\Big) +\nonumber\\
 &&+\,
  {\textstyle\frac{3}{14}} \Big(2 \sin2 \theta+ 7 \sin4 \theta\Big) \Big(C_{42}\cos 2 \psi -
     S_{42} \sin2 \psi\Big) +   \nonumber\\
     &&+\,
 {\textstyle\frac{60}{7}} \cos^2\theta \Big(-1 + 2\cos 2 \theta\Big) \Big(S_{43} \cos 3 \psi +
     C_{43} \sin3 \psi\Big) +
  24 \cos\theta^3 \sin\theta\Big(C_{44} \cos4\psi -  S_{44} \sin4 \psi\Big),
\nonumber\\
R^{-4}{\cal T}'_{2233} &=&
 {\textstyle\frac{1}{280}} C_{40} \Big(3 - 35 \cos 4 \theta\Big) +
  {\textstyle\frac{1}{2}} \sin4 \theta\Big(S_{41} \cos\psi + C_{41} \sin\psi\Big) -   {\textstyle\frac{3}{14}} \Big(1 + 7 \cos 4 \theta\Big) \Big(C_{42}\cos 2 \psi -  S_{42} \sin2 \psi\Big) + \nonumber\\
  &&+\,
  {\textstyle\frac{30}{7}} \sin4 \theta\Big(S_{43} \cos 3 \psi + C_{43} \sin3 \psi\Big) +
  6 \sin^22 \theta \Big(C_{44} \cos4\psi - S_{44} \sin4 \psi\Big),    \nonumber\\
R^{-4}{\cal T}'_{1222} &=& {\textstyle\frac{1}{14}} \Big(1- 7\cos 2 \theta\Big) \sin\theta\Big(C_{41} \cos\psi - S_{41} \sin\psi\Big) +
  {\textstyle\frac{3}{14}} \Big(3 \cos\theta - 7 \cos3 \theta\Big) \Big(S_{42}\cos 2 \psi +
     C_{42} \sin2 \psi\Big) - \nonumber\\
     &&-\,
  {\textstyle\frac{180}{7}} \cos^2\theta \sin\theta\Big(C_{43} \cos 3 \psi - S_{43} \sin3 \psi\Big) -
  24 \cos^3\theta \Big(S_{44} \cos4\psi + C_{44} \sin4 \psi\Big),
\nonumber\\
R^{-4}{\cal T}'_{1223} &=&
 {\textstyle\frac{1}{28}}\Big(-3 \cos\theta + 7 \cos3 \theta\Big) \Big(C_{41} \cos\psi -
     S_{41} \sin\psi\Big) +
  {\textstyle\frac{3}{14}} \Big(\sin\theta- 7 \sin3 \theta\Big) \Big(S_{42}\cos 2 \psi +
     C_{42} \sin2 \psi\Big) + \nonumber\\
     &&+\,
  {\textstyle\frac{15}{7}} \Big(\cos\theta + 3 \cos3 \theta\Big) \Big(C_{43} \cos 3 \psi -
     S_{43} \sin3 \psi\Big) -
  24 \cos^2\theta \sin\theta\Big(S_{44} \cos4\psi +   C_{44} \sin4 \psi\Big),     \nonumber\\
R^{-4}{\cal T}'_{1233} &=&
 {\textstyle\frac{1}{28}} \Big(3 \sin\theta+ 7 \sin3 \theta\Big) \Big(C_{41} \cos\psi -
     S_{41} \sin\psi\Big) +
  {\textstyle\frac{3}{14}} \Big(\cos\theta + 7 \cos3 \theta\Big) \Big(S_{42}\cos 2 \psi +
     C_{42} \sin2 \psi\Big) - \nonumber\\
     &&-\,
  {\textstyle\frac{15}{7}} \Big(\sin\theta- 3 \sin3 \theta\Big) \Big(C_{43} \cos 3 \psi -
     S_{43} \sin3 \psi\Big) -
  24 \cos\theta \sin^2\theta \Big(S_{44} \cos4\psi + C_{44} \sin4 \psi\Big),
\nonumber\\
R^{-4}{\cal T}'_{2333} &=& -{\textstyle\frac{1}{56}} C_{40}\Big (2 \sin2 \theta+ 7 \sin4 \theta\Big) -
  {\textstyle\frac{1}{14}} \Big(\cos2 \theta+ 7 \cos 4 \theta\Big)\Big(S_{41} \cos\psi +
     C_{41} \sin\psi\Big) +
     \nonumber\\
     &&+\,
  {\textstyle\frac{3}{14}} \Big(2 \sin2 \theta- 7 \sin4 \theta\Big) \Big(C_{42}\cos 2 \psi -
     S_{42} \sin2 \psi\Big) +
       \nonumber\\
     &&+\,
 {\textstyle\frac{60}{7}} \Big(1 + 2\cos 2 \theta\Big) \sin^2\theta \Big(S_{43} \cos 3 \psi +
     C_{43} \sin3 \psi\Big) +
  24 \cos\theta \sin^3\theta \Big(C_{44} \cos4\psi -  S_{44} \sin4 \psi\Big),
   \nonumber\\
R^{-4}{\cal T}'_{1333} &=& -{\textstyle\frac{1}{14}}
\cos\theta \Big(1 + 7\cos 2 \theta\Big) \Big(C_{41} \cos\psi -
     S_{41} \sin\psi\Big) +
  {\textstyle\frac{3}{14}} \Big(3 \sin\theta+ 7 \sin3 \theta\Big) \Big(S_{42}\cos 2 \psi +
     C_{42} \sin2 \psi\Big) + \nonumber\\
     &&+\,
  {\textstyle\frac{180}{7}} \cos\theta \sin^2\theta \Big(C_{43} \cos 3 \psi -
     S_{43} \sin3 \psi\Big) -
  24 \sin^3\theta \Big(S_{44} \cos4\psi + C_{44} \sin4 \psi\Big),
    \nonumber\\
R^{-4}{\cal T}'_{3333} &=&
 {\textstyle\frac{1}{280}} C_{40} \Big(9 + 20\cos 2 \theta+ 35 \cos 4 \theta\Big) -
  {\textstyle\frac{1}{14}} \Big(2 \sin2 \theta+ 7 \sin4 \theta\Big) \Big(C_{41}  \sin\psi +
     S_{41} \cos\psi \Big) + \nonumber\\
     &&+\,
  {\textstyle\frac{6}{7}} \Big(5 + 7\cos 2 \theta\Big) \sin^2\theta \Big(S_{42} \sin2 \psi - C_{42}\cos 2 \psi \Big) +
  {\textstyle\frac{240}{7}}  \cos\theta \sin^3\theta \Big(C_{43} \sin3 \psi +
     S_{43}  \cos 3 \psi\Big) +
  \nonumber\\
  &&+\,
  24 \sin^4\theta \Big(C_{44} \cos4\psi - S_{44}  \sin4 \psi\Big).
  \label{eq:spij4}
\end{eqnarray}

As a result, the relations between
$\big\{C'_{4k},S'_{4k}\big\}$ and ${\cal T}'{}^{<abcd>}$ are established to be
{}
\begin{eqnarray}
C'_{40} &=&  {\textstyle\frac{35}{8}} {\cal T}'_{3333}R^{-4},
\qquad C'_{41} = -{\textstyle\frac{7}{4}} {\cal T}'_{1333}R^{-4},
\qquad C'_{42} = {\textstyle\frac{7}{48}}\Big({\cal T}'_{1133}-{\cal T}'_{2233}\Big)R^{-4}, \nonumber\\
C'_{43} &=&  -{\textstyle\frac{7}{240}}\Big({\cal T}'_{1113}-3{\cal T}'_{1223}\Big)R^{-4},
\qquad
C'_{44} = {\textstyle\frac{1}{96}}\Big({\cal T}'_{1111}+{\cal T}'_{2222}-6{\cal T}'_{1122}\Big)R^{-4},
\qquad
S'_{41} =-{\textstyle\frac{7}{4}}{\cal T}'_{2333}R^{-4}
\nonumber\\[3pt]
S'_{42} &=&  {\textstyle\frac{7}{12}}{\cal T}'_{1223}R^{-4},
\qquad
S'_{43} =  -{\textstyle\frac{7}{240}}\Big(3{\cal T}'_{1123}-{\cal T}'_{2223}\Big)R^{-4},
\qquad
S'_{44} =   {\textstyle\frac{1}{48}}\Big({\cal T}'_{1112}-{\cal T}'_{1222}\Big)R^{-4}.~~
  \label{eq:spijT4=}
\end{eqnarray}

\subsubsection{Hexadecapole phase in terms of spherical harmonics}

Relationships (\ref{eq:spij4})--(\ref{eq:spijT4=}) allow us to express (\ref{eq:delta-eik-hexa-pr}) in terms of the spherical harmonics coefficients $C'_{4k},S'_{4k}$ as
{}
\begin{eqnarray}
\varphi_4(\vec r,\vec r_0)
&=&kr_g\Big(\frac{R_\oplus}{b}\Big)^4
\bigg\{48\Big\{C'_{44}\cos4\phi_\xi+S'_{44}\sin4\phi_\xi\Big\}
\Big\{1-(\vec k\cdot \vec n)\Big(1+{\textstyle\frac{1}{2}}\frac{b^2}{r^2}+{\textstyle\frac{3}{8}}
\frac{b^4}{r^4}+{\textstyle\frac{5}{16}} \frac{b^6}{r^6}\Big)\Big\}-\nonumber\\
&&\hskip 25pt -\,
 {\textstyle\frac{150}{7}}\Big\{C'_{43}\cos3\phi_\xi+S'_{43}\sin3\phi_\xi\Big\}\frac{b^7}{r^7} +
{\textstyle\frac{15}{2}} \Big\{C'_{42}\cos2\phi_\xi+S'_{42}\sin2\phi_\xi\Big\}\frac{b^6}{r^6}-\nonumber\\
&&\hskip 34pt -\,
2\Big\{ C'_{41}\cos\phi_\xi+S'_{41}\sin\phi_\xi\Big\} \frac{b^5}{r^5}\Big(1-
{\textstyle\frac{5}{4}} \frac{b^2}{r^2}\Big)+{\textstyle\frac{1}{4}} C'_{40} \frac{b^4}{r^4}(\vec k\cdot \vec n)\Big(1-{\textstyle\frac{5}{2}} \frac{b^2}{r^2}\Big)
\bigg\}\Big|^r_{r_0},
\label{eq:delta-eik-hexa-pr2+}
\end{eqnarray}
where $C'_{4 k}$ and $S'_{4 k}$ in the rotated coordinate system using (\ref{eq:spijT4=}) and (\ref{eq:spij4}) are given by\footnote{Note that the form of Eq.~(\ref{eq:delta-eik-hexa-pr2+}) may be generalized to any order $\ell$. Here we present the result that captures the contribution of the sectoral spherical harmonics $C'_{\ell\ell},S'_{\ell\ell}$ that has the form:
{}
\begin{eqnarray*}
\varphi_{\ell}(\vec r,\vec r_0)
&=&kr_g\Big(\frac{R_\oplus}{b}\Big)^\ell
\bigg\{(-1)^{\ell}(2\ell-2)!!\Big\{C'_{\ell\ell}\cos\ell\phi_\xi+S'_{\ell\ell}\sin\ell\phi_\xi\Big\}
\Big\{1-(\vec k\cdot \vec n)\sum_{k=0}^{\ell-1}\frac{(2k)!}{4^k(k!)^2}\Big(\frac{b}{r}\Big)^{2k}\Big\}
\bigg\}\Big|^r_{r_0},
\end{eqnarray*}
which extends our results from \cite{Turyshev-Toth:2021-STF-moments} (that were developed for the case when $b\ll r$) on the case with any relations between $b$ and $r$. The relevant work is currently underway; results, when available, will be reported.
}
{}
\begin{eqnarray}
C'_{40} &=& {\textstyle\frac{1}{64}} \Big(9 + 20 \cos 2\theta + 35 \cos 4 \theta\Big)C_{40}  -
  {\textstyle\frac{5}{16}} \Big(2 \sin2\theta + 7 \sin4 \theta\Big) \Big(C_{41} \sin\psi+S_{41} \cos\psi \Big) - \nonumber\\
  &&-\,
  {\textstyle\frac{15}{4}} \Big(5 + 7 \cos 2 \theta\Big) \sin^2\theta \Big(C_{42} \cos 2 \psi - S_{42} \sin2 \psi\Big) +
  150 \cos\theta \sin^3\theta \Big(C_{43} \sin3 \psi+S_{43} \cos 3 \psi\Big) + \nonumber\\
  &&+\,
  105 \sin^4\theta \Big(C_{44} \cos 4 \psi - S_{44} \sin4 \psi\Big),
\nonumber\\
C'_{41} &=& {\textstyle\frac{1}{8}} \cos\theta \Big(1 + 7 \cos 2 \theta\Big) \Big(C_{41} \cos\psi -  S_{41} \sin\psi\Big) -
  {\textstyle\frac{3}{8}} \Big(3 \sin\theta + 7 \sin3 \theta\Big) \Big(C_{42} \sin2 \psi+S_{42} \cos 2 \psi\Big) - \nonumber\\
  &&-\, 45 \cos\theta \sin^2\theta \Big(C_{43} \cos 3 \psi - S_{43} \sin3 \psi\Big) + 42 \sin^3\theta \Big(C_{44} \sin4 \psi+S_{44} \cos 4 \psi\Big),
\nonumber\\
C'_{42} &=& -{\textstyle\frac{1}{48}} \sin^2\theta \Big(5 + 7 \cos 2 \theta\Big) \,C_{40} +
  {\textstyle\frac{1}{48}} \Big(2 \sin2\theta - 7 \sin4 \theta\Big) \Big(C_{41} \sin\psi+S_{41} \cos\psi\Big) + \nonumber\\
     &&+\,
  {\textstyle\frac{1}{16}}\Big(5 + 4 \cos 2\theta +
     7 \cos 4 \theta\Big) \Big(C_{42} \cos 2 \psi - S_{42} \sin2 \psi\Big) -
  10 \cos^3\theta \sin\theta \Big(C_{43} \sin3 \psi+S_{43} \cos 3 \psi\Big) - \nonumber\\
  &&-\,
  {\textstyle\frac{7}{2}} \Big(3 + \cos 2 \theta \Big)\sin^2\theta \Big(C_{44} \cos 4 \psi - S_{44} \sin4 \psi\Big),
\nonumber\\
C'_{43} &=& -{\textstyle\frac{7}{80}}
     \cos\theta \sin^2\theta \Big(C_{41} \cos\psi - S_{41} \sin\psi\Big) -
  {\textstyle\frac{7}{80}} \Big(1 + 3 \cos 2 \theta\Big) \sin\theta \Big(C_{42} \sin2 \psi+S_{42} \cos 2 \psi\Big) + \nonumber\\
  &&+\,
  {\textstyle\frac{1}{16}}\Big(7 \cos\theta + 9 \cos 3 \theta\Big) \Big(C_{43} \cos 3 \psi -
     S_{43} \sin3 \psi\Big) -
  {\textstyle\frac{7}{40}}\Big(7 \sin\theta + 3 \sin3 \theta\Big) \Big(C_{44} \sin4 \psi+S_{44} \cos 4 \psi\Big),
  \nonumber\\
C'_{44} &=& {\textstyle\frac{1}{192}} \sin^4\theta\,C_{40}  +
  {\textstyle\frac{1}{48}} \cos\theta \sin^3\theta \Big(C_{41} \sin\psi+S_{41} \cos\psi \Big) -
  {\textstyle\frac{1}{32}} \Big(3 + \cos 2 \theta\Big) \sin^2\theta \Big(C_{42} \cos 2 \psi - S_{42} \sin2 \psi\Big) - \nonumber\\
  &&-\,
  {\textstyle\frac{5}{224}} \Big(14 \sin2\theta + \sin4 \theta\Big) \Big(C_{43} \sin3 \psi+S_{43} \cos 3 \psi\Big) +
  {\textstyle\frac{1}{64}} \Big(35 + 28 \cos 2\theta + \cos 4 \theta\Big) \Big(C_{44} \cos 4 \psi - S_{44} \sin4 \psi\Big),
\nonumber\\
S'_{41} &=& {\textstyle\frac{1}{32}} \Big(2 \sin2\theta + 7 \sin4 \theta\Big)C_{40}  +
  {\textstyle\frac{1}{8}} \Big(\cos 2\theta + 7 \cos 4 \theta \Big)\Big(C_{41} \sin\psi+S_{41} \cos\psi\Big) - \nonumber\\
     &&-\,
  {\textstyle\frac{3}{8}} \Big(2 \sin2\theta - 7 \sin4 \theta\Big) \Big(C_{42} \cos 2 \psi -
     S_{42} \sin2 \psi\Big) -
  15 \Big(1 + 2 \cos 2 \theta\Big) \sin^2\theta\Big(C_{43} \sin3 \psi+S_{43} \cos 3 \psi\Big) - \nonumber\\
  &&-\,
  42 \cos\theta \sin^3\theta \Big(C_{44} \cos 4 \psi - S_{44} \sin4 \psi\Big),
    \nonumber\\
S'_{42} &=& {\textstyle\frac{1}{48}} \Big(3 \sin\theta + 7 \sin3 \theta\Big) \Big(C_{41} \cos\psi - S_{41} \sin\psi\Big) +
  {\textstyle\frac{1}{8}} \Big(\cos\theta + 7 \cos 3 \theta\Big) \Big(C_{42} \sin2 \psi+S_{42} \cos 2 \psi\Big) - \nonumber\\
     &&-\,
  {\textstyle\frac{5}{4}} \Big(\sin\theta - 3 \sin3 \theta\Big) \Big(C_{43} \cos 3 \psi - S_{43} \sin3 \psi\Big) -
  14 \cos\theta \sin^2\theta \Big(C_{44} \sin4 \psi+S_{44} \cos 4 \psi\Big),
        \nonumber\\
S'_{43} &=&-{\textstyle\frac{7}{240}} \cos\theta \sin^3\theta \, C_{40}  - {\textstyle\frac{7}{240}}\Big(1 + 2 \cos 2 \theta\Big) \sin^2\theta \Big(C_{41} \sin\psi+S_{41} \cos\psi \Big) + \nonumber\\
  &&+\,
{\textstyle\frac{7}{20}} \cos^3\theta \sin\theta \Big(C_{42} \cos 2 \psi - S_{42} \sin2 \psi\Big) +
   {\textstyle\frac{1}{8}} \Big(7 \cos 2\theta + \cos 4 \theta\Big) \Big(C_{43} \sin3 \psi+S_{43} \cos 3 \psi\Big) + \nonumber\\
  &&+\,
  {\textstyle\frac{7}{80}} \Big(14 \sin2\theta + \sin4 \theta\Big) \Big(C_{44} \cos 4 \psi -
     S_{44} \sin4 \psi\Big),
     \nonumber\\
S'_{44} &=& -{\textstyle\frac{1}{48}}\sin^3\theta \Big(C_{41} \cos\psi - S_{41} \sin\psi\Big) -  {\textstyle\frac{1}{8}} \cos\theta \sin^2\theta \Big(C_{42} \sin2 \psi+S_{42} \cos 2 \psi \Big) + \nonumber\\
&&+\,
  {\textstyle\frac{5}{112}} \Big(7 \sin\theta + 3 \sin3 \theta \Big)\Big(C_{43} \cos 3 \psi - S_{43} \sin3 \psi\Big) + {\textstyle\frac{1}{8}} \Big(7 \cos\theta + \cos 3 \theta\Big) \Big(C_{44} \sin4 \psi+S_{44} \cos 4 \psi\Big).
  \label{eq:spCSij4}
\end{eqnarray}

\subsubsection{The case of an axisymmetric body}

For an axisymmetric body, all the spherical harmonic coefficients except for $C_{40}$ vanish, namely: $C_{41}=C_{42}=C_{43}=C_{41}=S_{41}=S_{42}=S_{43}=S_{43}=0$. In this case, expressions (\ref{eq:delta-eik-hexa-pr2+})--(\ref{eq:spCSij4}) take the form
{}
\begin{eqnarray}
\varphi_4(\vec r,\vec r_0)
&=&{\textstyle\frac{1}{4}} kr_gC_{40}R_\oplus^4
\Big\{
\cos 4\phi_\xi
\sin^4\theta
\Big\{\frac{1}{b^2}\Big(\frac{1}{r\big(r+({\vec k}\cdot{\vec r})\big)}-\frac{(\vec k\cdot\vec r)}{2r^3}\Big)-{\textstyle\frac{3}{8}}\frac{(\vec k\cdot\vec r)}{r^5}
\Big\}-\nonumber\\
&-&
{\textstyle\frac{5}{2}}\Big(
\sin^4\theta\sin^4 \phi_\xi -6
\sin^2\theta\cos^2\theta\sin^2 \phi_\xi + \cos^4 \theta \Big)
b^2\frac{(\vec k\cdot \vec r)}{r^7}+
\nonumber\\
&&
\hskip -40pt +\,
10\sin\theta\cos\theta  \sin\phi_\xi \Big(\cos^2\theta-\sin^2\theta\sin^2\phi_\xi\Big)
\frac{b^3}{r^7}+
2\sin \theta\cos\theta\sin\phi_\xi
\Big(3 - 7 \cos^2 \theta\Big)\frac{b}{r^5}
\Big\}\Big|^r_{r_0}.
\label{eq:delta-eik-hexa*all=8}
\end{eqnarray}
Again, using $J_4=-C_{40}$ and relying on the definitions of the unit vectors $\vec m,\vec k,\vec s$, we present (\ref{eq:delta-eik-hexa*all=8}) as
{}
\begin{eqnarray}
\varphi_4(\vec r,\vec r_0)
&=& {\textstyle\frac{1}{4}}kr_gJ_4R_\oplus^4 \Big\{
\Big(8(\vec s\cdot \vec m)^2\big(({\vec s}\cdot{\vec m})^2+({\vec s}\cdot{\vec k})^2-1\big)+\big(({\vec s}\cdot{\vec k})^2-1\big)
^2\Big)\frac{1}{b^2}\Big(\frac{(\vec k\cdot \vec r)}{2r^3}-\frac{1}{r\big(r+({\vec k}\cdot{\vec r})\big)}\Big)+
\nonumber\\
&+&
\Big(3(\vec s\cdot \vec m)^2
\big(({\vec s}\cdot{\vec m})^2+({\vec s}\cdot{\vec k})^2-1\big)+({\vec s}\cdot{\vec k})^2
\big(3-4({\vec s}\cdot{\vec k})^2\big)\Big)\frac{(\vec k\cdot\vec r)}{r^5}+\nonumber\\
&+&
{\textstyle\frac{5}{2}}\Big(\big(({\vec s}\cdot{\vec m})^2-({\vec s}
\cdot{\vec k})^2\big)^2-4({\vec s}\cdot{\vec m})^2({\vec s}\cdot{\vec k})^2\Big)b^2\frac{(\vec k\cdot\vec r)}{r^7}+\nonumber\\
&+&
10({\vec s}\cdot{\vec m})({\vec s}\cdot{\vec k})\big(({\vec s}\cdot{\vec k})^2-({\vec s}\cdot{\vec m})^2\big)\frac{b^3}{r^7}+2({\vec s}\cdot{\vec m})({\vec s}\cdot{\vec k})\big(3-7({\vec s}\cdot{\vec k})^2\big)\frac{b}{r^5}\Big\}\Big|^r_{r_0},
\label{eq:phase-sh-axisymJ4}
\end{eqnarray}
which agrees with the relevant $J_4$-part of (\ref{eq:phiE0}).

\subsection{Tidal and spin contributions to the phase shift}

Considering signal propagation in the vicinity of the Earth, we can now integrate the contributions to the total phase shift from the remaining two terms present in (\ref{eq:Psi+}), namely the tidal terms and the Earth's vector potential (\ref{eq:pot_loc-w_a+}), that are given by (\ref{eq:u-tidal-E}) and (\ref{eq:pot_loc-w_a+}), correspondingly.  Integration of these terms along the light path is straightforward, yielding the following result for the tidal term:
{}
\begin{eqnarray}
\varphi_G^{\tt tidal}(\vec x) &=&
-k\int^{\tau}_{\tau_0} \frac{2}{c^2}u^{\tt tidal}_{\rm E}(\tau')d\tau' \simeq-
k \sum_{b\not={\rm E}}
\frac{GM_b}{c^2r^3_{b{\rm E}}}\int^{\tau}_{\tau_0} \Big(3(\vec{n}^{}_{b{\rm E}}\cdot\vec{x})^2-\vec{x}^2\Big)
d\tau'\simeq\nonumber\\
&&\hskip -55pt \simeq\,
k \sum_{b\not={\rm E}}
\frac{GM_b}{c^2r^3_{b{\rm E}}}\Big\{\Big(3(\vec{n}^{}_{b{\rm E}}\cdot \vec m)^2-1\Big)b^2(\vec k\cdot \vec r)
+3(\vec{n}^{}_{b{\rm E}}\cdot \vec m)(\vec{n}^{}_{b{\rm E}}\cdot \vec k)b(\vec k\cdot \vec r)^2+
{\textstyle\frac{1}{3}}\Big(3(\vec{n}^{}_{b{\rm E}}\cdot \vec k)^2-1\Big)(\vec k\cdot \vec r)^3\Big\}\Big|^r_{r_0}\approx
\nonumber\\
&&\hskip -55pt \approx\,
k \sum_{b\not={\rm E}}
\frac{GM_b}{c^2r^3_{b{\rm E}}}
{\textstyle\frac{1}{3}}\Big(3(\vec{n}^{}_{b{\rm E}}\cdot \vec k)^2-1\Big)\Big((\vec k\cdot \vec r)^3-(\vec k\cdot \vec r_0)^3\Big).
\label{eq:phiTidal-ph}
\end{eqnarray}
Similarly, we integrate the phase term due to the Earth's rotation:
{}
\begin{eqnarray}
\varphi_G^{\tt S}(\vec x) &=&
-k\int^{\tau}_{\tau_0} \frac{4}{c^3}(k_\epsilon w^\epsilon_{\rm E}(\tau'))d\tau' =-
k\frac{2GM_\oplus}{c^3}
({\vec S}_\oplus\cdot[{\vec k}\times \vec b])\int^{\tau}_{\tau_0}\frac{d\tau'}{r^3} =
- k \frac{2GM_\oplus}{c^3}\frac{({\vec S}_\oplus\cdot[{\vec k}\times \vec m])}{b}\big(\vec k\cdot(\vec n-\vec n_0)\big).~~~~~~
\label{eq:phiS-ph}
\end{eqnarray}

We can now evaluate these terms for a typical GPS orbit with altitude of $d_{\rm GPS}=20,000$~km, so that $r_0=R_\oplus+d_{\rm GPS}$. The tidal term from (\ref{eq:phiTidal-ph}) is
{}
\begin{eqnarray}
\varphi_G^{\tt tidal}(\vec x) &\approx&
k \sum_{b\not={\rm E}}
\frac{GM_b}{c^2r^3_{b{\rm E}}}
{\textstyle\frac{1}{3}}\Big(3(\vec{n}^{}_{b{\rm E}}\cdot \vec k)^2-1\Big)\Big((\vec k\cdot \vec r)^3-(\vec k\cdot \vec r_0)^3\Big)=\nonumber\\
&&\hskip -50pt=\,
k \sum_{b\not={\rm E}}
\frac{GM_b}{3c^2}\frac{r^3_0}{r^3_{b{\rm E}}}
\Big(3(\vec{n}^{}_{b{\rm E}}\cdot \vec k)^2-1\Big)\Big(\frac{r^3}{r^3_0}(\vec k\cdot \vec n)^3-(\vec k\cdot \vec n_0)^3\Big)\approx
kc\Big(4.07\times 10^{-17}\, {\rm s}+1.84\times 10^{-17}\, {\rm s}\Big),
\label{eq:phiTidal-ph-est}
\end{eqnarray}
where the two numerical contributions are from the Moon and the Sun, respectively. Along similar lines, the phase contribution from the Earth's rotation (\ref{eq:phiS-ph}) may be at most
{}
\begin{eqnarray}
\varphi_G^{\tt S}(\vec x) &=&
- k \frac{2GM_\oplus}{c^3}\frac{\big({\vec S}_\oplus\cdot[{\vec k}\times \vec m]\big)}{b}\big(\vec k\cdot(\vec n-\vec n_0)\big)
\approx kc\Big(\frac{R_\oplus}{b}\Big)\big(\vec k\cdot(\vec n-\vec n_0)\big)\Big(1.52\times 10^{-17}\, {\rm s}\Big),
\label{eq:phiS-ph-est}
\end{eqnarray}
which may be insignificant in many scenarios, though its magnitude can be non-negligible for vertical transmissions.

\subsection{Evaluating the magnitudes of the various multipole terms}

Although we were able to develop analytical expressions for the gravitational phase shifts induced by the $\ell=2, 3, 4$ spherical harmonics (\ref{eq:phase-sh-quad-prime-h+}), (\ref{eq:delta-eik-octu*all-prime-h=}) and (\ref{eq:delta-eik-hexa-pr2+}), correspondingly, we recognize that based on the values of the spherical harmonic coefficients shown in Table~\ref{tab:sp-harmonics}, their individual contributions will be  very different.

The relative magnitudes of individual terms depend significantly on the location of the transmitter and receiver and the direction of transmission. The significance of these terms and their contributions to the phase shift is application-dependent. As a general observation, we note that the contributions of all but the quadrupole zonal harmonic $C_{20}=-J_2$ are small, with typical magnitudes of $kc\,{\cal O}(10^{-14}~{\rm s})$ or less. Therefore, for many applications accounting only for $J_2$ and $C_{22}, S_{22}$ may be sufficient, ignoring most of the tesseral, sectoral, and higher-order terms.

\begin{figure}
\includegraphics[width=0.49\linewidth]{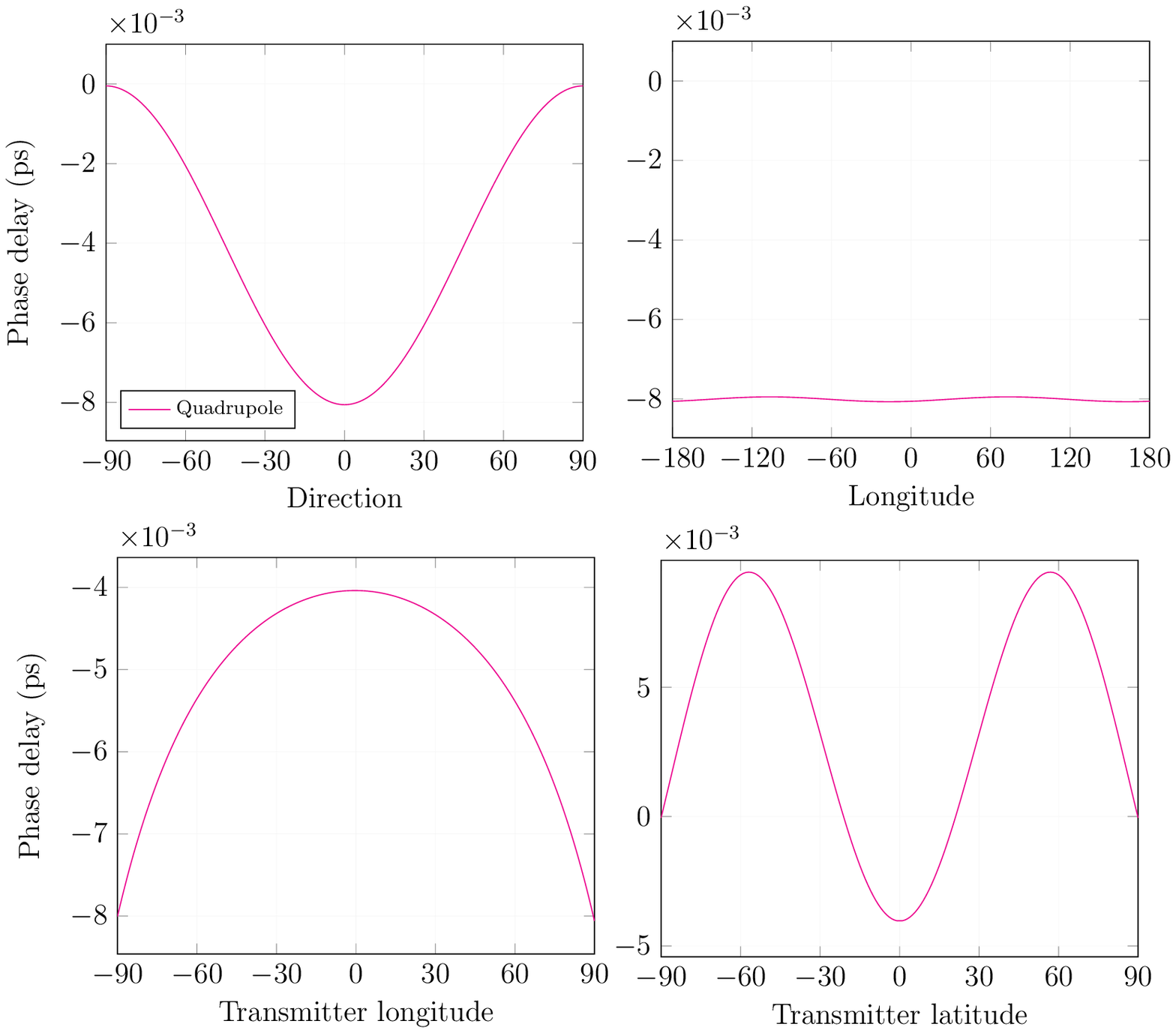}\hskip 0.019\linewidth\includegraphics[width=0.49\linewidth]{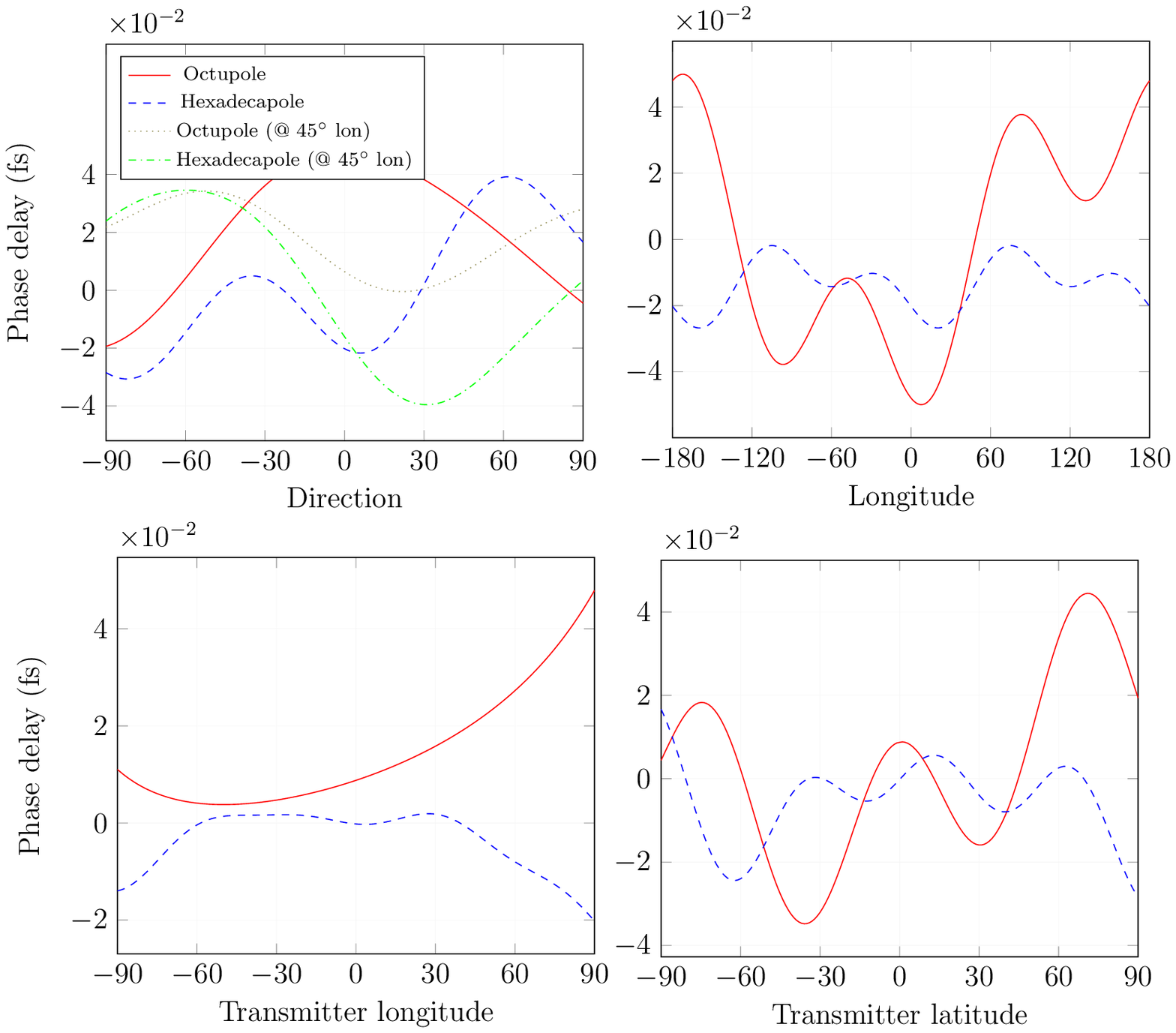}
\caption{\label{fig:plots}The quadrupole (left panel, values in picoseconds,  from (\ref{eq:phase-sh-quad-prime-h+})--(\ref{eq:spij})) and the octupole/hexadecapole (right panel, values in femtoseconds, from  (\ref{eq:delta-eik-octu*all-prime-h=})--(\ref{eq:spij3})  and (\ref{eq:delta-eik-hexa-pr2+})--(\ref{eq:spCSij4})) phase delay for various configurations involving a GPS satellite and a station on the surface of the Earth. Clockwise from top left: Transmitter on the horizon as seen by a receiving station at the equator at $0^\circ$ and $45^\circ$ longitude (the two quadrupole curves are identical) in various directions from south ($-90^\circ$) through east ($0^\circ$) to north ($90^\circ$); transmitter on the horizon, receiving station in the equatorial plane at various longitudes (the small modulation of the quadrupole is a numerical artifact); transmission from various celestial latitudes in the plane of the prime meridian to a receiver at the intersection of the equator and the prime meridian; and transmission from various celestial longitudes in the equatorial plane to a receiver at the intersection of the equator and the prime meridian.}
\end{figure}

Evaluating the phase delay using a ground-based station for a variety of scenarios yielded the results shown in Fig.~\ref{fig:plots}. As expected, the largest contribution is due to the quadrupole moment, but it remains small, never exceeding $\sim 0.01$~ps in magnitude. This can also be confirmed analytically for specific cases, as shown in Appendix~\ref{app:analest}. The contributions of the octupole and hexadecapole moments are much less, measured in hundredths of femtoseconds. For consistency, the cases depicted in Fig.~\ref{fig:plots} all involve ground-based stations. To assess the delay between two distant stations in space, the phase delays shown in the top row of images of Fig.~\ref{fig:plots} must be multiplied by two, to account for the incoming and outgoing leg of a transmission grazing the Earth's surface (similar to the situation discussed in \cite{Turyshev-Toth:2021-multipoles}). For signal paths with a larger impact parameter, the phase delay decreases, so the curves in the top row of Fig.~\ref{fig:plots} represent upper limits for such transmissions.

We may wonder why even the quadrupole contribution remains small, perhaps surprisingly small, in all the cases considered. When we look at Eqs.~(\ref{eq:phase-sh-quad-prime-h+}), (\ref{eq:delta-eik-octu*all-prime-h=}) and (\ref{eq:delta-eik-hexa-pr2+}), there are three competing factors at work, which are best understood if we recall that ultimately, all variable terms in these expressions, including the impact parameter $b$ defined in (\ref{eq:impact-par}), are functions of the vector quantities $\vec{r}_0$ and $\vec{r}$, and that moreover, these expressions, developed by integrating the eikonal equation, are themselves differences of values evaluated at $\vec{r}$ and $\vec{r}_0$.

To wit, when $b$ is small, it implies a near vertical transmission. In these cases, terms with $b$ in the numerator become insignificant, whereas $(\vec{k}\cdot\vec{n})\simeq 1$. Looking at, e.g., (\ref{eq:phase-sh-quad-prime-h+}), we can see how as a result, all terms vanish, or nearly vanish from the result. Conversely, a large $b$ implies transmission in the horizontal direction. Again looking at (\ref{eq:phase-sh-quad-prime-h+}), this implies that $(\vec{k}\cdot\vec{n})\ll 1$. Of the remaining terms, the coefficient 1 in the first term of (\ref{eq:phase-sh-quad-prime-h+}) does not depend on $\vec{r}$ so it is canceled when we compute the difference between $\vec{r}$ and $\vec{r}_0$. What remains, the namely the $C'_{21}$ and $S'_{21}$ terms, are small to begin with as these are the terms that include no contribution from the largest spherical harmonic coefficient $C_{20}$. As a result of this interplay between the value of $b$, the value of $(\vec{k}\cdot\vec{n})$, and the differencing between $\vec{r}$ and $\vec{r}_0$, the magnitude of Eq.~(\ref{eq:phase-sh-quad-prime-h+}) remains small. Similar behavior is exhibited by the octupole and hexadecapole expressions (\ref{eq:delta-eik-octu*all-prime-h=}) and (\ref{eq:delta-eik-hexa-pr2+}).

While these contributions are much too small to affect time synchronization with present-generation clocks, they will likely become significant in the near future, as clocks of even greater accuracy are deployed. Furthermore, while these phase shifts are insignificant for time synchronization, they represent a substantial contribution for phase-coherent transmissions and as such, they will have to be accounted for in any implementations or experiments that use phase coherent infrared or shorter wavelength signals. This includes experiments that rely on optical interferometry with signal paths in the gravitational field of the Earth.

Although the formalism that we introduced on these pages is aimed primarily at estimating the gravitational phase delay in the vicinity of the Earth, the methods are generic and can be readily applied to other gravitating bodies. We looked, in particular, at the quadrupole contribution to the Shapiro delay for a signal grazing the Sun. We found that if such a signal travels in the solar equatorial plane, the maximum phase delay due to the solar quadrupole moment (which is very small, $J_2\simeq -2.3\times 10^{-7}$ only) is less than 1.1~ps. The Sun has no measured octupole moment (it is ``north-south symmetric'' in addition to axisymmetry) and its hexadecapole moment contributes even less, at the sub-femtosecond level.

Finally, we looked at Jupiter, anticipating the possibility that future orbiters at Jupiter or one of its moons will utilize precision signals grazing the Jovian surface. Jupiter is not only massive but has substantial oblateness ($J_2=-1.474\times 10^{-2}$) and although it, too, is ``north-south symmetric'', its hexadecapole moment remains substantial as well. Indeed, we find that for a transmission grazing Jupiter's surface in its equatorial plane, the cumulative phase delay due to the planet's quadrupole moment can reach 70~ps, and even the hexadecapole moment can contribute more than 2~ps.

Coming back to the Earth, we also assessed the magnitudes of contributions due to spin and solar and lunar tides. For realistic signal paths within the vicinity of the Earth, these contributions remain very small: tidal contributions are of ${\cal O}(0.01~{\rm ps})$, whereas the spin contribution is less than 0.1~femptoseconds (fs). Comparatively, tidal contributions are approximately of the same magnitude as the contribution of the Earth's quadrupole moment, whereas the spin contribution is on the level of the octupole or hexadecapole moments.

\subsection{Relativistic gravitational phase shift}
\label{sec:phase-shift}

Based on the analysis in the earlier sections, we can now write the post-Minkowskian expresison for the phase of an EM wave that propagates in the vicinity of the extended and rotating gravitating body, such as the Earth. In the body's proper reference frame (a formulation that accounts for the presence of the external gravity field produced by the external bodies of the $N$-body system \cite{Turyshev:2012nw,Turyshev-Toth:2013}), collecting all the appropriate contributions  coming from the Earth's mass distribution $\varphi^{\rm E}_G$, Earth's rotation $\varphi^{\rm S}_G$, and external gravity $\varphi^{\rm tidal}_G$, the total phase Eq.~(\ref{eq:eq_eik-phi}) has the form:
{}
\begin{equation}
\varphi(t,{\vec x}) = \varphi_0+\int k_m dx^m+\varphi^{\tt E}_G(t,{\vec x})+\varphi^{\tt S}_G(t,{\vec x})+\varphi^{\rm tidal}_G(t,{\vec x})+{\cal O}(G^2).
\label{eq:eq_eik-phi+}
\end{equation}

We now have the ability to evaluate this expression at previously unavailable levels of accuracy using expressions for the quadrupole term, $\varphi_2$, from (\ref{eq:phase-sh-quad-prime-h+})--(\ref{eq:spij}), for the octupole term $\varphi_3$, from (\ref{eq:delta-eik-octu*all-prime-h=})--(\ref{eq:spij3}),  for the hexadecapole term, $\varphi_4$, from (\ref{eq:delta-eik-hexa-pr2+})--(\ref{eq:spCSij4}), and expressions (\ref{eq:phiTidal-ph}) and (\ref{eq:phiS-ph}) for the tidal, $\varphi^{\rm tidal}_G$, and rotational, $\varphi^{\tt S}_G$, terms, correspondingly.

The full available level of accuracy with all the terms shown  in (\ref{eq:eq_eik-phi+}), however, may be excessive in practical cases. For instance, taking into account the smallness of the spherical harmonics coefficients (as seen in Table~\ref{tab:sp-harmonics}), and assuming a time transfer accuracy of $10^{-12}$~s is acceptable, the result can be given as
{}
\begin{eqnarray}
\varphi(t,{\vec x}) &=& \varphi_0+k\Big\{c(t-t_0)-{\vec k}\cdot ({\vec r}-{\vec r}_0)-r_g\ln \Big[\frac{r+(\vec k\cdot \vec r)}{r_0+(\vec k\cdot \vec r_0)}\Big]+\nonumber\\
 &+&
{\textstyle\frac{1}{2}}r_g\Big(\frac{R_\oplus}{b}\Big)^2
\Big(\cos2\phi_\xi\Big[C_{20}\sin^2\theta
+2(1+\cos^2\theta)\big(S_{22}\sin2\psi-C_{22}\cos2\psi\big)\Big]-\nonumber\\
 &&\hskip 40pt-\,\,
4\sin2\phi_\xi\Big[
\cos\theta\big(S_{22}\cos2\psi+C_{22}\sin2\psi\big)\Big]\Big)\big(\vec k\cdot(\vec n-\vec n_0)\big)\Big\}+
kc\,{\cal O}(\ll 0.01~{\rm ps}),~~~~~
\label{eq:phase_t2}
\end{eqnarray}
where, in accordance with Table~\ref{tab:sp-harmonics},  we retained only the largest spherical harmonics contributing to the quadrupole term $\varphi_2$. As discussed in Sec~\ref{sec:compute-shift}, the  angles $\theta, \psi$ and $\phi_\xi$ are uniquely defined  in terms of the GCRS positions of the transmitter, $\vec r_0$, and the receiver, $\vec r$. (See the discussion of the analytical treatment of the relevant terms that is given in Appendix \ref{app:analest}). The order terms are due to the omitted quadrupole and higher-order multipole contributions.

From the result (\ref{eq:phase_t2}), the total time of the propagation of an electromagnetic signal from the point $(t_0 , \vec r_0)$ (transmitter) to the point $(t, \vec r)$ (receiver) with the help of (\ref{eq:rel2b}) and (\ref{eq:impact-dd}) is given by the following expression:
{}
\begin{eqnarray}
t -t_0&=& \frac{|{\vec r}-{\vec r}_0|}{c}+\frac{2GM_\oplus}{c^3}\Big\{\ln  \Big[\frac{r+r_0+|{\vec r}-{\vec r}_0|}{r+r_0-|{\vec r}-{\vec r}_0|}\Big]+\nonumber\\
 &+&
\Big(\cos2\phi_\xi\Big[(1+\cos^2\theta)\big(C_{22}\cos2\psi-S_{22}\sin2\psi\big)-{\textstyle\frac{1}{2}}C_{20}\sin^2\theta
\Big]+\nonumber\\
 &&\hskip 26pt+\,\,
2\sin2\phi_\xi\Big[
\cos\theta\big(C_{22}\sin2\psi+S_{22}\cos2\psi\big)\Big]\Big)  \frac{R_\oplus^2}{rr_0}\Big(\frac{1}{r}+\frac{1}{r_0}\Big)\frac{|\vec{r}-\vec{r}_0|}{1+(\vec n\cdot\vec n_0)}\Big\}+
{\cal O}\big(\hskip -2pt \ll 0.01~{\rm ps}\big),~~~~~
\label{eq:phase_t2tt}
\end{eqnarray}
where $|{\vec r}-{\vec r}_0|$ is the usual Euclidean distance between the points of emission, $\vec r_0(t_0)$, and reception, $\vec r(t)$. The logarithmic term is the well-known Shapiro time delay that may contribute up to 42.3 ps for terrestrial applications, while the terms with $C_{20}, C_{22}$ and $S_{22}$ spherical harmonics is the contribution of the Earth's quadrupole moment  to the relativistic time delay induced by the gravitational field of the extended Earth contributing periodic terms with magnitude of up to 0.01 ps. Due to their smallness, contributions of other multipole terms in (\ref{eq:eq_eik-phi+}) were neglected.

Eq.~(\ref{eq:phase_t2tt}) extends the formulation for the general-relativistic time delay. In addition to the classic Shapiro gravitational time delay \cite{Moyer:2003} due to a mass monopole (represented by the logarithmic term), it also includes contributions due to quadrupole moment of the extended Earth, $C_{20}=-J_2$ and $C_{22},S_{22}$ spherical harmonics (see Table~\ref{tab:sp-harmonics}). We exceeded our stated goal of modeling the delay to picosecond accuracy, as the terms due to the quadrupole moment contribute at the sub-picosecond level, at ${\cal O}(0.01~{\rm ps})$. Although these terms may not be relevant to current generation clocks, as more advanced future-generation clocks become available, these contributions will also become significant. In that case, additional terms may be included from the preceding derivation if even a greater accuracy is required.

\section{Conclusions and recommendations}
\label{sec:sonc}

Satellites in low-Earth orbit are affected by a broad spectrum of perturbations due to the Earth's gravity field. The largest of these perturbations are produced by the Earth's oblateness, $J_2$.  Beyond the oblateness, there exist much smaller undulations of higher order in the gravity field. These variations produce lesser, but certainly observable effects on low-Earth orbiters \cite{Montenbruck-Gill:2012,Will-2014}. The same gravity perturbations affect clocks and light propagation in the Earth's vicinity. However, most of the models capture only relativistic corrections due to the Earth's monopole potential, such as the Shapiro phase delay term -- the largest among the relevant gravitational effects. Beyond that, only the contribution of the Earth's oblateness, characterized by $J_2=-C_{20}$, was accounted for (e.g. \cite{Klioner:1991SvA,Klioner-Kopeikin:1992,Zschocke-Klioner:2010,Soffel-Han:2019}). No models capturing the contributions of other multipole moments present in the Earth's gravity potential were available.

In this paper, we addressed this challenge with the aim to obtain results in terms of the spherical harmonics. We studied the transformation between proper and coordinate time and the propagation of an EM wave in the vicinity of a gravitating body with a gravitational potential that deviates from a perfect monopole, such as the Earth. We found that at the present level of accuracy, more and more terms from the Earth's gravitational potential must be included in the model formulations. In fact, the lower the orbit of a satellite, more terms are needed.  Also, for high orbits, tidal gravity from the moon (and soon, that of the Sun) becomes significant and needs to be accounted for.

To study light propagation in the vicinity of the Earth, we represented the gravitational potential in terms of symmetric trace-free (STF) Cartesian tensor mass multipole moments. The multipole expansion is one of the most useful tools of physics, but its use in general relativity is difficult because of the nonlinearity of the theory and the tensorial character of the gravitational interaction \cite{Kopeikin:1997,Blanchet:2002}.
STF tensors offer a mathematically equivalent representation of the multipole mass moments of a gravitating body, but with significant practical advantages over spherical harmonics. First, the tensorial nature of the STF representation makes it possible to express relationships in a coordinate system independent fashion, leading to a remarkable form invariance that is preserved even in coordinate representations that are rotated relative to each other. Second, the relationship between the spherical harmonic coefficients and the STF tensor components expressed in a Cartesian coordinate system is linear and nondegenerate, which means it is always invertible. Finally, and perhaps most significantly, utilizing the STF tensor representation of the mass multipole moments allowed us to integrate the eikonal equation to all STF orders $\ell$.

With the solution of the eikonal equation at hand, we demonstrated a straightforward procedure to obtain the Cartesian STF components from spherical harmonic coefficients. We explicitly carried out this procedure in the cases of the quadrupole, octupole, and hexadecapole ($\ell=2,3,4$) cases. In all these cases, the form invariance of the result made it possible to express the corresponding phase shift in a remarkably simple, elegant form in terms of rotated spherical harmonic coefficients. The mechanics of the rotation, in turn, can be carried out by first obtaining STF tensor components in the original coordinate system, performing the rotation next, and finally by solving a linear system of equations that has the same form in the unrotated and rotated coordinate frames. Clearly, the same approach may be used to extend our results to any order $\ell$. The resulting equations are compact even for higher values of $\ell$, and are directly actionable.

Note that in this paper we dealt only with the STF mass moments that are used to represent the scalar external gravitational potential (\ref{eq:pot_w_0sh}) and the relevant scalar spherical harmonics. The same approach may be used to consider the contributions from the vector potential due to body's rotation (\ref{eq:pot_loc-w_a+}), and the relevant STF current moments and vector spherical harmonics.  As our objective was to consider measurements in the Earth's vicinity, any contribution from the Earth's vector potential is currently negligible, but may be addressed with the same tools presented here.

The numerical magnitudes of these corrections are small. Considering present-generation clocks, only the quadrupole term offers a significant contribution and only insofar as proper time to coordinate time conversions are concerned. However, the gravitational phase delay due to the quadrupole and higher order terms may become relevant with next generation clocks. These terms can also be very significant for phase coherent signaling at infrared or shorter wavelengths. We also applied our formalism for signals traversing in the vicinity of the Sun and, especially, Jupiter, and found more significant contributions, which may in the foreseeable future become relevant for deep space precision navigation and observations in the solar system.

The results presented here are new. They offer a comprehensive model for the gravitational phase shift of a EM wave as it propagates in the gravitational  field of the extended Earth.  As the performance of new generation of precision clocks increases, such results may have a wide range of practical applications, including clock synchronization, frequency transfer and interferometry. They may also lead to new uses including relativistic geodesy \cite{Svehla-book:2018}, quantum communication links \cite{Liao-etal:2017,Lu-etal:2022},  and various tests of fundamental physics \cite{Turyshev-etal:2007}.  These and other possibilities are currently being investigated. Results, when available, will be reported elsewhere.

\begin{acknowledgments}

We thank Yoaz E. Bar-Sever and William I. Bertiger of JPL who provided us with motivation, encouragement, and valuable comments of various topics discussed in this document. The work described here, in part, was carried out at the Jet Propulsion Laboratory, California Institute of Technology, under a contract with the National Aeronautics and Space Administration. VTT acknowledges the generous support of Plamen Vasilev and other Patreon patrons.
\end{acknowledgments}


\appendix

\section{Correspondence between the STF mass moments and spherical harmonics}
\label{sec:stf-sph-harm}

For practical applications, the potential $U(\vec r)$  is typically expanded in terms of spherical harmonics:
{}
\begin{eqnarray}
U(\vec r)&=&\frac{GM}{r}\Big(1+\sum_{\ell=2}^\infty\sum_{k=0}^{+\ell}\Big(\frac{R}{r}\Big)^\ell P_{\ell k}(\cos\theta)(C_{\ell k}\cos k\psi+S_{\ell k}\sin k\psi)\Big)+
{\cal O}(c^{-4}),
\label{eq:pot_w_0sh+}
\end{eqnarray}
where $P_{\ell k}$ are the associated Legendre polynomials \cite{Abramovitz-Stegun:1965}, while $C_{\ell k}$ and $S_{\ell k}$ are  the normalized spherical harmonic coefficients that characterize nonspherical contributions to the gravitational field.

To derive the relations between the Cartesian and spherical quadrupole ($\ell = 2$) moments explicitly, we can express the spherical harmonics in terms of Cartesian coordinates. For that we use (\ref{eq:pot_w_0sh}) and write
{}
\begin{eqnarray}
U^{[2]}(\vec r) &=&  \frac{G}{r^3} \Big(P_{20}C_{20} + P_{21}(C_{21} \cos \psi + S_{21}\sin \psi) + P_{22}(C_{22} \cos 2\psi + S_{22} \sin 2\psi)\Big).
\label{eq:U2-sphar}
\end{eqnarray}
Using a spherical coordinate system $ (x = r \sin \theta \cos \psi, y = r \sin \theta \sin \psi, z = r \cos \theta)$, we have
$r^2P_{20} = (2z^2 -x^2 -y^2)/2$,
$r^2P_{21} \cos \psi = 3xz$,
$r^2P_{21} \sin \psi = 3yz$,
$r^2P_{22} \cos 2\psi = 3(x^2 - y^2)$,
$r^2P_{22} \sin 2\psi = 6xy$. Substituting these expressions in (\ref{eq:U2-sphar}), we get
{}
\begin{eqnarray}
U^{[2]}(\vec r) &=&  \frac{GM}{r^5} \Big(C_{20} {\textstyle\frac{1}{2}}(2z^2 - x^2- y^2) + 3C_{21} xz+ 3S_{21}yz + 3C_{22}(x^2-y^2)+6S_{22} xy\Big).
\label{eq:U2-sphar2}
\end{eqnarray}

From (\ref{eq:pot_w_0STF2}), we have the same quantity expressed via the components of the STF quadrupole moment ${\cal T}_{<ab>}$:
{}
\begin{eqnarray}
U^{[2]}(\vec r)&=&GM \frac{3{\cal T}^{<ab>}}{2r^5}x^ax^b=
GM \frac{3}{2r^5}\Big({\cal T}_{11}x^2+2{\cal T}_{12}xy+2{\cal T}_{13}xz+2{\cal T}_{23}yz+{\cal T}_{22}y^2+{\cal T}_{33}z^2\Big).~~~
\label{eq:pot_w_0STF2*}
\end{eqnarray}

Equating the terms with the same powers of $x,y,z$  between the from of the potential in terms of spherical harmonics present in (\ref{eq:U2-sphar2}) and that expressed via the STF mass quadrupole in (\ref{eq:pot_w_0STF2*}) yields the following relations:
{}
\begin{eqnarray}
{\cal T}_{11}&=&\Big(-{\textstyle\frac{1}{3}}C_{20}+2C_{22}\Big)R^2, \qquad  {\cal T}_{12}=2S_{22}R^2, \nonumber\\
{\cal T}_{22}&=&\Big(-{\textstyle\frac{1}{3}}C_{20}-2C_{22}\Big)R^2, \qquad {\cal T}_{13}=C_{21}R^2,\nonumber\\
{\cal T}_{33}&=&{\textstyle\frac{2}{3}} C_{20} R^2, \qquad\qquad \qquad\quad~~
{\cal T}_{23}=S_{21}R^2.
\label{eq:sp-harm2*}
\end{eqnarray}

Following the same approach, we can establish the corresponding relationships between STF multipole moments at any order $\ell$ and the appropriate  spherical harmonics coefficients, see (\ref{eq:sp-harm3}), (\ref{eq:sp-harm4}), and the relevant discussion in \cite{Turyshev-Toth:2021-STF-moments}.

\section{Useful relations for some STF orders}
\label{sec:cases}

We derive several low order terms in (\ref{eq:eik-ph2*}). First, we recognize that with $\vec k$ being constant, the two-dimensional vector $\vec b$ and the one-dimensional quantity $\tau$, from (\ref{eq:x-Newt*=0})--(\ref{eq:b0}), may be treated as two independent variables, yielding $dx^a=db^a+k^ad\tau$ and also ${\partial }/{\partial x^a}={\partial}/{\partial b^a}+k^a {\partial}/{\partial \tau}$, where differentiation with respect to $\vec{b}$ is carried out in two dimensions only, which is indicated by the hatted notation. Then, to compute the needed partial derivatives in (\ref{eq:eik-ph2*}), with respect to the vector of the impact parameter, $\hat\partial_a\equiv \partial/\partial b^a\equiv(\partial/\partial b^x,\partial/\partial b^y,0)$ in our chosen Cartesian coordinate system, we may formally write
{}
\begin{eqnarray}
\frac{\partial x^a}{\partial x^b}= \delta^a_b= \Big\{{\hat\partial}_b+k_b \frac{\partial}{\partial \tau}\Big\}\Big\{b^a+k^a \tau +{\cal O}(r_g)\Big\}=\hat\partial_b b^a+k^a k_b+{\cal O}(r_g).
\label{eq:eik-exp+}
\end{eqnarray}
By re-arranging the terms in this identity, we obtain the following useful  expression (see also \cite{Kopeikin:1997,Soffel-Han:2019}):
{}
\begin{eqnarray}
\hat\partial_b b^a&=&
\delta^a_b -k^a k_b.
\label{eq:eik-exp}
\end{eqnarray}
This result essentially is the projection operator onto the plane perpendicular to $\vec k$, namely $P^{ab}=\delta^{ab}-k^ak^b$; this plane, given either GCRS (\ref{eq:k+b}) or light ray (\ref{eq:note-k}) parameterizations, is the plane of the impact parameter $\vec b$ and where $ b^a$ is the  $a$-th component of the vector impact parameter.

To evaluate (\ref{eq:eik-ph2*}), we need to compute the following sets of derivatives:
{}
\begin{eqnarray}
 \partial_{<a_1}... \partial_{a_\ell>}\ln k\Big({\sqrt{b^2+\tau^2}+\tau}\Big)
 ~~~{\rm and}~~~
  \sum_{p=1}^\ell \frac{\ell!}{p!(\ell-p)!}k_{<a_1}...k_{a_p} \partial_{a_{p+1}}... \partial_{a_\ell>}
  \Big\{\frac{\partial^{p-1}}{\partial \tau^{p-1}}  \frac{1}{\sqrt{b^2+\tau^2}}\Big\}.
\label{eq:eik-phA0+*}
\end{eqnarray}

The first type of the  derivatives needed to compute the terms with $\ell=1,2,3,4$ are
{}
\begin{eqnarray}
\hat\partial_a \ln k\Big(\sqrt{b^2+\tau^2}+\tau\Big)&=& \frac{1}{\sqrt{b^2+\tau^2}+\tau}\frac{b_a}{\sqrt{b^2+\tau^2}},
\label{eq:dab510}\\
\hat\partial^2_{ab} \ln k\Big(\sqrt{b^2+\tau^2}+\tau\Big)&=& \frac{1}{\sqrt{b^2+\tau^2}+\tau}\frac{1}{\sqrt{b^2+\tau^2}}\Big\{P_{ab}-\frac{1}{\sqrt{b^2+\tau^2}}\Big(\frac{1}{\sqrt{b^2+\tau^2}+\tau}+\frac{1}{\sqrt{b^2+\tau^2}}\Big)b_ab_b\Big\},
\label{eq:dab51}\\
\hat\partial^3_{abc} \ln k\Big(\sqrt{b^2+\tau^2}+\tau\Big)&=& -\frac{P_{ab}b_c+P_{ac}b_b+P_{bc}b_a}{\big(\sqrt{b^2+\tau^2}+\tau \big)\big(b^2+\tau^2\big)}\frac{}{}
\Big(\frac{1}{\sqrt{b^2+\tau^2}+\tau}+\frac{1}{\sqrt{b^2+\tau^2}}\Big)+\nonumber\\
&&\hskip -50pt +\,
\frac{b_ab_bb_c}{\big(\sqrt{b^2+\tau^2}+\tau\big)(b^2+\tau^2)^\frac{3}{2}}\Big\{\frac{2}{(\sqrt{b^2+\tau^2}+\tau)^2}+\frac{3}{\big(\sqrt{b^2+\tau^2}+\tau\big)\sqrt{b^2+\tau^2}}+\frac{3}{b^2+\tau^2}\Big\},
\label{eq:dab53}\\
\hat\partial^4_{abcd} \ln k\Big(\sqrt{b^2+\tau^2}+\tau\Big)&=& -\frac{(P_{ab}P_{cd}+P_{ac}P_{bd}+P_{ad}P_{bc})}{\big(\sqrt{b^2+\tau^2}+\tau \big)\big(b^2+\tau^2\big)}\frac{}{}
\Big(\frac{1}{\sqrt{b^2+\tau^2}+\tau}+\frac{1}{\sqrt{b^2+\tau^2}}\Big)+\nonumber\\
&&\hskip -100pt +\,
\Big(P_{ab}b_cb_d+P_{ac}b_bb_d+P_{bc}b_ab_d+P_{ad}b_bb_c+P_{bd}b_ab_c+P_{cd}b_ab_b\Big)\times \nonumber\\
&&\hskip -50pt \times\,
\frac{1}{\big(\sqrt{b^2+\tau^2}+\tau\big)(b^2+\tau^2)^\frac{3}{2}}\Big\{\frac{2}{(\sqrt{b^2+\tau^2}+\tau)^2}+\frac{3}{\big(\sqrt{b^2+\tau^2}+\tau\big)\sqrt{b^2+\tau^2}}+\frac{3}{b^2+\tau^2}\Big\}-\nonumber\\
&&\hskip -100pt -\,
\frac{b_ab_bb_cb_d}{\big(\sqrt{b^2+\tau^2}+\tau\big)(b^2+\tau^2)^2}\Big\{\frac{6}{(\sqrt{b^2+\tau^2}+\tau)^3}+\frac{12}{\big(\sqrt{b^2+\tau^2}+\tau\big)^2\sqrt{b^2+\tau^2}}+\nonumber\\
&&\hskip 20pt +\,
\frac{15}{\big(\sqrt{b^2+\tau^2}+\tau\big)(b^2+\tau^2)}+
\frac{15}{(b^2+\tau^2)^\frac{3}{2}}\Big\}.
\label{eq:dab54}
\end{eqnarray}

We also need the following derivatives for $\ell=2,3,4$:
\vskip -12pt
\begin{eqnarray}
  \sum_{p=1}^2 \frac{2!}{p!(2-p)!}k_{<a_1}...k_{a_p} \hat\partial_{a_{p+1}}... \hat\partial_{a_2>}
\frac{\partial^{p-1}}{\partial \tau^{p-1}}  \frac{1}{\sqrt{b^2+\tau^2}}&=&
-\frac{2k_ab_b+\tau k_ak_b}{(b^2+\tau^2)^\frac{3}{2}},
 \label{eq:dab61}\\
   \sum_{p=1}^3 \frac{3!}{p!(3-p)!}k_{<a_1}...k_{a_p} \hat\partial_{a_{p+1}}... \hat\partial_{a_3>}
\frac{\partial^{p-1}}{\partial \tau^{p-1}}  \frac{1}{\sqrt{b^2+\tau^2}}&=&
\nonumber\\
&&\hskip -170pt =\,
3k_a\Big(\frac{3b_bb_c}{(b^2+\tau^2)^\frac{5}{2}}-\frac{P_{bc}}{(b^2+\tau^2)^\frac{3}{2}}\Big)+\frac{9k_ak_bb_c \, \tau}{(b^2+\tau^2)^\frac{5}{2}}+k_ak_bk_c\Big(\frac{3\tau^2}{(b^2+\tau^2)^\frac{5}{2}}-\frac{1}{(b^2+\tau^2)^\frac{3}{2}}\Big),
 \label{eq:dab63}\\
   \sum_{p=1}^4 \frac{4!}{p!(4-p)!}k_{<a_1}...k_{a_p} \hat\partial_{a_{p+1}}... \hat\partial_{a_4>}
\frac{\partial^{p-1}}{\partial \tau^{p-1}}  \frac{1}{\sqrt{b^2+\tau^2}}&=&
\nonumber\\
&&\hskip -200pt =\,
4k_a\Big(\frac{3}{(b^2+\tau^2)^\frac{5}{2}}(P_{dc}b_b+P_{bd}b_c+P_{bc}b_d)-\frac{15b_bb_cb_d}{(b^2+\tau^2)^\frac{7}{2}}
\Big)+18k_ak_b \,\tau \Big(\frac{P_{cd}}{(b^2+\tau^2)^\frac{5}{2}} -\frac{5b_cb_d}{(b^2+\tau^2)^\frac{7}{2}}
\Big)+
\nonumber\\
&&\hskip -170pt +\,
4k_ak_bk_c\Big(\frac{3b_d}{(b^2+\tau^2)^\frac{5}{2}}-\frac{15\tau^2\, b_d}{(b^2+\tau^2)^\frac{7}{2}}
\Big)+
k_ak_bk_ck_d\Big(\frac{9\tau}{(b^2+\tau^2)^\frac{5}{2}}-\frac{15\tau^3}{(b^2+\tau^2)^\frac{7}{2}}
\Big).
 \label{eq:dab64}
\end{eqnarray}
Clearly, the same expressions may be used to describe the terms that depend on $\tau_0$.

\section{Gravitational phase delay for an axisymmetric body}
\label{app:axisymm}

In the case of an axisymmetric body (i.e., the Sun), its external gravitational potential is reduced to the $k=0$ zonal harmonics, where we keep on the the terms of $J_\ell=-C_{\ell 0}$ with all other terms in the expression for the potential  (\ref{eq:pot_w_0sh}) vanish, i.e., $C_{\ell k}=S_{\ell k}=0$. As a result, the   gravitational potential of an axisymmetric body may be expressed in terms of the usual dimensionless multipole moments $J_\ell$ (see discussion in \cite{Roxburgh:2001,LePoncinLafitte:2007tx}):
{}
\begin{eqnarray}
U(\vec x)&=&\frac{GM}{r}\Big\{1-\sum_{\ell=2}^\infty J_\ell \Big(\frac{R}{r}\Big)^\ell P_\ell\Big(\frac{{\vec k}_3\cdot{\vec x}}{r}\Big)\Big\}+
{\cal O}(c^{-4}),
\label{eq:pot_stis}
\end{eqnarray}
where ${\vec k}_3$ denotes the unit vector along the $x^3$-axis, $P_\ell$ are the Legendre polynomials. Furthermore, in the case of an axisymmetric and rotating body with ``north-south symmetry'', the expression (\ref{eq:pot_stis}) contains only the even moments or $\ell=2,4,6,8...$  \cite{Roxburgh:2001}. Below, we will not impose the ``north-south symmetry'' restriction and will keep the terms of all the orders $\ell$.

Following \cite{Linet_2002}, we take into account the identity
{}
\begin{eqnarray}
\frac{\partial^\ell}{\partial z^\ell}\Big(\frac{1}{r}\Big)&=&\frac{(-1)^\ell \ell!}{r^{1+\ell}} P_\ell \Big(\frac{{\vec k}_3\cdot{\vec x}}{r}\Big), \qquad z=x^3,
\label{eq:iden}
\end{eqnarray}
and present  $U$ as the following expansion in a series of derivatives of $1/r$:
{}
\begin{eqnarray}
U(\vec x)&=&GM\Big\{\frac{1}{r}-\sum_{\ell=2}^\infty \frac{(-1)^\ell}{ \ell!} J_\ell  R^\ell\frac{\partial^\ell}{\partial z^\ell}\Big(\frac{1}{r}\Big)\Big\}+
{\cal O}(c^{-4}).
\label{eq:pot_st3}
\end{eqnarray}
As we shall see below, this form is much more convenient  for the computation of integrals involving  $U$.

Here we develop an expression for the eikonal phase in the case of an axisymmetric body, with its potential given by (\ref{eq:pot_st3}). In this case, the decomposition of the post-Newtonian potential takes the from
{}
\begin{eqnarray}
\frac{2U}{c^2}&=&
r_g\Big\{\frac{1}{r}-\sum_{\ell=2}^\infty \frac{(-1)^\ell}{ \ell!} J_\ell  R^\ell\frac{\partial^\ell}{\partial s^\ell}\Big(\frac{1}{r}\Big)\Big\}+{\cal O}(r_g^2).
\label{eq:U-c2]=}
\end{eqnarray}

Following the approach demonstrated in \cite{Turyshev-Toth:2021-multipoles} (see Appendix therein), we compute the leading term of this expansion. For that, we define the vector $\vec s$ to be a unit vector in the direction of the axis of rotation. Remembering that $r=\sqrt{b^2+\tau^2}+{\cal O}(r_g)$ from (\ref{eq:b0}), we evaluate directional derivatives ${\partial}/{\partial s}$ along  $\vec s= \vec k_3$, which have the form
{}
\begin{eqnarray}
\frac{\partial}{\partial { s}}=({\vec s}\cdot{\vec \nabla})=\Big({\vec s}\cdot\frac{\partial}{\partial {\vec r}}\Big).
\label{eq:dir-s}
\end{eqnarray}
This relation allows us to compute the relevant partial derivatives for the leading terms in (\ref{eq:pot_st3}):
{}
\begin{eqnarray}
\frac{\partial}{\partial { s}} \, \frac{1}{r}&=&-\frac{({\vec s}\cdot{\vec r})}{r^3}, \qquad \frac{\partial^2}{\partial { s}^2} \, \frac{1}{r}=\frac{3({\vec s}\cdot{\vec r})^2}{r^5}-\frac{1}{r^3},
\qquad \frac{\partial^3}{\partial { s}^3} \, \frac{1}{r}=-3\Big(\frac{5({\vec s}\cdot{\vec r})^3}{r^7}-\frac{3({\vec s}\cdot{\vec r})}{r^5}\Big),
\label{eq:dir-s1}\\
\frac{\partial^4}{\partial { s^4}} \, \frac{1}{r}&=&3\Big(\frac{35({\vec s}\cdot{\vec r})^4}{r^9}-\frac{30({\vec s}\cdot{\vec r})^2}{r^7}+\frac{3}{r^5}\Big).
\label{eq:dir-s2}
\end{eqnarray}

Using these expressions in (\ref{eq:U-c2]=}) and defining $r_g={2G M_\oplus}/{c^2}$, we have
{}
\begin{eqnarray}
\frac{2U_{\rm E}({\vec b},\tau)}{c^2}&=&
r_g \Big\{\frac{1}{r}-\Big\{J_2R^2_\oplus\frac{1}{2}\Big(\frac{3({\vec s}\cdot{\vec r})^2}{r^5}-\frac{1}{r^3}\Big)+
J_3R^3_\oplus\frac{1}{2}\Big(\frac{5({\vec s}\cdot{\vec r})^3}{r^7}-\frac{3({\vec s}\cdot{\vec r})}{r^5}\Big)+\nonumber\\
&+&
J_4R^4_\oplus\frac{1}{8}\Big(\frac{35({\vec s}\cdot{\vec r})^4}{r^9}-\frac{30({\vec s}\cdot{\vec r})^2}{r^7}+\frac{3}{r^5}\Big)+
\sum_{\ell=5}^\infty \frac{(-1)^\ell}{ \ell!} J_\ell  R^\ell_\oplus\frac{\partial^\ell}{\partial s^\ell}\Big(\frac{1}{r}\Big)\Big\}\Big\}.
\label{eq:V-sr-m2A3}
\end{eqnarray}

This expression represents the gravitational potential in terms of the zonal harmonics projected on the trajectory of the photon propagation. We substitute (\ref{eq:V-sr-m2A3}) into expression (\ref{eq:Psi+}) and integrate it. As a result, we have the following expression for the gravitational phase shift induced by the lowest order gravitational multipoles, i.e., $\ell=0, 2,3,4$:
{}
\begin{eqnarray}
\varphi_G^{\tt E}(\vec r,\vec r_0) &=&
-k\int^{\tau}_{\tau_0} \frac{2}{c^2}U_{\rm E}(\tau')d\tau' = -
kr_g \bigg\{\ln \Big[\frac{r+(\vec k\cdot \vec r)}{r_0+(\vec k\cdot \vec r_0)}\Big]+\nonumber\\\
&&\hskip -50pt +\,
{\textstyle\frac{1}{2}}J_2R_\oplus^2 \Big\{\big(2({\vec s}\cdot{\vec m})^2+({\vec s}\cdot{\vec k})^2-1\big)
\Big(\frac{1}{r\big(r+({\vec k}\cdot{\vec r})\big)}-\frac{1}{b^2}\Big)+
\big(({\vec s}\cdot{\vec k})^2-({\vec s}\cdot{\vec m})^2\big)\frac{(\vec k\cdot\vec r)}{r^3}+
2({\vec s}\cdot{\vec m}) ({\vec s}\cdot{\vec k})\frac{b}{r^3}\Big\}\Big|^r_{r_0}-\nonumber\\
&&\hskip -50pt -\,
{\textstyle\frac{1}{3}}J_3R_\oplus^3 \Big\{
(\vec s\cdot \vec m)\big(4({\vec s}\cdot{\vec m})^2+3({\vec s}\cdot{\vec k})^2-3\big)
\Big(\frac{1}{b^3}-\frac{1}{br\big(r+({\vec k}\cdot{\vec r})\big)}+\frac{1}{2b}\frac{(\vec k\cdot \vec r)}{r^3}\Big)+
\nonumber\\
&&\hskip -10pt +\,
{\textstyle\frac{3}{2}}(\vec s\cdot \vec m)
\big(({\vec s}\cdot{\vec m})^2-3({\vec s}\cdot{\vec k})^2\big)\frac{b(\vec k\cdot\vec r)}{r^5}+
{\textstyle\frac{3}{2}}(\vec s\cdot \vec k)\big(({\vec s}\cdot{\vec k})^2-3({\vec s}
\cdot{\vec m})^2\big)\frac{b^2}{r^5}+{\textstyle\frac{1}{2}}({\vec s}\cdot{\vec k})\big(3-5({\vec s}\cdot{\vec k})^2\big)\frac{1}{r^3}\Big\}\Big|^r_{r_0}
-\nonumber\\
&&\hskip -50pt -\,
{\textstyle\frac{1}{4}}J_4R_\oplus^4 \Big\{
\Big(8(\vec s\cdot \vec m)^2\big(({\vec s}\cdot{\vec m})^2+({\vec s}\cdot{\vec k})^2-1\big)+\big(({\vec s}\cdot{\vec k})^2-1\big)
^2\Big)\Big(\frac{1}{b^4}-\frac{1}{b^2r\big(r+({\vec k}\cdot{\vec r})\big)}+\frac{1}{2b^2}\frac{(\vec k\cdot \vec r)}{r^3}\Big)+
\nonumber\\
&&\hskip -10pt +\,
\Big(3(\vec s\cdot \vec m)^2
\big(({\vec s}\cdot{\vec m})^2+({\vec s}\cdot{\vec k})^2-1\big)+({\vec s}\cdot{\vec k})^2
\big(3-4({\vec s}\cdot{\vec k})^2\big)\Big)\frac{(\vec k\cdot\vec r)}{r^5}+\nonumber\\
&&\hskip -10pt +\,
{\textstyle\frac{5}{2}}\Big(\big(({\vec s}\cdot{\vec m})^2-({\vec s}
\cdot{\vec k})^2\big)^2-4({\vec s}\cdot{\vec m})^2({\vec s}\cdot{\vec k})^2\Big)b^2\frac{(\vec k\cdot\vec r)}{r^7}+\nonumber\\
&&\hskip -10pt +\,
10({\vec s}\cdot{\vec m})({\vec s}\cdot{\vec k})\big(({\vec s}\cdot{\vec k})^2-({\vec s}\cdot{\vec m})^2\big)\frac{b^3}{r^7}+2({\vec s}\cdot{\vec m})({\vec s}\cdot{\vec k})\big(3-7({\vec s}\cdot{\vec k})^2\big)\frac{b}{r^5}\Big\}\Big|^r_{r_0}
+{\cal O}(J_5)  \bigg\}.
\label{eq:phiE0}
\end{eqnarray}

Note that a similar result for the quadrupole $J_2$ term  was obtained in \cite{Klioner:1991SvA,Klioner-Kopeikin:1992,Zschocke-Klioner:2010,Soffel-Han:2019}. Expression (\ref{eq:phiE0}) extends all the previous computations to the higher order terms including $J_3$ and $J_4$. In fact, this result is new. It generalizes a similar result obtained in \cite{Turyshev-Toth:2021-multipoles} (see Appendix therein) that was derived  for all orders of the zonal harmonics, $\ell$, in the case when the transmitter and receiver are at a very large distance from the gravitating body, $b\ll r$.

\section{Analytical estimates of gravitational phase shift terms}
\label{app:analest}

Contributions from the Earth's monopole (i.e., the Shapiro term) and quadrupole moments to the overall gravitational phase shift, as given by (\ref{eq:eq_eik-phi+}), can also be estimated analytically in specific cases.

\subsubsection{The monopole contribution  to the gravitational phase shift}

Considering the Shapiro term in the gravitational phase shift, we rely on the following exact relationships:
{}
\begin{eqnarray}
r+({\vec k}\cdot{\vec r})=\frac{(|{\vec r}-{\vec r}_0|+r)^2-r_0^2}{2|{\vec r}-{\vec r}_0|}, \qquad \qquad
r_0+({\vec k}\cdot{\vec r}_0)=\frac{r^2-(|{\vec r}-{\vec r}_0|-r_0)^2}{2|{\vec r}-{\vec r}_0|},
\label{eq:rel2}
\end{eqnarray}
which yield the following form for the Shapiro phase shift in (\ref{eq:phase_t2}):
{}
\begin{eqnarray}
\varphi_0(\vec r,\vec r_0)= kr_g \ln \Big[\frac{r+(\vec k\cdot \vec r)}{r_0+(\vec k\cdot \vec r_0)}\Big]=k
\frac{2GM}{c^2} \ln \Big[\frac{r+r_0+|{\vec r}-{\vec r}_0|}{r+r_0-|{\vec r}-{\vec r}_0|}\Big].
\label{eq:rel2b}
\end{eqnarray}

Expression (\ref{eq:rel2b}) can be used to conveniently evaluate two transmission scenarios. In the case of a horizontal transmission, when $r\simeq r_0$ and $(\vec n \cdot \vec n_0)=\cos2\alpha$, thus, $|{\vec r}-{\vec r}_0|\simeq 2r_0\sin\alpha$, the following approximation for the associated gravitational time delay (derived as usual $\Delta t= \varphi_0(\vec r,\vec r_0)/kc$) in the case of small $\alpha$ is valid:
{}
\begin{eqnarray}
\frac{2GM_\oplus}{c^3} \ln \Big[\frac{r+r_0+|{\vec r}-{\vec r}_0|}{r+r_0-|{\vec r}-{\vec r}_0|}\Big]\simeq \frac{4GM_\oplus}{c^3} \alpha \simeq \alpha \, 5.92\times 10^{-11} ~ {\rm s}.
\label{eq:impact-dhor0}
\end{eqnarray}
Similarly, in the case of a vertical transmission from a GPS spacecraft at the zenith to a ground-based receiver, when $r=r_0-h$ with $h\gg R_\oplus$, and $(\vec n \cdot \vec n_0)\simeq 1$, and, thus, $|{\vec r}-{\vec r}_0|\simeq h$, from (\ref{eq:rel2b}) we have
{}
\begin{eqnarray}
\frac{2GM_\oplus}{c^3} \ln \Big[\frac{r+r_0+|{\vec r}-{\vec r}_0|}{r+r_0-|{\vec r}-{\vec r}_0|}\Big]\simeq \frac{2GM_\oplus}{c^3} \ln \Big[1+\frac{h}{R_\oplus}\Big] \simeq  4.23\times 10^{-11} ~ {\rm s}.
\label{eq:impact-dvert}
\end{eqnarray}
The estimates (\ref{eq:impact-dhor0}) and (\ref{eq:impact-dvert}) demonstrate the significance of the Shapiro term when picosecond accuracy is desired.

\subsubsection{The quadrupole term in the gravitational phase shift}

We consider the $\ell=2$ contribution to the relativistic phase given by (\ref{eq:phase-sh-quad-prime-h+}) and, by explicitly expanding each of the terms, we present it in the following equivalent form:
{}
\begin{eqnarray}
\varphi_2(\vec r,\vec r_0)
&=& -kr_g  \bigg\{2\Big\{C'_{22}\cos2\phi_\xi+ S'_{22}\sin2\phi_\xi\Big\}
\Big(\frac{R_\oplus}{b}\Big)^2 \Big({\vec k}\cdot({\vec n}-{\vec n}_0)\Big)+\nonumber\\
&&\hskip -50pt+ \,
\Big\{C'_{22}\cos2\phi_\xi+ S'_{22}\sin2\phi_\xi-{\textstyle\frac{1}{2}} C'_{20}\Big\}
R^2_\oplus\Big( \vec k\cdot\Big( \frac{\vec n}{r^2}-\frac{\vec n_0}{r^2_0}\Big)\Big)
-\Big\{C'_{21} \cos\phi_\xi+
S'_{21}\sin\phi_\xi\Big\}
R^2_\oplus b\Big(\frac{1}{r^3}-\frac{1}{r^3_0}\Big)
\bigg\}.~~
\label{eq:phase-quad+}
\end{eqnarray}

We begin by evaluating the first term in this expression.  For that, using the definitions for the vectors $\vec k$ and $\vec b$ from (\ref{eq:k+b}), we obtain the following expression for the multiplier of the first term in (\ref{eq:phase-quad+}) that is explicitly expressed as a function of the transmitter and  receiver position vectors:
{}
\begin{eqnarray}
\Big(\frac{R_\oplus}{b}\Big)^2 \Big({\vec k}\cdot({\vec n}-{\vec n}_0)\Big)= \frac{R_\oplus^2}{rr_0}\Big(\frac{1}{r}+\frac{1}{r_0}\Big)\frac{|\vec{r}-\vec{r}_0|}{1+(\vec n\cdot\vec n_0)}\equiv
 \frac{R_\oplus^2}{rr_0}\Big(\frac{1}{r}+\frac{1}{r_0}\Big)\frac{(r+r_0)}{1+(\vec n\cdot\vec n_0)}\Big(1-\frac{2rr_0}{(r+r_0)^2}\Big(1+(\vec n\cdot\vec n_0)\Big)\Big)^\frac{1}{2},~~
\label{eq:impact-dd}
\end{eqnarray}
where for GPS transmissions in the Earth's vicinity expression $(\vec n\cdot\vec n_0)$ is never vanishes, i.e., $(\vec n\cdot\vec n_0)\not=0$.

We can now evaluate expression (\ref{eq:impact-dd}) for two specific transmission scenarios. In the case of a horizontal transmission, the following approximation for small $\alpha$ is valid:
{}
\begin{eqnarray}
\Big(\frac{R_\oplus}{b}\Big)^2 \Big({\vec k}\cdot({\vec n}-{\vec n}_0)\Big)&\simeq &
 \frac{R_\oplus^2}{r_0^2}\frac{4\sin\alpha}{1+\cos2\alpha}\simeq\frac{R_\oplus^2}{r_0^2}2\alpha.
\label{eq:impact-dhor}
\end{eqnarray}
This expression may be used to estimate the magnitude of the relevant contribution to  (\ref{eq:phase-quad+}):
{}
\begin{eqnarray}
2kr_g \Big\{C'_{22}\cos2\phi_\xi+ S'_{22}\sin2\phi_\xi\Big\}
\Big(\frac{R_\oplus}{b}\Big)^2 \Big({\vec k}\cdot({\vec n}-{\vec n}_0)\Big)&=&
4kr_g \Big\{C'_{22}\cos2\phi_\xi+ S'_{22}\sin2\phi_\xi\Big\}
\frac{\alpha R_\oplus^2}{(R_\oplus+h)^2} \lesssim
\nonumber\\
&\lesssim &
k c \Big(\alpha
 \Big\{C'_{22}\cos2\phi_\xi+ S'_{22}\sin2\phi_\xi\Big\}  ~ 6.80\times 10^{-12}~{\rm s}\Big).~~~~~
\label{eq:impact-dhor2}
\end{eqnarray}
Given the fact that the terms $C'_{22}$ and $S'_{22}$ are at least $\sim 1.08\times 10^{-3}$, this results suggest that the relevant contribution of the term (\ref{eq:impact-dhor2}) is much less than $\alpha\, 7.67\times 10^{-15}$~s, which is too small to consider.

Similarly, in the case of a vertical transmission, we have
{}
\begin{eqnarray}
\Big(\frac{R_\oplus}{b}\Big)^2 \Big({\vec k}\cdot({\vec n}-{\vec n}_0)\Big)&\simeq&
{\textstyle\frac{1}{2}}\Big(1+
 \frac{R_\oplus}{R_\oplus+h}\Big)\frac{h}{R_\oplus+h}\simeq {\textstyle\frac{1}{2}},
\label{eq:impact-dd2}
\end{eqnarray}
which yields
{}
\begin{eqnarray}
2kr_g \Big\{C'_{22}\cos2\phi_\xi+ S'_{22}\sin2\phi_\xi\Big\}
\Big(\frac{R_\oplus}{b}\Big)^2 \Big({\vec k}\cdot({\vec n}-{\vec n}_0)\Big)&\simeq&
kr_g \Big\{C'_{22}\cos2\phi_\xi+ S'_{22}\sin2\phi_\xi\Big\} \lesssim
\nonumber\\
&\lesssim &
k c \Big( \Big\{C'_{22}\cos2\phi_\xi+ S'_{22}\sin2\phi_\xi\Big\}  ~ 2.96\times 10^{-11}~{\rm s}\Big).~~~~~~
\label{eq:impact-dhors}
\end{eqnarray}
Similarly to (\ref{eq:impact-dhor2}), taking into account the magnitudes of the terms $C'_{22}$ and $S'_{22}$ from (\ref{eq:spij}),   this term is evaluated would contribute to the time delay the term on the order of $\sim8.00\times 10^{-15}$~s, which is also too small to consider. In fact, considering various transmission architectures and the angles $\theta$ and $\psi$ involved in the definitions of $C'_{22}$ and $S'_{22}$,  from (\ref{eq:spij}), we  estimate that both  of these  transmission regimes result in small corrections on the order of $\sim$0.01 ps.

Now we will evaluate the second term in (\ref{eq:phase-quad+}). Again using the expression for  $\vec k$ from (\ref{eq:k+b}) and expressing $|\vec r- \vec r_0|$, we have the following expression for the multiplier of the second term in (\ref{eq:phase-quad+}):
  {}
\begin{eqnarray}
R^2_\oplus\Big( \vec k\cdot\Big( \frac{\vec n}{r^2}-\frac{\vec n_0}{r^2_0}\Big)\Big)&=&
\frac{R^2_\oplus}{rr_0}\frac{ 1+\Big(1-\frac{r^2+r_0^2}{rr_0}\Big)(\vec n\cdot\vec n_0)}{\Big(1-\frac{2rr_0}{(r+r_0)^2}\Big(1+(\vec n\cdot\vec n_0)\Big)\Big)^\frac{1}{2}}.
\label{eq:impact-er2}
\end{eqnarray}

Evaluating this result for the case of horizontal transmission, we see that (\ref{eq:impact-er2}) yields the approximate expression that is identical to  (\ref{eq:impact-dhor}). In the case of  vertical transmission,  the magnitude of the result is twice that of (\ref{eq:impact-er2}). Thus, we can see that the contribution in (\ref{eq:phase-quad+}) from both transmission cases would be below $10^{-15}$~s, which is  negligible.

Finally, relying on the same approach as above by using expressions for the vectors $\vec k$ and $\vec b$ from (\ref{eq:k+b}), we see that $ b=|[\vec r\times \vec r_0]|/|\vec r- \vec r_0|$, that allows us to develop the following expression for the multiplier of the third term in (\ref{eq:phase-quad+}):
 {}
\begin{eqnarray}
R^2_\oplus b\Big(\frac{1}{r^3}-\frac{1}{r^3_0}\Big)&=&
R^2_\oplus\frac{rr_0}{r+r_0}\frac{|[\vec n\times \vec n_0]|}{\Big(1-\frac{2rr_0}{(r+r_0)^2}\Big(1+(\vec n\cdot\vec n_0)\Big)\Big)^\frac{1}{2}}\Big(\frac{1}{r^3}-\frac{1}{r^3_0}\Big).
\label{eq:impact-s2}
\end{eqnarray}
We can see that this expression provides a negligible contribution for both transmission cases either horizontal or vertical. For the horizontal transition, this is due to the fact that for $r\simeq r_0$, we have  $r^{-3}-r^{-3}_0\simeq0$. In the case of  the vertical transmission, $(\vec n \cdot \vec n_0)\simeq1$ and thus $[\vec n\times \vec n_0]\simeq 0$. In all the intermediate cases, the contribution of this term to (\ref{eq:phase-quad+}) is well below $10^{-15}$ s and, thus, this term may be neglected for present day terrestrial GPS applications.

\end{document}